\documentclass[12pt]{article}
\pdfoutput=1
\usepackage[nosort]{cite}
\usepackage[height=8.85in,width=6.45in]{geometry}
\usepackage{pifont}

\usepackage{times}
\usepackage{tikz}
\usepackage{fix-cm}
\usepackage{comment}
\usepackage{booktabs}
\usepackage[utf8]{inputenc}
\usepackage{amsmath}
\usepackage{amssymb}
\usepackage{mathtools}
\numberwithin{equation}{section}
\usepackage{slashed}
\usepackage{braket}
\usepackage{xcolor}
\usepackage{url}
\usepackage{hyperref}
\usepackage{cite}
\usepackage{graphicx}
\usepackage{float}
\usepackage{xfrac}
\usepackage{courier}
\usepackage{bm}
\usepackage{caption}
\usepackage{subcaption}
\usepackage{enumitem}
\usepackage{footmisc}
\usepackage{blkarray}
\usepackage{arydshln}
\usepackage{dsfont}
\usepackage{calc}
\usepackage{accents}

\renewcommand{\title}[1]{\vbox{\center\LARGE{#1}}\vspace{5mm}}
\renewcommand{\author}[1]{\vbox{\center#1}\vspace{5mm}}
\newcommand{\address}[1]{\vbox{\center\em#1}}

\newcommand{\sps}{\mathrm{spin}\text{-}s}
\makeatletter
\newsavebox{\@brx}
\makeatother

\begin{document}

\begin{titlepage}

\title{Defect Anomalies, a Spin-Flux Duality, and Boson-Kondo Problems}

\author{Zohar Komargodski${}^1$, Fedor K. Popov${}^1$, Brandon C.\ Rayhaun${}^{2,3}$}

\address{${}^{1}$Simons Center for Geometry and Physics, Stony Brook University, Stony Brook, NY\\
${}^{2}$C.\ N.\ Yang Institute for Theoretical Physics, Stony Brook University, Stony Brook, NY\\
${}^{3}$School of Natural Sciences, Institute for Advanced Study, Princeton, NJ 
 }

\abstract

We show that the infrared phases of certain line defects in 2+1d quantum field theories are determined by anomalies, including anomalies in the space of defect coupling constants, together with a symmetry-refined corollary of the $g$-theorem.

As an example, we prove that the spin-$\sfrac{1}{2}$ impurities in the 2+1d critical $O(2)$ and $O(3)$ models---known respectively as the Halon and Boson-Kondo defects---flow to non-trivial conformal line operators in the IR, and we supply evidence that the same extends to all spin $s$. 
We also argue that, under particle/vortex duality,  the Halon impurity is exchanged with the $\pi$-flux vortex line leading to \emph{spin-flux duality}, a proposal which we test with a detailed matching of symmetries, anomalies, and phases.  

Finally, we write down quantum lattice Hamiltonians which can be used to test our predictions, and give an argument on the lattice in favor of spin-flux duality.

\end{titlepage}

\eject

\setcounter{tocdepth}{2}
\tableofcontents

\section{Introduction}

There is a long and fruitful history of using symmetries and  anomalies to constrain the dynamics of quantum field theories and lattice models. The idea of anomaly matching goes back to 't Hooft \cite{tHooft:1979rat}. A theory with an 't Hooft anomaly cannot be trivially gapped at long distances while preserving the symmetry. Anomalies are  relatively tractable and thus furnish   non-trivial tests of many ideas about strong coupling dynamics.

Various generalizations of the notions of symmetry and anomaly have been considered in recent years.
One direction to emerge from these developments is a renewed interest in the interplay between symmetries and topological aspects of parameter spaces of quantum field theories and lattice models \cite{Gaiotto:2017yup,Kikuchi:2017pcp,Tanizaki:2018xto,Karasik:2019bxn,Cordova:2019jnf,Cordova:2019uob,Hsin:2020cgg,Kapustin:2020eby,Kapustin:2022apy,Artymowicz:2023erv,Choi:2022odr,Wen:2021gwc,Beaudry:2023qyg,Ohyama:2022cib,Ohyama:2023suc,Qi:2023ysw,Shiozaki:2023xky,Sommer:2024dtb,Sommer:2024lzp,Brennan:2024tlw}. Of particular salience for our work is the notion of an anomaly in the space of coupling constants \cite{Gaiotto:2017yup,Kikuchi:2017pcp,Tanizaki:2018xto,Karasik:2019bxn,Cordova:2019jnf,Cordova:2019uob} (also called ``global inconsistency''). 
These are quantum field theoretic generalizations of the notion of Berry phase \cite{Berry:1984jv}, and they have interesting implications for the low-energy physics, similar to 't Hooft anomalies.

Extended operators (to be contrasted with local operators) have also attracted attention lately. (We will use the terms ``extended operators,'' ``defects,'' and ``impurities'' interchangeably.) They have numerous applications. For example, Wilson lines and 't Hooft lines  \cite{Wilson:1974sk,tHooft:1977nqb} are important in diagnosing phases of gauge theories. Defects are also used for modeling impurities in condensed matter physics (this is a vast field with many  different examples, see for instance~\cite{Andrei:2018die} for a few references and many additional references below). They have interesting recent appearances in quantum information theory as well (see for example~\cite{garratt2023measurements,Lee:2023fsk}). Moreover, the collection of extended objects is part of the defining data of a quantum field theory, and is therefore a crucial part of our quest to understand what quantum field theory is. 

Quite a bit is known about the space of local operators, but comparatively little is known about extended operators, with the exception of topological defects, which have been studied a lot in recent years due to their relationship with generalized symmetries \cite{Gaiotto:2014kfa}. Non-topological defects are equally if not more more important, albeit, under far less control. 

Another non-perturbative property of quantum field theory and defects in quantum field theory is the irreversibility of the Renormalization Group (RG) flow \cite{Zamolodchikov:1986gt,Cardy:1988cwa,Komargodski:2011vj,Komargodski:2011xv,Affleck:1992ng,Affleck:1991tk,Jafferis:2011zi,Casini:2012ei,Myers:2010tj,Elvang:2012st,Fluder:2020pym,Cuomo:2021rkm,Friedan:2003yc,Wang:2021mdq,Gaiotto:2014gha,Casini:2018nym,Casini:2022bsu,Casini:2023kyj}.
It is proven in many, but not all cases. In certain situations, RG flows on defects are known to be irreversible. 

The goal of this paper is to weave these various threads together.
We will combine  constraints from anomalies and anomalies in the space of (defect) coupling constants, as well as the irreversibility of the renormalization group flow,
to paint compelling pictures about the phase diagrams of defects in well-known lattice models and quantum field theories. See e.g.\ \cite{Liu:2021nck,Aharony:2022ntz,Delmastro:2022pfo,Brennan:2022tyl,Antinucci:2024izg,Brennan:2025acl} for previous work which studies the interplay between symmetries, anomalies, and defects.

Our main focus is what one might call Boson-Kondo defects, which are essentially spin impurities in various magnets at criticality. For some of the earlier literature on such magnetic impurities see~\cite{sachdev1999quantum,Vojta_2000,zhu2002critical,sachdev2004quantum,vojta2006impurity}.
The original Kondo problem \cite{Kondo:1964nea,anderson1970poor,Wilson:1974mb} was historically important because it became a paradigmatic example of strong coupling physics, influencing many areas of research.  More recently, attention has turned to defects in 2+1 dimensions, including bosonic analogues, such as the Bose–Kondo problem, on which we focus  here. The impurity spin couples not to a Fermi sea but to gapless bosonic modes—e.g.\ critical spin fluctuations  in 2+1 dimensional quantum magnets. This problem captures the interplay between impurity dynamics and critical bosonic baths.

We will also discuss vortex loops in gauge theories. Vortex loops are defined by fixing the gauge field holonomy at the impurity. 
These are important defects in spin liquid phases~\cite{He:2016mbo}, in supersymmetric theories~\cite{Drukker:2012sr,Assel:2015oxa}, and elsewhere. Conceptually, vortex loops describe the insertion of Aharonov-Bohm solenoids and we investigate their fate in the infrared.
We will study the anomalies and deformations of vortex loop defects and compare them to Boson-Kondo defects.

Similarly to  bulk phases of many body systems, line defects can be in one of several phases. A {\bf trivial} phase of a defect---in which it flows to the identity operator at long distances---is analogous to a trivially gapped phase in the bulk. A {\bf TQM$_k$} phase (topological quantum mechanics with $k$ ground states)---in which the line flows to a direct sum of $k$ superselection sectors/$k$ copies of the identity operator---is analogous to spontaneous symmetry breaking in the bulk. A {\bf topological} phase---in which the defect flows to a topological line operator---is analogous to topological order in the bulk. Finally, a {\bf conformal} phase---where the system flows to a non-trivial defect conformal field theory---is analogous to a gapless bulk phase. See Section \ref{subsec:defects/RGflows} for further details on this classification.

Some effort has gone into simulations of Boson-Kondo defects in an attempt to understand which of the above phases they realize at long distances.
In this work, we show that anomalies, in conjunction with results about the irreversibility of the renormalization group flow, in some cases {\it determine} the answer, and in other cases, as we said, paint a compelling picture for the phase diagram.

\subsection{Boson-Kondo defects in the \texorpdfstring{$O(2)$}{O(2)} and \texorpdfstring{$O(3)$}{O(3)} models}
Now we briefly summarize our results.
Our first example involves the 2+1d $O(3)$ Wilson-Fisher model at criticality. Denote the order parameter $\phi^a$ with $a=1,2,3$. We deform the bulk Hamiltonian as
\begin{align}
    H_{\mathcal{D}} = H_{O(3)}+h\phi^a(t,\vec{x}=0) S_a+\gamma S_z,
\end{align} 
where $S^a$ acts on the Hilbert space of an impurity spin in the $2s+1$ dimensional representation, $h$ is the coupling between the impurity and the bulk, and $\gamma$ is an external magnetic field acting on the impurity.
We can also represent this in a continuum language, starting from the 
critical $O(3)$ model in 2+1d, 
\begin{align}\label{eqn:O(3)model}
    S_{O(3)}=\int d^3x \left( \frac12\partial_\mu\phi^a \partial^\mu\phi^a -\frac{\lambda}{4!} (\phi^a\phi^a)^2\right),
\end{align}
and adding a coupling%
\begin{align}\label{eqn:spinimpurity}
    S= S_{O(3)} + S_{\sps} + \int d\tau\left( h\phi^a(\tau,\vec{0}) S_a(\tau)+\gamma S_z(\tau)\right).
\end{align}
Here, $S_{\sps}$ is the quantum mechanical action of a spin $s$ representation, see Section \ref{subsec:quantuminterlude}.

Another point of view on Equation \eqref{eqn:spinimpurity} is that it represents the insertion of a line defect
\begin{equation}\label{linedefect}
\operatorname{Tr}_sP e^{ \int_\gamma \left(h \phi^a T_a+\gamma T_z\right)}~,
\end{equation}
with $T_a$ matrices in the spin $s$ representation of $\mathfrak{su}(2)$, and $P$ denoting path ordering. This is essentially a non-Abelian scalar line, technically similar to a Wilson line.

For generic values of $\gamma$, the defect preserves, in particular, a $U(1)$ symmetry which simultaneously rotates the $\phi^a$ and the spin $S_a$ of the impurity particle in the $xy$ plane. We are interested in the interplay of this $U(1)$ symmetry with the topology of the parameter space of the defect corresponding to the coupling $\gamma$. 
The first claim is that the $\gamma\to +\infty$ and $\gamma\to -\infty$ impurities are in a different $U(1)$ SPT phase,
\begin{align}\label{eqn:O(3)anomalydefectcoupling}
  \lim_{\gamma\to\infty} \frac{Z_{\gamma}[A]}{Z_{-\gamma}[A]} = \exp\left( 2is\int d\tau  A_\tau(\tau,\vec{0})   \right),
\end{align}
where $A$ is a background gauge field for the $U(1)$ symmetry.\footnote{Recall that 0+1d $U(1)$ SPT phases are labeled by $H^1(U(1),U(1))\cong \mathbb{Z}$, which is the level of 1d Chern-Simons term for the background gauge field.}
This SPT jump already has some implications, for instance: \\

\noindent\textbf{Claim:} \emph{For every spin $s$, as one tunes the defect parameter $\gamma$, one must encounter at least one diabolical value $\gamma_\ast$ for which the line defect \eqref{eqn:spinimpurity} is non-trivial (i.e.\ not simply the identity line operator) in the IR.} \\

\noindent We think of Equation \eqref{eqn:O(3)anomalydefectcoupling} as an anomaly in the space of defect coupling constants.

When $s$ is a half-integer, one can make a much stronger statement. Setting $\gamma_\ast=0$, the symmetry preserved by the impurity enhances to include an $SO(3)$ which is realized anomalously on the defect in the sense that the endpoint operators transform in a projective representation (see Section \ref{subsec:defectanomalies} for a review of defect 't Hooft anomalies). Thus, we can immediately conclude that $\gamma_\ast=0$ is a diabolical point for half-integer $s$. That is, at $\gamma_\ast=0$ we must have either a TQM phase or a conformal defect.\footnote{A topological phase of the line defect is ruled out because the $O(3)$ model is not known to possess any one-form symmetries.} Note that for half-integer $s$, the SPT phase jumps by an odd integer and hence the $U(1)$ charge is a half-integer in the phases with $\gamma\neq 0$, i.e.\ the impurity has fractionalized charge, unlike any bulk excitation. 

As we explain in Section \ref{sec:magnets}, if we combine the observations thus far with a further application of (a refinement of) the $g$-theorem \cite{Cuomo:2021rkm}, Equation \eqref{genco}, we arrive at the following:\\

\noindent \textbf{Claim}: \emph{The $s=\sfrac12$ spin impurity with $\gamma_\ast=0$ in the $O(3)$ model---a.k.a.\ the Boson-Kondo defect---defines a non-trivial \underline{conformal} defect in the IR with $1<g<2$ (the defect entropy is $0<s<\log 2$).} \\

\noindent Spin impurities have also been studied in a large $s$ expansion, which is essentially an expansion in large quantum numbers \cite{Cuomo:2022xgw} (see also~\cite{Beccaria:2022bcr,Rodriguez-Gomez:2022gbz,Nahum:2022fqw,Giombi:2022anm,Rodriguez-Gomez:2022xwm}), where it was found that the defect with $\gamma_\ast=0$ is  conformal for $s$ large enough. In light of the new input coming from anomalies and the $g$-theorem, the most economic conjecture is that this behavior holds for all $s$:\\

\noindent \textbf{Conjecture}: \emph{When $\gamma_\ast=0$, the spin impurity \eqref{eqn:spinimpurity} in the $O(3)$ model defines a non-trivial conformal line defect for every spin $s$.}\\

\begin{figure}
\begin{center}
\input{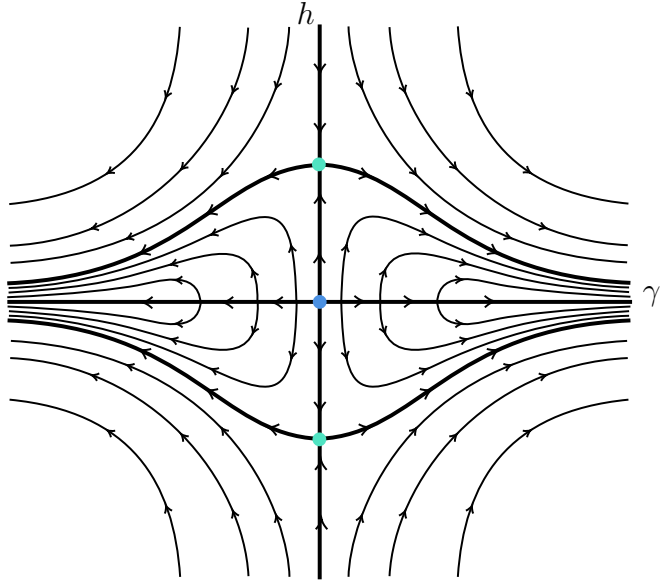}
\caption{RG flow diagram of a spin-$s$ impurity in the $O(3)$ model. The fixed point at the origin represents a decoupled quantum mechanical spin-$s$ qudit (TQM$_{2s+1}$), while the remaining two fixed points with $h_\ast \neq 0$ are conformal.}\label{fig:RGflow}
\end{center}
\end{figure}

\noindent A sketch of the RG flows for spin impurities in the $O(3)$ model is given in Figure \ref{fig:RGflow}.
Our proposals should be possible to test using the lattice Hamiltonian for the defected $O(3)$ model that we provide in Section \ref{subsec:O(3)bosonkondoproblem}. For some simulations see~\cite{wang2006high,luscher2005long,hoglund2007anomalous,shinkevich2011spin}, among many others.

The reasoning above applies also to spin impurities in the critical $O(2)$ model (the XY model). We take the complex order parameter of the $O(2)$ transition, $\Phi$, and couple it to a spin impurity.
The corresponding action is
\begin{align}\label{eqn:O(2)spinimpurity}
    S=S_{O(2)}+S_{\sps}+\int d\tau \left( h \Phi(\tau,\vec{0})S_-(\tau)+ h \Phi^\ast(\tau,\vec{0}) S_+(\tau)+ \gamma S_z(\tau)\right),
\end{align}
with $S_{O(2)}$ the usual $O(2)$ model tuned to the critical point,
\begin{align}
    S_{O(2)}= \int d^2xd\tau \left(\partial_\mu \Phi^\ast \partial^\mu \Phi-\lambda |\Phi|^4\right).
\end{align}
Again, using anomalies and the $g$-theorem, we prove in Section \ref{subsec:XY} that the $\gamma_\ast=0$ spin impurity with $s=\sfrac12$  flows to a non-trivial conformal line defect in the $O(2)$ model. Following \cite{Whitsitt:2017pmf,Chen:2018xqa}  we call it the ``Halon'' defect. As in the previous example, in the phases with $\gamma\neq0$ it carries half-integer $U(1)$ charge, unlike any bulk excitation, i.e.\ it is fractionalized.

\subsection{A spin-flux duality}
Since the $O(2)$ model is particle-vortex dual to the Abelian-Higgs model \cite{Peskin:1977kp,Dasgupta:1981zz,Nguyen:1999zn,Kajantie:2004vy,Seiberg:2016gmd,Karch:2016sxi},
\begin{align}
    S=\int d^3x \left(-\frac{1}{4e^2}f^2_{\mu\nu}+ \left|D_\mu\Phi\right|^2 - V(|\Phi|^2)\right),
\end{align}
we can ask what the dual description is of the Halon.

We propose that the dual is a vortex line. Vortex lines are disorder operators, defined by imposing singular boundary conditions for the fundamental fields in the path integral. In the present case, we impose that the holonomy of the gauge field $a$, taken with respect to small loops $C_l$ of length $l$ about $x=y=0$, is fixed, 
\begin{align}\label{eqn:vortexline}
    \lim_{l\to 0}\,\exp\left(i\oint_{C_l} a\right)=e^{i\alpha}.
\end{align}
We think of $\alpha$ as a parameter of the defect; clearly, it is $2\pi$-periodic. The vortex line defect is symmetric under the magnetic $U(1)$ zero-form symmetry, in the sense that the topological surfaces which implement the symmetry admit  topological point junctions with \eqref{eqn:vortexline}, as in Figure \ref{parfusion}. 

The main point is that as we change $\alpha\to\alpha+2\pi$, the SPT phase on the vortex line for the magnetic $U(1)$ zero-form symmetry jumps by one unit, 
\begin{align}
    Z_{\alpha+2\pi}[A]=Z_\alpha[A]\exp\left(-i\int d\tau A_\tau \right),
\end{align}
where $Z_\alpha[A]$ is the partition function in the presence of a background gauge field $A$ and a vortex line with holonomy $\alpha$. We again think of this as an anomaly in the space of defect couplings. This anomaly implies that there is at least one diabolical point on the $\alpha$ circle. The proof that the SPT phase jumps is simply because $\alpha=2\pi$ and $\alpha=0$ differ by a (local) monopole operator, hence one unit of $U(1)$ magnetic charge. 

At $\alpha=0,\pi$, the $U(1)$ symmetry of the vortex line enhances to an $O(2)$ due to charge conjugation symmetry. At $\alpha_\ast=\pi$, this $O(2)$ acts projectively on the endpoints of the line. In particular, there is a degenerate doublet of ``half-monopoles'', i.e.\ monopole operators with magnetic $U(1)$ charges given by $\pm\sfrac12$. Thus, from the 't Hooft anomaly, we can conclude that $\alpha_\ast=\pi$ is certainly a diabolical point in parameter space. In fact, we have the following conjecture.\\ 

\noindent\textbf{Conjecture (Spin-Flux Duality)}: The $\alpha_\ast=\pi$ vortex line of the Abelian-Higgs model \eqref{eqn:vortexline} is dual to the Halon spin impurity of the $O(2)$ model \eqref{eqn:O(2)spinimpurity} with $s=\sfrac12$ and $\gamma_\ast=0$.
 The deformation $\gamma S_z$ corresponds to the deformation $\delta\alpha = \alpha - \pi$. For $\alpha\neq0,\pi$, there is an RG flow with $\alpha=0$ an attractive fixed point and $\alpha=\pi$ a repulsive one. \\

\noindent An immediate corollary of the spin-flux duality conjecture, combined with  our theorem about the Halon, is that the $\alpha_\ast=\pi$ vortex line defines a non-trivial conformal defect.

We test this proposal by matching symmetries, anomalies, and phases on both sides of the duality. (In particular, that the deformation $\delta \alpha$ leads to a trivial vortex loop makes sense in light of~\cite{upcoming}, where it is shown that the currents tend to create opposing magnetic fields that screen the solenoid, unlike in supersymmetric theories.) We also present concrete lattice models, in Section \ref{subsubsec:xylattice} and Section \ref{subsec:vortexlattice}, that realize the Halon and the vortex loop, respectively. We show that the lattice models realize the  anomalies predicted by the continuum, and  should therefore allow for a test of our proposal, in principle. It should also be possible to use the $O(2)$ invariant defect bootstrap~\cite{Gimenez_Grau_2022}.

\subsection{Stable non-simple defects in the Ising CFT}

Finally, let us summarize some defect anomalies in the 2+1d Ising CFT. 
Consider the Ising order parameter $\phi$ at criticality and couple it to the $z$-component of a spin impurity: $H_{\mathcal{D}} = H_{\mathrm{Ising}}+ \gamma \phi S_z+\cdots$. The two components of the spin are decoupled and this clearly flows to a non-simple conformal defect
\begin{equation}\label{dsumintro}
    \mathcal{D}[+]\oplus \mathcal{D}[-],
\end{equation}
where $\mathcal{D}[\pm]$ are the pinning field defects, about which a lot is already known both from analytical and numerical work 
\cite{Vojta_2000,Sachdev_2003,Florens_2006,Assaad_2013,Parisen_Toldin_2017,allais2014magneticdefectlinecritical,Cuomo_2022,Gimenez-Grau:2022ebb,Popov:2022nfq,Franchi_2022,Rodriguez_Gomez_2022,Gimenez_Grau_2022,Giombi:2022vnz,Bianchi_2023,Pannell_2023,Hu_2024,Zhou_2024,Dey:2024ilw,Popov:2025cha,barrat2025linedefectcorrelatorsfermionic,coordinate}. 

If we add a perturbation that allows spin transitions, for instance  $\delta H_{\mathcal{D}} = h S_x$, the defect becomes trivial. However, we prove that any deformation preserving $\mathbb{Z}^T_2\times \mathbb{Z}_2$, realized in a particular way on the impurity spin and in the usual way on the Ising order parameter, cannot bring the defect to a trivial one. This is proven by showing that in the presence of the defect there is an anomaly for $\mathbb{Z}^T_2\times \mathbb{Z}_2 $ (of course there is no such anomaly in the bulk critical Ising model without a defect).

The deformation $\delta H_\mathcal{D} = hS_x$ is indeed not invariant under our $\mathbb{Z}^T_2\times \mathbb{Z}_2 $, consistent with the anomaly.
This gives an example of stable spontaneous symmetry breaking on a line defect immersed in an interacting bulk bath.\footnote{\label{SSBline}When we say ``spontaneous symmetry breaking'' on a line defect, there is of course a certain abuse of terminology since there is no large volume limit on a point like impurity in space. What is often meant by this terminology is that the non-simple line defect is RG stable under symmetric perturbations.} Note that the fact that the relevant deformation $hS_x$  leads to a trivial defect is compatible with the $g$-theorem, because $g(\mathcal{D}[+]\oplus\mathcal{D}[-])>1$ (as we will explain in Section \ref{subsec:Ising}). On the other hand, for the analogous spin-$\sfrac12$ impurity in the $O(N)$ model, $g(\mathcal{D}[\hat{h}]\oplus \mathcal{D}[-\hat{h}])<1$ when $N$ is large enough, and thus we might expect that all $\mathbb{Z}_2$-symmetric operators (including $S_x$) are irrelevant.\footnote{This defect can be realized in a classical statistical mechanics system, showing that spontaneous symmetry breaking on a 1d line is possible when it is embedded into a gapless bath, contrary to the situation for isolated 1d systems \cite{peierls1936ising}.}

\subsection{Outline}

The outline of this paper is as follows. In Section \ref{sec:generaltechniques} we define the precise possible infrared limits of defects in a CFT, their 't Hooft anomalies, and anomalies in defect parameter  spaces. We also discuss a symmetry-refined corollary of the irreversibility theorem. In Section \ref{sec:magnets}, we discuss in detail magnetic defects in the Ising, XY, and $O(3)$ models, including their lattice realizations and our predictions for their infrared limits. Finally in Section \ref{sec:vortexlines}, we comment on vortex lines and our proposed duality with the Halon. We also include a lattice realization of the vortex loop defect.\\

\noindent\textbf{Note added:} We would like to thank Ryan Lanzetta, Shang Liu, and Max Metlitski for coordinating the submission of their related  paper \cite{coordinate} to the arXiv with us. While this manuscript was in its final stages of preparation, the papers \cite{Choi:2025ebk,Wen:2025xka,Copetti:2025sym} appeared, which also discuss the notion of an anomaly in the space of defect couplings.

\section{General Techniques} \label{sec:generaltechniques}

In this section, we lay out the tools that we will use throughout the rest of the paper. In particular, in Section \ref{subsec:defects/RGflows}, we review the basics of defects in quantum field theory and their renormalization group (RG) flows. In Section \ref{subsec:defectanomalies}, we describe 't Hooft anomalies of bulk global symmetries in the presence of symmetry-preserving line defects, and explain some of their physical consequences. In Section \ref{subsec:anomaliesinspaceofdefectcouplings}, we introduce the distinct (but related) notion of an ``anomaly in the space of defect couplings'', using the 1+1d compact free boson CFT as an illustrative example. 

\subsection{Defects and their RG flows}\label{subsec:defects/RGflows}
Consider a quantum theory in $d+1$ spacetime dimensions.
A defect $\mathcal{D}$ is an extended operator with support on a submanifold of spacetime dimension $p+1$. 
A useful perspective is to orient one of the dimensions of the defect in the time direction. Then, the defect can be viewed as a modification of the Hilbert space $\mathcal{H}$ and the Hamiltonian $H$ of the theory,
\begin{equation}\label{Hami} \mathcal{H}\to \mathcal{H}_{\mathcal{D}}, \ \ \ \ \ H\to H_{\mathcal{D}}. \end{equation}

The defect $\mathcal{D}$ is not entirely unambiguous, e.g.\ the zero-point energy of $H_{\mathcal{D}}$ is scheme dependent, corresponding to the counterterm $\mathcal{D}\to \mathcal{D} e^{i \int dtd^{p}x  \Lambda}$. Other counterterms exist as well, sensitive to the curvature of the extended operator $\mathcal{D}$. 

For a straight $\mathcal{D}$ with $p\geq 0$, the normalization of $\mathcal{D}$ is physical. Indeed, we cannot multiply $\mathcal{D}$ by, say, $\sqrt2$, since we would lose the physical meaning of a Hilbert space in~\eqref{Hami}. On the other hand, multiplying $\mathcal{D}$ by an integer $n$ can be interpreted as taking a direct sum of $\mathcal{D}$ with itself $n$ times (see below), and hence changes the defect. The normalization of $\mathcal{D}$ will be of interest later.

An important allowed operation is the direct sum, 
\begin{equation}\label{directsum} \mathcal{D}_1\oplus \mathcal{D}_2\oplus \cdots~\end{equation}
corresponding to defect superselection sectors. (The same defect can appear multiple times in \eqref{directsum}.) By definition, the Hilbert space and Hamiltonian associated to the defect $\mathcal{D}_1\oplus\mathcal{D}_2\oplus\cdots$ are  
\begin{align}
   \mathcal{H}_{\mathcal{D}_1\oplus\mathcal{D}_2\oplus\cdots}\equiv \mathcal{H}_{\mathcal{D}_1}\oplus \mathcal{H}_{\mathcal{D}_2}\oplus \cdots, \ \ \ \ H_{\mathcal{D}_1\oplus\mathcal{D}_2\oplus \cdots} \equiv \left(\begin{array}{ccc} H_{\mathcal{D}_1} & 0 & 0 \\ 0 & H_{\mathcal{D}_2} & 0\\
   0 & 0 & \ddots \end{array}\right).
\end{align}
Another allowed operation is  defect fusion  which corresponds to bringing two defects in parallel to each other, leading  to some new effective defect \begin{equation}\mathcal{D}_1\otimes \mathcal{D}_2\end{equation} with certain (exponential) singularities associated to the fusion \cite{Bachas:2007td,Bachas:2013ora,Konechny:2015qla,Diatlyk:2024zkk,Cuomo:2024psk}. (The fusion could also contain the trivial defect.)

The case of a gapped bulk $H$ reduces to the study of defects in topological quantum field theory (TQFT), which is a mature subject by now. Our interest in this problem is mainly when $H$ is gapless, i.e.\ when it defines a conformal field theory (CFT). Very little is known about the space of extended operators in gapless theories (with the exception of topological defects, which have been studied intensely following the introduction of the notion of generalized global symmetries \cite{Gaiotto:2014kfa}). On the other hand, they are important probes of the dynamics and appear commonly in simulations and experiments.  

\begin{figure}
    \begin{center}
        \tikzset{every picture/.style={line width=0.75pt}} 

\begin{tikzpicture}[x=0.75pt,y=0.75pt,yscale=-1,xscale=1,scale=.9]

\draw  [fill={rgb, 255:red, 184; green, 233; blue, 134 }  ,fill opacity=1 ][line width=1.5]  (9.08,138.21) .. controls (9.08,64.64) and (68.72,5) .. (142.29,5) .. controls (215.86,5) and (275.5,64.64) .. (275.5,138.21) .. controls (275.5,211.78) and (215.86,271.42) .. (142.29,271.42) .. controls (68.72,271.42) and (9.08,211.78) .. (9.08,138.21) -- cycle ;
\draw  [fill={rgb, 255:red, 80; green, 227; blue, 194 }  ,fill opacity=1 ] (25,154.67) .. controls (25,98.48) and (70.56,52.92) .. (126.75,52.92) .. controls (182.94,52.92) and (228.5,98.48) .. (228.5,154.67) .. controls (228.5,210.87) and (182.94,256.42) .. (126.75,256.42) .. controls (70.56,256.42) and (25,210.87) .. (25,154.67) -- cycle ;
\draw  [fill={rgb, 255:red, 74; green, 144; blue, 226 }  ,fill opacity=1 ] (38,175.25) .. controls (38,138.66) and (67.66,109) .. (104.25,109) .. controls (140.84,109) and (170.5,138.66) .. (170.5,175.25) .. controls (170.5,211.84) and (140.84,241.5) .. (104.25,241.5) .. controls (67.66,241.5) and (38,211.84) .. (38,175.25) -- cycle ;
\draw  [fill={rgb, 255:red, 65; green, 117; blue, 5 }  ,fill opacity=1 ] (50,188.75) .. controls (50,167.9) and (66.9,151) .. (87.75,151) .. controls (108.6,151) and (125.5,167.9) .. (125.5,188.75) .. controls (125.5,209.6) and (108.6,226.5) .. (87.75,226.5) .. controls (66.9,226.5) and (50,209.6) .. (50,188.75) -- cycle ;

\draw (111,24) node [anchor=north west][inner sep=0.75pt]   [align=left] {conformal};
\draw (92,79) node [anchor=north west][inner sep=0.75pt]   [align=left] {topological};
\draw (74,128) node [anchor=north west][inner sep=0.75pt]   [align=left] {TQM$_k$};
\draw (64,181) node [anchor=north west][inner sep=0.75pt]   [align=left] {trivial};

\end{tikzpicture}
        \caption{The hierarchy of possible IR behaviors of line defects.}\label{fig:IRpossibilities}
    \end{center}
\end{figure}
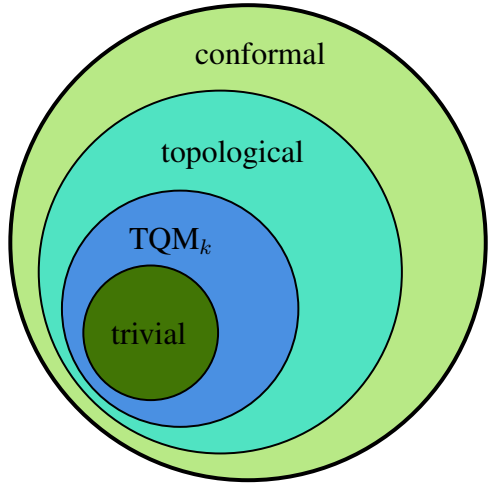

We now describe in more detail the case of $p=0$, when the impurity is a point in space and $\mathcal{D}$ is a line operator. Even if $H$ is fine tuned to a fixed point, there is a renormalization group flow on $\mathcal{D}$,
\begin{equation}\label{eqn:RGflow}\mathcal{D}_{\rm UV}\longrightarrow \mathcal{D}_{\rm IR}~.\end{equation}
A common problem is, given a $\mathcal{D}_{\mathrm{UV}}$,  to determine
the IR fate of the RG flow in \eqref{eqn:RGflow}. There are several possibilities, that are summarized in Figure \ref{fig:IRpossibilities}.\footnote{It may be the case that the RG flow does not terminate on any kind of healthy endpoint, whether it be trivial, TQM, topological, or conformal. Such a scenario might be referred to as \textbf{runaway}, see e.g.\ \cite{Cuomo_2022} for an example in free theory with linear coupling on a defect. This somewhat exotic behavior is not expected to arise in any of the examples we encounter in this paper.}
\begin{enumerate}[label=\arabic*)]
    \item The line may flow to the unit operator, 
    \begin{align}\label{eqn:trivialdefect}
        \mathcal{D}_{\mathrm{IR}}=\mathds{1},
    \end{align}
    in which case it is transparent to an observer situated at long distances away from the defect. We say that the defect is \textbf{trivial} in the IR.
    \item The line may flow to multiple copies of the unit operator, 
    \begin{align}\label{eqn:decoupleddefect}
        \mathcal{D}_{\mathrm{IR}}=\mathds{1}\oplus\mathds{1}\oplus\cdots\oplus\mathds{1}.
    \end{align}
    In this scenario, the defect supports a decoupled TQFT in 0+1-dimensions, in the sense that it has multiple topological local operators on its worldline, but it does not interact in an interesting way with the bulk. We  say that the defect is \textbf{TQM$_k$} (topological quantum mechanics with $k$ ground states) in the IR. (Technically, a defect which is trivial in the IR is TQM$_1$.)
    \item The line may flow to a (not necessarily simple) topological operator, 
    \begin{align}\label{eqn:topologicalIR}
        \mathcal{D}_{\mathrm{IR}}=\mathcal{L}_1\oplus \mathcal{L}_2\oplus\cdots\mathcal{L}_n.
    \end{align}
    Naturally, we say that it is \textbf{topological} in the IR. (Technically, a defect which is trivial or TQM$_k$ in the IR is also topological.)
    \item The line may become \textbf{conformal} in the IR, in which case it defines a non-trivial defect conformal field theory. A conformal defect preserves  $SO(1,2)\times SO(d)$, which is a subgroup of the conformal symmetry $SO(d+1,2)$ of $H$.\footnote{Some defects may preserve $SO(1,2)$ but not the transverse rotations $SO(d)$; we do not discuss such defects here.} (Technically, a defect which is trivial, TQM$_k$, or topological in the IR is also conformal. We will use the terminology ``non-trivial conformal line'' to refer to a conformal defect which does not fit into any of the three previous scenarios.)
    The conformal line may be non-simple if there are multiple defect operators of dimension $0$. 
\end{enumerate}

Characterizing the endpoint of a defect RG flow, even at the coarse level described above, is  a challenging task. Occasionally, techniques like monotonicity theorems and anomaly matching can be effectively brought to bear on this problem. We review some of these tools in the next two subsections. 

\subsection{Defect 't Hooft anomalies}\label{subsec:defectanomalies}

We are by now accustomed to the idea that a quantum system in $d+1$ spacetime dimensions can have a 't Hooft anomaly. Importantly, there are possible contributions to anomalies which come from defects, as we now review.

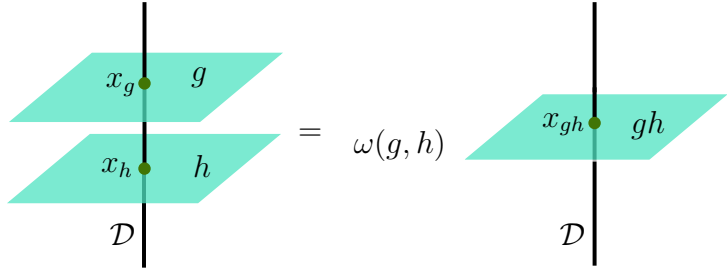
\begin{figure}
    \begin{center}
\tikzset{every picture/.style={line width=0.75pt}} 

\begin{tikzpicture}[x=0.75pt,y=0.75pt,yscale=-1,xscale=1]

\draw [line width=1.5]    (71.75,89) -- (71.75,139) ;
\draw  [color={rgb, 255:red, 80; green, 227; blue, 194 }  ,draw opacity=.75 ][fill={rgb, 255:red, 80; green, 227; blue, 194 }  ,fill opacity=.75 ] (44.65,72) -- (139.5,72) -- (98.85,106) -- (4,106) -- cycle ;
\draw [line width=1.5]    (72,48) -- (72,89) ;
\draw  [color={rgb, 255:red, 80; green, 227; blue, 194 }  ,draw opacity=.75 ][fill={rgb, 255:red, 80; green, 227; blue, 194 }  ,fill opacity=.75 ] (45.65,31) -- (140.5,31) -- (99.85,65) -- (5,65) -- cycle ;
\draw [line width=1.5]    (72,5) -- (72,48) ;
\draw  [color={rgb, 255:red, 65; green, 117; blue, 5 }  ,draw opacity=1 ][fill={rgb, 255:red, 65; green, 117; blue, 5 }  ,fill opacity=1 ] (69.38,46) .. controls (69.38,44.55) and (70.55,43.38) .. (72,43.38) .. controls (73.45,43.38) and (74.63,44.55) .. (74.63,46) .. controls (74.63,47.45) and (73.45,48.63) .. (72,48.63) .. controls (70.55,48.63) and (69.38,47.45) .. (69.38,46) -- cycle ;
\draw  [color={rgb, 255:red, 65; green, 117; blue, 5 }  ,draw opacity=1 ][fill={rgb, 255:red, 65; green, 117; blue, 5 }  ,fill opacity=1 ] (69.38,89) .. controls (69.38,87.55) and (70.55,86.38) .. (72,86.38) .. controls (73.45,86.38) and (74.63,87.55) .. (74.63,89) .. controls (74.63,90.45) and (73.45,91.63) .. (72,91.63) .. controls (70.55,91.63) and (69.38,90.45) .. (69.38,89) -- cycle ;
\draw [line width=1.5]    (299,48) -- (299,115) -- (299,138) ;
\draw  [color={rgb, 255:red, 255; green, 255; blue, 255 }  ,draw opacity=.75 ][fill={rgb, 255:red, 80; green, 227; blue, 194 }  ,fill opacity=.75 ] (272.65,51) -- (367.5,51) -- (326.85,85) -- (232,85) -- cycle ;
\draw [line width=1.5]    (299,5) -- (299,66) ;
\draw  [color={rgb, 255:red, 65; green, 117; blue, 5 }  ,draw opacity=1 ][fill={rgb, 255:red, 65; green, 117; blue, 5 }  ,fill opacity=1 ] (296.38,66) .. controls (296.38,64.55) and (297.55,63.38) .. (299,63.38) .. controls (300.45,63.38) and (301.63,64.55) .. (301.63,66) .. controls (301.63,67.45) and (300.45,68.63) .. (299,68.63) .. controls (297.55,68.63) and (296.38,67.45) .. (296.38,66) -- cycle ;

\draw (94,36.4) node [anchor=north west][inner sep=0.75pt]    {$g$};
\draw (95,80.4) node [anchor=north west][inner sep=0.75pt]    {$h$};
\draw (316,58.4) node [anchor=north west][inner sep=0.75pt]    {$gh$};
\draw (148,67.4) node [anchor=north west][inner sep=0.75pt]    {$=$};
\draw (176,67.4) node [anchor=north west][inner sep=0.75pt]    {$\omega ( g,h)$};
\draw (51,39.4) node [anchor=north west][inner sep=0.75pt]    {$x_{g}$};
\draw (48.65,82.4) node [anchor=north west][inner sep=0.75pt]    {$x_{h}$};
\draw (271,59.4) node [anchor=north west][inner sep=0.75pt]    {$x_{gh}$};
\draw (280,115) node [anchor=north west][inner sep=0.75pt]    {$\mathcal{D}$};
\draw (53,115) node [anchor=north west][inner sep=0.75pt]    {$\mathcal{D}$};

\end{tikzpicture}
\caption{The fusion of symmetry operators $g$ and $h$ becomes anomalous due to the existence of the line defect $\mathcal{D}$. Here, $\omega\in H^2(G,U(1))$ is a 2-cocycle characterizing the anomaly.}
  \label{parfusion}
    \end{center}
\end{figure}

Consider a bulk theory $H$ with a faithfully acting unitary zero-form symmetry $G$. 
Assume all interactions preserve $G$, and that the line defect is neutral under any 1-form symmetries.
Consider the topological codimension-1 operator corresponding to the element $g$. While in the bulk the fusion of parallel symmetry operators proceeds via the group law, if they pierce the line defect a phase can appear, see Figure \ref{parfusion}. 

Indeed, in the absence of the line defect, no phases can arise in the fusion of two topological symmetry operators, simply because such a phase would spoil the Hilbert space interpretation of the composite operator, as described in the paragraph above Equation \eqref{directsum}. On the other hand, when the line defect is introduced, one must choose, for each $g\in G$, a topological point junction $x_g$ where the line and $g$ operator intersect; unlike defects with $p+1\geq 1$ spacetime dimensions, points can be rescaled by arbitrary complex numbers, and hence their fusion can lead to phases, e.g.\ $x_g\otimes x_h = \omega(g,h)x_{gh}$.

If $\omega(g,h)$ is a nontrivial element of $H^2(G,U(1))$ then there is a \emph{defect 't Hooft anomaly}. 
The reason that $\omega$ defines a 2-cocycle is that associativity of the fusion of symmetry operators in the presence of the line defect enforces $\omega(h,k)\omega(g,hk)=\omega(gh,k)\omega(g,h)$. However, only the cohomology class of $\omega$ is physical since one is always free to rescale the choice of point junctions, $x_g'\equiv \alpha(g) x_g$, which has the effect of modifying $\omega$ by an exact 2-cocycle, $\omega'(g,h) =\frac{\alpha(g)\alpha(h)}{\alpha(gh)}\omega(g,h)$.

As in quantum mechanics, local operators on the line defect transform in $G$ representations since the phases $\omega(g,h)$ cancel when acting on the defect operators from above and below. However, endpoint operators are generally in projective representations of $G$, with projective class specified by the defect 't Hooft anomaly $\omega$. (See e.g.\ \cite{Kobayashi:2025pxs} for a recent review of projective representations geared towards physicists.) In quantum mechanics, the endpoints are the states in the Hilbert space (which indeed transform under projective representations of the symmetry). Here the endpoint operators are ``quench operators'', or ``defect-creation operators''.

It should be interesting to recast the statements made thus far in the framework of the SymTFT, following e.g.\ \cite{Bartsch:2022mpm,Bartsch:2022ytj,Bhardwaj:2023wzd,Bhardwaj:2023ayw,Bartsch:2023pzl,Bartsch:2023wvv,Copetti:2024onh,Cordova:2024iti,Choi:2024tri,Choi:2024wfm,Bhardwaj:2024igy}, which would allow for accessing more exotic symmetries.

Since $H^2(G,U(1))$ is discrete, we cannot deform classes $\omega \in H^2(G,U(1))$ continuously. Thus, if the endpoint operators are in a nontrivial projective representation in $\mathcal D_{\rm UV}$, they should remain so in the infrared. Anomaly matching therefore predicts the following.\\

\noindent\textbf{Claim.} \emph{If a line defect $\mathcal{D}_{\mathrm{UV}}$ has a defect 't Hooft anomaly $\omega\in H^2(G,U(1))$ which is cohomologically non-trivial, then the defect is non-trivial in the IR, $\mathcal{D}_{\mathrm{IR}}\neq \mathds{1}$.}\\

\noindent In particular, if there is a defect anomaly, the infrared limit $\mathcal{D}_{\rm IR}$ either has to be coupled to the bulk (as a topological or conformal defect) or decoupled from the bulk with a nontrivial 0+1-dimensional TQFT carrying a nontrivial projective representation of $G$ (i.e.\ TQM$_k$ for some $k$).

Now we turn to another important property of line defects. One can define the entropy of the impurity at zero temperature. It is essentially the normalization of $\mathcal{D}$ computed as explained in detail in~\cite{Cuomo:2021rkm}.
It is scale-dependent, $s(\mu)$, where $\mu$ is the resolution scale. If the defect is conformal then $s(\mu)=s$ is scale independent and represents the (logarithm of the) effective number of degrees of freedom of the conformal impurity.
An important fact is that 
\begin{equation}\label{mono}
\frac{d s(\mu)}{ d\log \mu} \geq 0~, 
\end{equation}
which implies that the effective  number of degrees of freedom on a point-like impurity can only decrease as we coarse grain. This also implies, if there is any RG flow at all, a strict inequality \begin{equation}\label{strictin}
s_{\rm UV}-s_{\rm IR}>0~,\end{equation} 
with $s_{\rm UV}$ and $s_{\rm IR}$ standing for the entropy of the conformal impurity at short and long distances, respectively. 
We can combine the above two constraints to obtain the following refinement of the $g$-theorem.  \\

\noindent\textbf{Claim.} \emph{Suppose that $\mathcal{D}_{\mathrm{UV}}$ carries a defect anomaly $\omega\in H^2(G,U(1))$, and let $\dim R_\omega$ be the dimension of the smallest projective representation of $G$ in the class $\omega$. If } 
\begin{equation}\label{genco}s_{\rm UV}\leq \log \dim R_\omega~,\end{equation}
\emph{then the line defect is not trivial and not TQM in the IR. In particular, it must be topological or conformal.}

\subsection{Anomalies in the Space of Defect Couplings: an Example in the Compact Boson}\label{subsec:anomaliesinspaceofdefectcouplings}

In addition to the more standard defect 't Hooft anomalies described in the previous subsection, we will also discuss the notion of an \emph{anomaly in the space of defect couplings} \cite{Cordova:2019jnf,Cordova:2019uob}. Instead of developing a general theory, we will illustrate the main features in an example: namely,  boundary conditions of the 1+1d compact boson CFT (see also Appendix \ref{app:furtherexamples} for further examples). The anomaly we find in the simple setting of the free boson, which is closely related to that of a quantum mechanical particle on a circle (see \cite{Gaiotto:2017yup,Cordova:2019jnf} for reviews), is nearly identical to the ones we will encounter later in the $O(2)$ model and its particle/vortex dual, the Abelian-Higgs model. 

Recall that the action of the 1+1d compact boson is given by 
\begin{align}
    S[\phi]=\frac{R^2}{8\pi} \int \, d\phi\wedge \star d\phi,
\end{align}
where $\phi\cong \phi+2\pi$ is a periodic scalar field. We let $\tau$ and $x$ denote the spacetime coordinates and, for concreteness, place the theory on an annulus $\Sigma^{(2)}:=S^1_\tau \times I_x$ defined by $\tau\cong \tau+\beta$ and $0\leq x \leq \ell$. At the two boundaries of the interval, we impose Neumann boundary conditions,
\begin{align}
    \star d\phi\vert_{x=0}=0, \quad  \star d\phi\vert_{x=\ell}=0.
\end{align}
Such boundary conditions are conformal and belong to a continuous family parametrized by a periodic parameter $\eta\cong \eta+2\pi$. We keep the boundary at $x=0$ untouched, but deform the boundary at $x=\ell$ to an arbitrary point on this conformal manifold,
\begin{align}
    S_\eta[\phi] = \frac{R^2}{8\pi}\int_{\Sigma^{(2)}}   \,d\phi\wedge\star d\phi-\frac{i}{2\pi}\int_{x=\ell}\, \eta \,d\phi.
\end{align}
The reason that $\eta$ is a $2\pi$-periodic parameter is that $\int_{x=\ell}d\phi$ is valued in $2\pi \mathbb{Z}$. Indeed, it measures the number of times that $\phi$ winds around the thermal circle. Another way to arrive at the same conclusion is to observe that this family of boundary conditions can be obtained by fusing the topological line operator corresponding to the winding symmetry,
\begin{align}\label{eqn:U(1)w}
    U(1)^{(0)}_w: \ \ \mathcal{L}_\eta^{(w)}[L^{(1)}]= \exp\left(i \eta  \int_{L^{(1)}} \frac{d\phi}{2\pi}\right),
\end{align}
onto the boundary; hence, $\eta$ is periodic simply because the winding symmetry group is $U(1)$. 

Note that $T$-duality exchanges Neumann boundary conditions with Dirichlet boundary conditions. Thus, everything we say below applies to Dirichlet boundary conditions as well.

\subsubsection{The anomaly}

The starting point of our discussion is the observation that Neumann boundary conditions preserve the bulk $U(1)^{(0)}_m$ momentum symmetry, which acts by shifting the boson,
\begin{align}\label{eqn:U(1)m}
    U(1)^{(0)}_m:\quad \phi\to \phi + \alpha.
\end{align}
We use the notation $\mathcal{L}_\alpha^{(m)}$ to represent the topological line operator which implements the symmetry in Equation \eqref{eqn:U(1)m}.
The main observation we make in this subsection is that there is a tension between the $2\pi$-periodicity of the boundary coupling $\eta$ and the preservation of the $U(1)_m^{(0)}$ symmetry. We refer to this as an anomaly in the space of boundary couplings, in the spirit of \cite{Cordova:2019jnf, Cordova:2019uob}, where the authors introduced such anomalies in the absence of defects and boundaries.

In the standard situation of a mixed anomaly between two global symmetries, one can diagnose the anomaly by coupling one of the symmetries to a background gauge field and noticing that the other symmetry is violated. We may proceed similarly here by coupling to a background gauge field $A$ for the $U(1)_m^{(0)}$ symmetry, 
\begin{align}
    S_\eta[\phi,A]=\frac{R^2}{8\pi}\int_{\Sigma^{(2)}}(d\phi - A) \wedge *(d\phi - A)
    -\frac{i}{2\pi}\int_{x=\ell}~\eta(d\phi-A).
\end{align}
This action action is invariant under background gauge transformations, 
\begin{align}\label{eqn:gaugetransformcompactboson}
    \phi\to\phi+\alpha, \ \ \ \ \ A\to A+d\alpha,
\end{align}
but at the price of violating the $2\pi$-periodicity of the boundary coupling $\eta$, 
\begin{align}\label{eqn:anomalyI}
    Z_{\eta+2\pi}[A]=Z_\eta[A]\exp\left(-i\int_{x=\ell}A\right), \ \ \ \ Z_{\eta}[A]=\int\mathcal{D}\phi~ e^{-S_\eta[\phi,A]}.
\end{align}
If one attempts to remove the offending boundary counterterm $\frac{i}{2\pi}\int \eta A$, then one succeeds in maintaining the $2\pi$-periodicity of $\eta$, but at the price of violating the gauge invariance of Equation \eqref{eqn:gaugetransformcompactboson}. This is our first indication that there is an anomaly in the space of boundary couplings.

We may also diagnose the anomaly the other way around. That is, we may allow $\eta$ to depend non-trivially on $\tau$ (which we think of in the same spirit as coupling to background gauge fields) and observe that the $U(1)_m^{(0)}$ symmetry is violated. Indeed, under a shift symmetry $\phi\to\phi+\alpha$, one sees that the partition function is no longer invariant,
\begin{align}\label{eqn:anomalyII}
    Z_{\eta(\tau)}\to \exp\left(-\frac{i}{2\pi}\int_{x=\ell}d\tau ~\dot\eta(\tau)\,\alpha\right)Z_{\eta(\tau)}, \ \ \ \ \ \ \ Z_{\eta(\tau)}\equiv \int \mathcal{D}\phi e^{-S_{\eta(\tau)}[\phi]},
\end{align}
where $\int d\tau ~\dot\eta(\tau)$ is the winding number of $\eta(\tau)$ times $2\pi$.

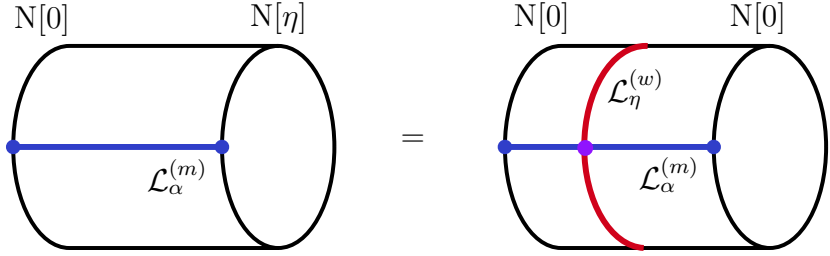
\begin{figure}
    \begin{center}
\tikzset{every picture/.style={line width=0.75pt}} 

\begin{tikzpicture}[x=0.75pt,y=0.75pt,yscale=-1,xscale=1]

\draw  [line width=1.5]  (147.65,129) -- (42.35,129) .. controls (26.69,129) and (14,106.17) .. (14,78) .. controls (14,49.83) and (26.69,27) .. (42.35,27) -- (147.65,27)(176,78) .. controls (176,106.17) and (163.31,129) .. (147.65,129) .. controls (131.99,129) and (119.3,106.17) .. (119.3,78) .. controls (119.3,49.83) and (131.99,27) .. (147.65,27) .. controls (163.31,27) and (176,49.83) .. (176,78) ;
\draw [color={rgb, 255:red, 48; green, 62; blue, 201 }  ,draw opacity=1 ][line width=2.25]    (14,78) -- (119.3,78) ;
\draw [shift={(119.3,78)}, rotate = 0] [color={rgb, 255:red, 48; green, 62; blue, 201 }  ,draw opacity=1 ][fill={rgb, 255:red, 48; green, 62; blue, 201 }  ,fill opacity=1 ][line width=2.25]      (0, 0) circle [x radius= 2.14, y radius= 2.14]   ;
\draw [shift={(14,78)}, rotate = 0] [color={rgb, 255:red, 48; green, 62; blue, 201 }  ,draw opacity=1 ][fill={rgb, 255:red, 48; green, 62; blue, 201 }  ,fill opacity=1 ][line width=2.25]      (0, 0) circle [x radius= 2.14, y radius= 2.14]   ;
\draw  [line width=1.5]  (395.65,129) -- (290.35,129) .. controls (274.69,129) and (262,106.17) .. (262,78) .. controls (262,49.83) and (274.69,27) .. (290.35,27) -- (395.65,27)(424,78) .. controls (424,106.17) and (411.31,129) .. (395.65,129) .. controls (379.99,129) and (367.3,106.17) .. (367.3,78) .. controls (367.3,49.83) and (379.99,27) .. (395.65,27) .. controls (411.31,27) and (424,49.83) .. (424,78) ;
\draw [color={rgb, 255:red, 48; green, 62; blue, 201 }  ,draw opacity=1 ][line width=2.25]    (262,78) -- (367.3,78) ;
\draw [shift={(367.3,78)}, rotate = 0] [color={rgb, 255:red, 48; green, 62; blue, 201 }  ,draw opacity=1 ][fill={rgb, 255:red, 48; green, 62; blue, 201 }  ,fill opacity=1 ][line width=2.25]      (0, 0) circle [x radius= 2.14, y radius= 2.14]   ;
\draw [shift={(262,78)}, rotate = 0] [color={rgb, 255:red, 48; green, 62; blue, 201 }  ,draw opacity=1 ][fill={rgb, 255:red, 48; green, 62; blue, 201 }  ,fill opacity=1 ][line width=2.25]      (0, 0) circle [x radius= 2.14, y radius= 2.14]   ;
\draw  [draw opacity=0][line width=2.25]  (331.99,128.97) .. controls (315.34,128.1) and (302,105.61) .. (302,78) .. controls (302,49.83) and (315.88,27) .. (333,27) .. controls (333.33,27) and (333.66,27.01) .. (333.99,27.03) -- (333,78) -- cycle ; \draw  [color={rgb, 255:red, 208; green, 2; blue, 27 }  ,draw opacity=1 ][line width=2.25]  (331.99,128.97) .. controls (315.34,128.1) and (302,105.61) .. (302,78) .. controls (302,49.83) and (315.88,27) .. (333,27) .. controls (333.33,27) and (333.66,27.01) .. (333.99,27.03) ;  
\draw  [color={rgb, 255:red, 144; green, 19; blue, 254 }  ,draw opacity=1 ][fill={rgb, 255:red, 144; green, 19; blue, 254 }  ,fill opacity=1 ] (299,78.5) .. controls (299,76.57) and (300.57,75) .. (302.5,75) .. controls (304.43,75) and (306,76.57) .. (306,78.5) .. controls (306,80.43) and (304.43,82) .. (302.5,82) .. controls (300.57,82) and (299,80.43) .. (299,78.5) -- cycle ;

\draw (132,3.4) node [anchor=north west][inner sep=0.75pt]    {$\mathrm{N}[ \eta ]$};
\draw (13,4.4) node [anchor=north west][inner sep=0.75pt]    {$\mathrm{N}[ 0]$};
\draw (374,3.4) node [anchor=north west][inner sep=0.75pt]    {$\mathrm{N}[ 0]$};
\draw (265,3.4) node [anchor=north west][inner sep=0.75pt]    {$\mathrm{N}[ 0]$};
\draw (208,70.4) node [anchor=north west][inner sep=0.75pt]    {$=$};
\draw (80,82.4) node [anchor=north west][inner sep=0.75pt]    {$\mathcal{L}_{\alpha }^{( m)}$};
\draw (328,81.4) node [anchor=north west][inner sep=0.75pt]    {$\mathcal{L}_{\alpha }^{( m)}$};
\draw (312,37.4) node [anchor=north west][inner sep=0.75pt]    {$\mathcal{L}_{\eta }^{( w)}$};

\end{tikzpicture}
\caption{The annulus partition function $Z_\eta[A]$ in the presence of a flat background gauge field $A$ with $\alpha=\int\limits_{x=\ell} A$. Here, $\mathrm{N}[\eta]$ denotes the Neumann boundary condition with coupling $\eta$. }\label{fig:annuluspf}
\end{center}
\end{figure}

The phase in Equation \eqref{eqn:anomalyII} admits a useful interpretation in terms of topological operators. To describe this, we continue to allow $\eta(\tau)$ to have a $\tau$-dependent profile, but now stretch a topological line operator corresponding to the $U(1)_m^{(0)}$ symmetry from one boundary to the other, as in the left of Figure \ref{fig:annuluspf}. The $\tau$-dependence of the boundary coupling modifies the equations of motion so that the $U(1)_m^{(0)}$ current receives a contribution from the boundary. In particular, the correct operator in the presence of the boundary coupling is
\begin{align}
    \mathcal{L}_\alpha^{(m)}[L^{(1)}]=\exp\left(-i\frac{\alpha\eta(\tau_{\mathrm{b}})}{2\pi}+ i\frac{\alpha R^2}{4\pi} \int_{L^{(1)}}\star d\phi  \right),
\end{align}
where $\tau_{\mathrm{b}}$ is the $\tau$-coordinate of the $x=\ell$ boundary point on which $L^{(1)}$ terminates.

We now ask what happens to the partition function as one sweeps the topological line operator one full revolution around the thermal circle. To answer this, we imagine deforming the line $L^{(1)}$ slightly near the $x=\ell$ boundary so that it now terminates on a new point $\tau_{\mathrm{b}}'$. Let us call this new line $L^{(1)}_+$, and the old line before deforming $L^{(1)}_-$. To compare these two insertions, we can evaluate 
\begin{align}\label{eqn:localsweep}
    \mathcal{L}_\alpha^{(m)}[L^{(1)}_+]\mathcal{L}_\alpha^{(m)}[L^{(1)}_-]^{-1}=  \mathcal{L}^{(m)}_\alpha[\delta L^{(1)}]=\exp\left(-i\frac{\alpha}{2\pi}(\eta(\tau_{\mathrm{b}}')-\eta(\tau_{\mathrm{b}}))\right),
\end{align}
where $\delta L^{(1)}$ is a small arc which stretches from the point $\tau_{\mathrm{b}}$ on the $x=\ell$ boundary to the point $\tau_{\mathrm{b}}'$. To evaluate $\mathcal{L}_\alpha^{(m)}[\delta L^{(1)}]$, we have deformed $\delta L^{(1)}$ by pushing it onto the boundary where one finds that the integral over $\star d\phi$ vanishes due to the Neumann boundary conditions, so that the only contributions are from the endpoints of $\delta L^{(1)}$. In particular, if we sweep all the way around the thermal circle by integrating \eqref{eqn:localsweep}, we find that the partition function  returns to itself up to precisely the phase in \eqref{eqn:anomalyII}.

A few remarks are in order. First, when $A$ is a flat background gauge field, we may invoke Poincaré duality and reinterpret the partition function $Z_\eta[A]$ as an annulus partition function twisted by a $U(1)_m^{(0)}$ topological line defect which stretches from one boundary to the other, as in the left of Figure \ref{fig:annuluspf}. (The value of the boundary Wilson line $\alpha\equiv \int_{x=\ell} A$ determines which element of $U(1)_m^{(0)}$ is inserted into the annulus.) The reason that $U(1)_m^{(0)}$ topological lines are able to terminate topologically on the Neumann boundaries is that they are $U(1)_m^{(0)}$-symmetric, see e.g.\ \cite{Choi:2023xjw} for a discussion.  Thus, an anomaly in the space of boundary couplings can be characterized as an alteration of the global structure of the boundary conformal manifold when symmetry line defects are allowed to terminate topologically on the boundary conditions parametrized by the couplings. This characterization should facilitate an eventual generalization to non-invertible symmetries.

Second, as we commented earlier, a Neumann boundary condition with generic $\eta$ can be obtained by fusing the $U(1)_w^{(0)}$ winding topological line  in Equation \eqref{eqn:U(1)w} onto the basic Neumann boundary condition with $\eta=0$, as in Figure \ref{fig:annuluspf}. Thus,  $Z_\eta[A]$ is equivalent to an annulus partition function with $\eta=0$ boundary conditions imposed at both ends of the interval,  with a $U(1)_m^{(0)}$ line stretching from one end of the interval to the other, and a $U(1)_w^{(0)}$ line wrapped around the thermal circle. As a consequence, the anomaly described by Equation \eqref{eqn:anomalyI} can be thought of as a kind of reinterpretation of the standard mixed 't Hooft anomaly between the momentum and winding symmetries of a compact boson.

Lastly, we note that this anomaly is robust to deformations of the bulk, so long as they preserve at least a $\mathbb{Z}_N^{(0)}$ subgroup of $U(1)_m^{(0)}$. For example, one can deform by a potential, 
\begin{align}\label{eqn:bosondeformation}
    S_\eta[\phi]=\frac{R^2}{8\pi}\int d\phi\wedge\star d\phi- \int d^2 x\,  V(\phi) -\frac{i}{2\pi}\int d\tau ~\eta d\phi,
\end{align}
with $V(\phi+2\pi/N)=V(\phi)$, like $V(\phi)\propto\sin(N\phi)$. One may even add vertex operators to the action which completely destroy the $U(1)_w^{(0)}$ winding symmetry, or couple in new degrees of freedom entirely, and the anomaly will persist. For example, we will still have 
\begin{align}
    Z_{\eta+2\pi}[\mathcal{L}^{(m)}_{e^{2\pi i/N}}]=Z_\eta[\mathcal{L}^{(m)}_{e^{2\pi i/N}}]e^{-2\pi i/N},
\end{align}
where $Z_\eta[\mathcal{L}_\alpha^{(m)}]$ denotes the annulus partition function with the line $\mathcal{L}^{(m)}_\alpha$ stretching in the interval direction. In particular, we see that $\eta$ is extended from a $2\pi$-periodic parameter to a $2\pi N$-periodic parameter, analogously to Equation \eqref{eqn:anomalyI}. Similarly, under a $\phi\to\phi+2\pi/N$ shift symmetry, the partition function transforms as 
\begin{align}
    Z_{\eta(\tau)}\to \exp\left(-\frac{i}{N}\int_{x=\ell}d\tau~\dot{\eta}(\tau)\right)Z_{\eta(\tau)},	\end{align}
generalizing Equation \eqref{eqn:anomalyII}. 

Thus, even though in certain situations the anomaly can be reinterpreted as a standard mixed 't Hooft anomaly between two global symmetries, it is a more general phenomenon. Because of this robustness to deformations, we will be able to extract non-trivial physics from the anomaly even in models which are more complicated than the free boson, and in particular are not exactly solvable.

\subsubsection{Physical consequences}\label{subsubsec:physicalconsequences}

In broad terms, the existence of an anomaly generally implies that some aspect of the physics must remain non-trivial at long distances. Let us unpack what this means in the case of an anomaly in the space of boundary couplings.

\subsubsection*{Dimensional reduction}

First, suppose we consider the family $\mathcal{T}_\eta$ of 0+1d quantum mechanics theories obtained by dimensionally reducing the free compact boson (or one of its $\mathbb{Z}_N^{(0)}$ momentum preserving deformations, as in Equation \eqref{eqn:bosondeformation}) on an interval with the $\eta=0$ Neumann boundary condition imposed at one end of the interval and a Neumann boundary condition with generic $\eta$ imposed at the other. The anomaly allows us to conclude that $\mathcal{T}_\eta$ must be non-trivial, in the sense of having a degenerate ground space, for some diabolical value $\eta_\ast$ of the coupling. 

One way to see this is as follows. The boundary coupling descends in the dimensionally reduced theory to an ordinary bulk coupling. Moreover, because the boundary conditions are symmetric, the dimensionally reduced theory will also inherit at least a $\mathbb{Z}_N^{(0)}$ symmetry. The anomaly in the space of boundary couplings in 1+1d is recapitulated in the 0+1d quantum mechanics as an ordinary anomaly in the space of coupling constants (without any defects). Thus, one concludes using the same logic as \cite{Cordova:2019jnf,Cordova:2019uob} that one must encounter a discontinuity in the ground space as one tunes $\eta$ from $0$ to $2\pi$, i.e.\ there must be level crossing.

Indeed, we can confirm this prediction in the case of the compact free boson. A straightforward Kaluza-Klein reduction reveals that the model reduces to a 1d quantum mechanics theory of a particle on a circle, defined by the action
\begin{align}\label{eqn:dimred}
    S_{\eta}\xrightarrow{\ell\ll \beta} \int d\tau  \left( \frac{\ell R^2}{8\pi} \dot{q}^2  -\frac{i}{2\pi} \eta \dot{q}\right)
\end{align}
where $q$ can be identified with the zero-mode of the 1+1d field $\phi$. It is known that this model has a doubly degenerate ground space at $\eta_\ast=\pi$ (see e.g.\ \cite{Gaiotto:2017yup}), confirming the non-triviality of the dimensional reduction.

\subsubsection*{Level crossing}

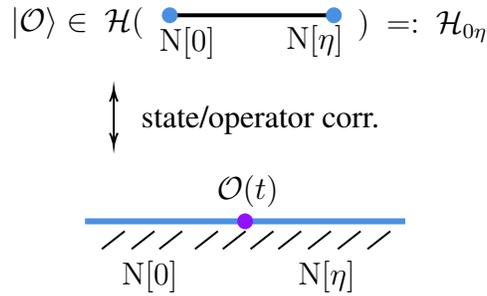
\begin{figure}
\begin{center}
\tikzset{every picture/.style={line width=0.75pt}} 

\begin{tikzpicture}[x=0.75pt,y=0.75pt,yscale=-1,xscale=1]

\draw [color={rgb, 255:red, 74; green, 144; blue, 226 }  ,draw opacity=1 ][line width=2.25]    (55,128) -- (216.5,128) ;
\draw  [color={rgb, 255:red, 144; green, 19; blue, 254 }  ,draw opacity=1 ][fill={rgb, 255:red, 144; green, 19; blue, 254 }  ,fill opacity=1 ] (132.25,128) .. controls (132.25,126.07) and (133.82,124.5) .. (135.75,124.5) .. controls (137.68,124.5) and (139.25,126.07) .. (139.25,128) .. controls (139.25,129.93) and (137.68,131.5) .. (135.75,131.5) .. controls (133.82,131.5) and (132.25,129.93) .. (132.25,128) -- cycle ;
\draw    (75,133) -- (63.5,142) ;
\draw    (90,133) -- (78.5,142) ;
\draw    (104,133) -- (92.5,142) ;
\draw    (122,133) -- (110.5,142) ;
\draw    (136,133) -- (124.5,142) ;
\draw    (150,133) -- (138.5,142) ;
\draw    (164,133) -- (152.5,142) ;
\draw    (179,133) -- (167.5,142) ;
\draw    (194,133) -- (182.5,142) ;
\draw    (209,133) -- (197.5,142) ;
\draw    (69.5,63) -- (69.5,88) ;
\draw [shift={(69.5,90)}, rotate = 270] [color={rgb, 255:red, 0; green, 0; blue, 0 }  ][line width=0.75]    (6.56,-1.97) .. controls (4.17,-0.84) and (1.99,-0.18) .. (0,0) .. controls (1.99,0.18) and (4.17,0.84) .. (6.56,1.97)   ;
\draw [shift={(69.5,61)}, rotate = 90] [color={rgb, 255:red, 0; green, 0; blue, 0 }  ][line width=0.75]    (6.56,-1.97) .. controls (4.17,-0.84) and (1.99,-0.18) .. (0,0) .. controls (1.99,0.18) and (4.17,0.84) .. (6.56,1.97)   ;
\draw [line width=1.5]    (99.5,24) -- (177,24) ;
\draw  [color={rgb, 255:red, 74; green, 144; blue, 226 }  ,draw opacity=1 ][fill={rgb, 255:red, 74; green, 144; blue, 226 }  ,fill opacity=1 ] (177,24) .. controls (177,22.07) and (178.57,20.5) .. (180.5,20.5) .. controls (182.43,20.5) and (184,22.07) .. (184,24) .. controls (184,25.93) and (182.43,27.5) .. (180.5,27.5) .. controls (178.57,27.5) and (177,25.93) .. (177,24) -- cycle ;
\draw  [color={rgb, 255:red, 74; green, 144; blue, 226 }  ,draw opacity=1 ][fill={rgb, 255:red, 74; green, 144; blue, 226 }  ,fill opacity=1 ] (94.25,24) .. controls (94.25,22.07) and (95.82,20.5) .. (97.75,20.5) .. controls (99.68,20.5) and (101.25,22.07) .. (101.25,24) .. controls (101.25,25.93) and (99.68,27.5) .. (97.75,27.5) .. controls (95.82,27.5) and (94.25,25.93) .. (94.25,24) -- cycle ;

\draw (120,103.4) node [anchor=north west][inner sep=0.75pt]    {$\mathcal{O}( t)$};
\draw (82,68) node [anchor=north west][inner sep=0.75pt]   [align=left] {state/operator corr.};
\draw (72,146.4) node [anchor=north west][inner sep=0.75pt]    {$\mathrm{N}[ 0]$};
\draw (161,146.4) node [anchor=north west][inner sep=0.75pt]    {$\mathrm{N}[ \eta ]$};
\draw (16,18.4) node [anchor=north west][inner sep=0.75pt]    {$|\mathcal{O} \rangle \in \ \mathcal{H}($};
\draw (191,19.4) node [anchor=north west][inner sep=0.75pt]    {$) \ =:\ \mathcal{H}_{0\eta }$};
\draw (91,28.4) node [anchor=north west][inner sep=0.75pt]    {$\mathrm{N}[ 0]$};
\draw (155,27.4) node [anchor=north west][inner sep=0.75pt]    {$\mathrm{N}[ \eta ]$};

\end{tikzpicture}
\caption{By the state/operator correspondence, states $|\mathcal{O}\rangle$ in the interval Hilbert space can be identified with boundary changing local operators $\mathcal{O}(\tau)$.}\label{fig:state/operatorcorrespondenceboundaries}
\end{center}
\end{figure}

There is a way to phrase the discussion above in terms of 1+1-dimensional physics. The annulus partition function can be formulated as a trace over the Hilbert space $\mathcal{H}_{0\eta}$ of states on the interval with the $\eta=0$ Neumann boundary condition imposed at one end and a Neumann boundary condition with general $\eta$ imposed at the other,
\begin{align}
    Z_\eta = \mathrm{Tr}_{\mathcal{H}_{0\eta}}e^{-\beta H}.
\end{align}
When the theory is conformal, by the state/operator correspondence, states in $\mathcal{H}_{0\eta}$ can be identified with boundary-changing local operators between the two Neumann boundary conditions, as in Figure \ref{fig:state/operatorcorrespondenceboundaries}.

One can zoom in on the ground space on the interval by taking $\ell\ll \beta$. If the ground space is trivial and separated by a gap from the rest of the excited states for all $\eta$, then we cannot reproduce the anomalies of Equation \eqref{eqn:anomalyI} and \eqref{eqn:anomalyII}, as the partition function should be a smooth function of $\eta$. Thus, there should be level-crossing in $\mathcal{H}_{0\eta}$ as one tunes $\eta$.

\begin{figure}
\begin{center}
\tikzset{every picture/.style={line width=0.75pt}} 

\begin{tikzpicture}[x=0.75pt,y=0.75pt,yscale=-1,xscale=1]

\draw  [line width=1.5]  (31,85.75) .. controls (31,57.72) and (53.72,35) .. (81.75,35) .. controls (109.78,35) and (132.5,57.72) .. (132.5,85.75) .. controls (132.5,113.78) and (109.78,136.5) .. (81.75,136.5) .. controls (53.72,136.5) and (31,113.78) .. (31,85.75) -- cycle ;
\draw [color={rgb, 255:red, 74; green, 144; blue, 226 }  ,draw opacity=1 ]   (81.75,35) .. controls (122.5,10) and (93.5,73) .. (119.5,65) .. controls (145.5,57) and (148.89,108.46) .. (121.5,104) .. controls (94.11,99.54) and (142.5,159) .. (81.75,137) ;
\draw [color={rgb, 255:red, 126; green, 211; blue, 33 }  ,draw opacity=1 ]   (81.75,35) .. controls (16.5,15) and (84.5,52) .. (34.5,68) .. controls (-15.5,84) and (72.5,92) .. (52.5,116) .. controls (32.5,140) and (67.5,160) .. (81.75,137) ;
\draw  [color={rgb, 255:red, 0; green, 0; blue, 0 }  ,draw opacity=1 ][fill={rgb, 255:red, 0; green, 0; blue, 0 }  ,fill opacity=1 ] (78.25,35) .. controls (78.25,33.07) and (79.82,31.5) .. (81.75,31.5) .. controls (83.68,31.5) and (85.25,33.07) .. (85.25,35) .. controls (85.25,36.93) and (83.68,38.5) .. (81.75,38.5) .. controls (79.82,38.5) and (78.25,36.93) .. (78.25,35) -- cycle ;
\draw  [color={rgb, 255:red, 0; green, 0; blue, 0 }  ,draw opacity=1 ][fill={rgb, 255:red, 0; green, 0; blue, 0 }  ,fill opacity=1 ] (78.25,137) .. controls (78.25,135.07) and (79.82,133.5) .. (81.75,133.5) .. controls (83.68,133.5) and (85.25,135.07) .. (85.25,137) .. controls (85.25,138.93) and (83.68,140.5) .. (81.75,140.5) .. controls (79.82,140.5) and (78.25,138.93) .. (78.25,137) -- cycle ;

\draw (61,9.4) node [anchor=north west][inner sep=0.75pt]    {$\eta =\pi $};
\draw (63,145.4) node [anchor=north west][inner sep=0.75pt]    {$\eta =0$};

\end{tikzpicture}
\caption{The states in $\mathcal{H}_{0\pi}$ are doubly degenerate because for every string configuration which stretches in one direction around the target circle (e.g.\ blue) there is another with the same energy which stretches in the other direction (e.g.\ green).}\label{fig:doublydegeneratestringstates}
\end{center}
\end{figure}
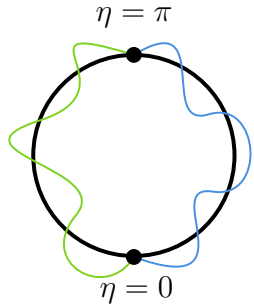

Again, this level crossing can be confirmed  to happen at $\eta_\ast=\pi$ by explicit computation of the annulus partition function in the case of a compact free boson. We can argue for this in words as follows. By T-duality, it suffices to check that level crossing occurs for Dirichlet boundary conditions instead of Neumann boundary conditions. In this case, the interval Hilbert space can be thought of as the space of open string states which stretch from a D-brane at one point on a target circle to a D-brane at another point. The special case that $\eta_\ast=\pi$ corresponds to the situation that the two D-branes are placed at antipodal points on the target space circle. When this happens, there are two ground states: one which corresponds to the string which stretches clockwise along the target space circle, and one where it stretches counter-clockwise. See Figure \ref{fig:doublydegeneratestringstates}.

\subsubsection*{Boundary-changing local operators}

When one has a continuous coupling constant $\eta$ in a quantum field theory, one can define an interface between the theory with coupling constant $\eta_0$ and the theory with coupling constant $\eta_1$ by allowing $\eta$ to smoothly interpolate from $\eta_0$ to $\eta_1$ along a codimension-1 locus of spacetime. In order for the interface so-obtained to be universal (i.e.\ completely determined by the UV theory), the coupling must vary over a distance scale which is larger than the UV cutoff. In the case that the coupling is periodic, we may obtain an interface between the theory with coupling constant $\eta_0$ and itself by allowing $\eta$ to quickly wind from $\eta_0$ to $\eta_0+2\pi k$ along a codimension-1 locus of spacetime. It is then interesting to ask whether the interface so-defined is the identity or constitutes a non-trivial defect.

If a \emph{defect} in a quantum field theory is controlled by a continuous defect coupling constant $\eta$, then we may similarly define codimension-1 (with respect to the defect worldvolume) interfaces between the defect with coupling $\eta_0$ and the defect with coupling $\eta_1$. In the present situation, we are studying boundary conditions of a 1+1d quantum field theory, so we obtain boundary-changing local operators by allowing the boundary coupling $\eta$ to smoothly jump around a point on the boundary; alternatively, we obtain local operators on a single boundary condition if we take $\eta$ to wind from $\eta_0$ to $\eta_0+2\pi k$. What can we say about such boundary/boundary-changing local operators?

Suppose for simplicity that we work in a deformation of the compact boson in which the full $U(1)_m^{(0)}$ is preserved. There are analogous statements one can make if the $U(1)_m^{(0)}$ is broken down to a $\mathbb{Z}_N^{(0)}$ subgroup. The anomaly as expressed in Equation \eqref{eqn:anomalyII} suggests that the boundary local operator defined by a coupling constant $\eta(\tau)$ carries charge $k$ under the bulk $U(1)_m^{(0)}$ symmetry if $\eta$ jumps from $\eta_0$ to $\eta_0-2\pi k$. Indeed, this can be deduced by using the physical interpretation of \eqref{eqn:anomalyII} explained above in terms of sweeping a topological line operator around the thermal circle: the phase one gets when one pulls a topological line operator through a boundary excitation is precisely its charge. This in particular implies that it cannot be the identity operator. 

We can again explicitly verify these predictions in the case of the 1+1d compact boson. Suppose we place the theory on the upper half-plane with $\tau$ parametrizing the boundary. We choose the following regularized profile for the boundary coupling,
\begin{align}\label{eqn:spatialprofiletheta}
    \eta_\epsilon(\tau) = \begin{cases}
        \eta_0, & \tau \leq 0 \\
        \eta_0 - \frac{2\pi k}{\epsilon}\tau,& 0\leq \tau \leq \epsilon \\
        \eta_0-2\pi k,& \tau\geq \epsilon,
    \end{cases} 
\end{align}
and take $\epsilon\to0$ at the end of the calculation. It is a straightforward exercise to show that the boundary action evaluates to 
\begin{align}
    \lim_{\epsilon\to 0}\int_{-\infty}^{\infty} d\tau \eta_\epsilon(\tau)\dot\phi(\tau,0) = \int\limits _{-\infty}^0 \eta_0\dot\phi(\tau,0) + \int\limits_{0}^\infty \left(\eta_0 - 2 \pi k\right)\dot\phi(\tau,0)  +  2\pi k\phi(0,0).
\end{align}
In particular, the partition function becomes 
\begin{align}
    Z_{\eta(\tau)}=\int \mathcal{D}\phi e^{ik\phi(0,0)}e^{-S_{\eta_0}[\phi]},
\end{align}
i.e.\ we encounter an insertion of a vertex operator, which clearly carries charge $k$ under the bulk $U(1)_m^{(0)}$ symmetry. 

\subsubsection{Relationship to defect 't Hooft anomalies}

Anomalies in the space of defect couplings allow one to argue that the physics of the defect is non-trivial for some value of the coupling. Their virtue is that they are robust to deformations of the theory which break much of the symmetry, but their drawback is that, on their own, they cannot tell you \emph{which} value of the defect coupling leads to non-trivial physics. However, in special situations, the symmetries of the problem will enhance at some specific value $\eta_\ast$ of the defect coupling, and one can use the more conventional anomalies reviewed in \S\ref{subsec:defectanomalies} to argue that the defect with coupling $\eta_\ast$ is the one which is non-trivial at long distances.

In the case of the free boson, one has an additional discrete charge conjugation symmetry in the bulk that we have not considered yet,
\begin{align}
    \mathbb{Z}_2^{C}: \ \phi\to -\phi,
\end{align} 
which combines with the $U(1)_m^{(0)}$ momentum symmetry to generate the group $O(2)$. This symmetry is completely non-anomalous in the bulk.

It is elementary to see that charge conjugation maps the boundary coupling as $\eta\to-\eta$, so the Neumann boundary condition breaks $C$ for generic values of $\eta$. However, the Neumann boundary conditions with $\eta=0$ and $\eta=\pi$ preserve charge conjugation (remembering that $\eta\cong \eta+2\pi$), and we can ask what the implications are of this additional symmetry for the physics at these points in the boundary conformal manifold.

The main claim is that the $O(2)$ symmetry remains non-anomalous when both boundaries are taken to have $\eta=0$, but anomalous (on the boundary) when one of the boundaries is taken to have $\eta=\pi$. In particular, the $O(2)$ anomaly allows us to conclude that $\eta_\ast=\pi$ is the ``diabolical point'' in the boundary conformal manifold guaranteed by the anomaly in the space of defect couplings.

Following the discussion in Section \ref{subsec:defectanomalies}, we expect that the anomaly is characterized by a class in $H^2(O(2),U(1))\cong \mathbb{Z}_2$. There are a variety of ways that we can diagnose this $\mathbb{Z}_2$-valued anomaly.

One way to proceed is to dimensionally reduce the theory on an interval down to 0+1d. There, the boundary anomaly we are after reveals itself as a standard 't Hooft anomaly of the quantum mechanics theory, and can therefore be determined by computing whether the Hilbert space transforms linearly (no anomaly) or projectively (anomaly) under the $O(2)$ symmetry. As we stated earlier, Equation \eqref{eqn:dimred}, the theory obtained by dimensional reduction is the particle on a circle with flux $\eta$, which is known to have a non-anomalously realized $O(2)$ symmetry at $\eta=0$ and an anomalously realized $O(2)$ symmetry at $\eta=\pi$, see e.g.\ \cite{Gaiotto:2017yup} for a review. 

In more intrinsically 1+1d terms, the Hilbert space of boundary local operators on the $\eta=0$ Neumann boundary transforms linearly under $O(2)$, whereas the space of boundary-changing local operators between $\mathrm{N}[0]$ and $\mathrm{N}[\pi]$ transforms projectively.\footnote{In the picture of \cite{Choi:2024tri,Choi:2024wfm} (see also \cite{Copetti:2024dcz,Cordova:2024vsq,Cordova:2024iti}), the two boundaries $\mathrm{N}[0]$ and $\mathrm{N}[\pi]$ transform in two inequivalent module categories of $O(2)$, and hence the boundary tube algebra which characterizes the action of $O(2)$ on interval states is the twisted group algebra $\mathbb{C}^\psi[O(2)]$, where $\psi$ is a representative 2-cocycle of the non-trivial class in $H^2(O(2),U(1))$. See the discussion around Equation 3.18 of \cite{Choi:2024tri}.}

We can also diagnose the anomaly directly from the path integral by coupling to a background gauge field $A$ for $U(1)_m^{(0)}$ and determining whether the partition function remains $C$-invariant (up to flipping the sign of $A$). When both boundaries are $\mathrm{N}[0]$, we see that it does remain $C$-invariant,
\begin{align}
    Z_{\eta=0}[A]\xrightarrow{C} Z_{\eta=0}[-A],
\end{align}
whereas when one boundary is taken to be $\mathrm{N}[\pi]$, we encounter a boundary Wilson line in the background gauge field
\begin{align}
    Z_\eta[A]\xrightarrow{C} Z_\eta[-A]\exp\left(i\int_{x=\ell}A\right),
\end{align}
signaling the presence of an anomaly.

\section{Three Examples in Magnets} \label{sec:magnets}

Next, we apply the tools described in the previous section to learn things about magnetic defects in three 2+1d critical points: the $O(1)$ WF fixed point (the Ising model), the $O(2)$ fixed point (the XY model) and the $O(3)$ fixed point. These three examples are the simplest magnetic defects in the literature and they occur in many important lattice models. (We will review some of it below.)
We use~\eqref{genco}, along with anomaly arguments, to make predictions about the infrared limit of these defects. 

\subsection{A Quantum Mechanics Interlude}
\label{subsec:quantuminterlude}
All of our constructions in this section involve coupling the bulk theory to a spin-$s$ representation of $SU(2)$ on a line defect in spacetime (i.e.\ a point in space). Therefore, we spend some time reviewing this quantum mechanical model.

\subsubsection{The model}
We implement a quantum mechanical model whose Hilbert space is the spin-$s$ representation by using a first order action (with a constraint) for a vector $z=(z_1,z_2)$ of complex scalar fields,
\begin{equation}\label{bis}S_{\sps}= \int d\tau z^\dagger \dot z~, \quad z^\dagger z = 2 s~.
\end{equation}
This is known as the coadjoint orbit method.
Alternatively, one may trade the constraint for a $U(1)$ gauge field Lagrange multiplier  $\lambda(\tau)$, 
\begin{align}\label{eqn:lmultiplier}
    S_{\sps}'=\int d\tau\left(z^\dagger(\partial_t-i\lambda)z+2is\lambda \right).
\end{align}
The action $S'_{\sps}$ goes over to $S_{\sps}$ upon integrating out $\lambda$. The action~\eqref{bis} is invariant under the gauge transformation
\begin{align}\label{eqn:gaugetransformation}
    z(\tau)\to e^{i\alpha(\tau)} z(\tau), \ \ \ \ \ \ \ \alpha(\tau)\in\mathbb{R}/2\pi \mathbb{Z}, 
\end{align} 
 and requiring that $e^{-S_{\mathrm{\sps}}}$ is well-defined leads to the requirement that $2s$ is a positive integer. Indeed, if we put the theory on $S^1$, we can compute that
\begin{equation}\label{variation}\delta_\alpha S_{\sps} = i\int d\tau  \dot \alpha z^\dagger z = 2s i \int d\tau  \dot \alpha = 4\pi i s W,\end{equation}
where $W\in\mathbb{Z}$ is the winding number of $\alpha$ around $S^1$.
The alternative action in Equation \eqref{eqn:lmultiplier} is invariant under \eqref{eqn:gaugetransformation} if one simultaneously transforms $\lambda$ like a $U(1)$ gauge field, 
\begin{align}\label{eqn:gaugetransformation2}
    \lambda(\tau) \to \lambda(\tau)-\dot{\alpha}(\tau),
\end{align}
and from this point of view, the integer quantization condition follows simply because $2s$ appears as a 1d Chern-Simons term for $\lambda$.

Let us now quantize the theory. The canonical commutation relations are
\begin{align}
    [z_i,z_j^\dagger]=\delta_{ij},
\end{align}
and so one may build up a Hilbert space by considering states of the form $(z_1^\dagger)^a (z_2^\dagger)^b|0\rangle$, where $|0\rangle$ is a vacuum which is annihilated by $z_1$ and $z_2$. However, we must impose the constraint $z^\dagger z=2s$ from Equation \eqref{bis}, which leads us to work with the subspace $\mathcal{H}$ spanned by states satisfying $a+b=2s$. A convenient basis is
\begin{align}\label{eqn:constrainedhilbertspace}    \mathcal{H}=\mathrm{span}_{\mathbb{C}}\{|m\rangle \mid m\in -s,-s+1,\dots,s\}
\end{align}
where we have defined 
\begin{align}\label{eqn:Szeigenstates}
    |m\rangle\equiv \frac{1}{\sqrt{(s+m)!(s-m)!}}(z_1^\dagger)^{s+m}(z_2^\dagger)^{s-m}|0\rangle.
\end{align}

Thus this first order quantum mechanical model has a $2s+1$ dimensional Hilbert space. The spin operators which generate the action of $SU(2)$ are given by 
\begin{align}
    S^a= z^\dagger \frac{\sigma^a}{2} z,
\end{align} 
and they satisfy the usual commutation relations $[S^a,S^b]=i\epsilon^{abc} S^c$. Since $z^\dagger$ and $z$ are canonically conjugate, when we later deform the action by insertions of $S^a$, we always adopt the ordering convention that $z^\dagger$ is slightly later in time than $z$. (The constraint $z^\dagger z =2s$ is interpreted similarly.)
As the labeling suggests, the $|m\rangle$  are eigenvectors of $S_z=\frac12(z_1^\dagger z_1-z_2^\dagger z_2)$ with eigenvalue $m$.

For later, we will need to discuss what happens when the line is half-infinite. 
From~\eqref{variation}, if we have an endpoint, gauge invariance requires that we attach an operator like  
\begin{equation}
z_1^{2s}(0)  e^{-\int\limits_0^\infty d\tau z^\dagger \dot z} ~.
\end{equation}
Of course, the set of endpoint operators is isomorphic to the Hilbert space of the model, and indeed we can replace the endpoint $z_1^{2s}$ by any other expression with charge $2s$ under the gauge symmetry. A convenient choice is to label the set of endpoints by the $2s+1$ states
$z_1^{2s}, z_1^{2s-1}z_2,\dots, z_2^{2s}$, i.e.\ the operators corresponding to the states $|m\rangle$ from Equation \eqref{eqn:Szeigenstates}.

\subsubsection{'t Hooft anomalies}\label{subsubsec:spinsthooftanomalies}
Next, we describe the symmetries and anomalies of the model~\eqref{bis} (or equivalently \eqref{eqn:lmultiplier}). Later, when we couple $S_{\sps}$ to higher-dimensional QFTs, the anomalies discussed here will be reinterpreted as defect 't Hooft anomalies, as reviewed in Section \ref{subsec:defectanomalies}.

Naively, there is an $SU(2)$ symmetry of \eqref{eqn:lmultiplier} which rotates the $z$ fields while leaving $\lambda$ invariant,
\begin{align}
    z\to Uz, \ \ \ \ \  \lambda \to \lambda.
\end{align}
However, the transformation by the central element $U=-1$ of $SU(2)$ can be undone by a gauge transformation using \eqref{eqn:gaugetransformation} and \eqref{eqn:gaugetransformation2}, and so the true global symmetry is $SO(3)\cong SU(2)/\{\pm 1\}$.

Another way to see that the true symmetry of the theory is $SO(3)$ is to note that the gauge invariant operators $S^a$ transform under this symmetry. On the other hand, we see from \eqref{eqn:constrainedhilbertspace} that the states/endpoints transform \emph{projectively} under $SO(3)$ for half-integer $s$. 
Therefore there is a nontrivial anomaly valued in $H^2(SO(3),U(1))=\mathbb{Z}_2$ when $s$ is a half-integer. 

The model also admits time reversal symmetry, which we can take to act as
\begin{equation}\label{Taction}
\mathcal{T}: z_1\to z_2^*~,\quad z_2\to -z_{1}^*\end{equation} 
accompanied by time reflection $t\to-t$. It is straightforward to check that 
\begin{equation}\label{timespin}\mathcal{T} S^a \mathcal{T}^{-1} = - S^a~,\end{equation} 
therefore, this choice of time reversal symmetry commutes with $SO(3)$.

Let us  determine how $\mathcal{T}$ acts on states. Because $\mathcal{T}$ flips the spin, the action has to be of the form $\mathcal{T} |m\rangle = \alpha_m \ket{-m}$ with some phase $\alpha_m$. Assuming this form and using  $\mathcal{T}S_+=-S_-\mathcal{T}$, 
 a short calculation shows that $\alpha_{m \pm 1} = - \alpha_m$.
Thus the phases alternate,
$\alpha_m = (-1)^{s - m} \alpha_s$, 
and since $\mathcal{T}$ is only determined up to phase, we are free to choose $\alpha_s$. We take $\alpha_s=1$, which leads to 
\begin{align}\label{eqn:Tactionmstates}
    \mathcal{T}|m\rangle =(-1)^{s-m}\ket{-m}.
\end{align}
In particular, we see that, on states, 
\begin{equation}\label{eqn:tactiononstates}
\mathcal{T}^2=(-1)^{2s}~,\quad \mathcal{T} e^{i\alpha^a S^a} \mathcal{ T}^{-1} =e^{i\alpha^a S^a}~.  
\end{equation}
The first equation represents a central extension implying that we have Kramers doublets for half-integer spin.\footnote{Sometimes, the fact that $\mathcal{T}^2=(-1)^{2s}$ is interpreted by saying that two time reflections is equivalent to a $2\pi$ rotation.}

More abstractly, the possible anomalies of $SO(N)\times\mathbb{Z}_2^T$ are classified by 
\begin{align}
    H^2(SO(N)\times \mathbb{Z}_2^T,U_T(1))\cong \mathbb{Z}_2\times \mathbb{Z}_2,
\end{align}
where $U_T(1)$ indicates that time reversal acts by complex conjugation on $U(1)$.
If we work in a basis where the $SO(N)$ generators commute with time reversal, as in Equation \eqref{eqn:tactiononstates}, the first $\mathbb{Z}_2$  specifies whether the $SO(N)$ is realized projectively on the Hilbert space, and the second $\mathbb{Z}_2$ determines whether time reversal squares to $1$ or $-1$. Therefore, the preceding discussion can be summarized by saying that the anomaly of $S_{\sps}$ corresponds to the class 
\begin{align}\label{anomaliesspin}
    (2s,2s)\in H^2(SO(3)\times\mathbb{Z}_2^T,U_T(1)).
\end{align}
In total, we see that for half-integer spin there is both an $SO(3)$ and time reversal anomaly, while for integer spin there is no anomaly. 

\subsubsection{An anomaly in the space of coupling constants}\label{subsubsec:spinsanomalycouplings}

It will also be useful to note that this model possesses an anomaly in the space of its coupling constants, in the sense of \cite{Cordova:2019jnf,Cordova:2019uob}. When we couple to bulk QFTs in later sections, the anomaly described here will be promoted to an anomaly in the space of \emph{defect} couplings, as reviewed in Section \ref{subsec:anomaliesinspaceofdefectcouplings}.

To expose this anomaly, we deform the action \eqref{bis} (or \eqref{eqn:lmultiplier}) as 
\begin{align}
    S=S_{\sps}+\gamma \int d\tau S_z(\tau).
\end{align}
In the limit that $\gamma\to\pm\infty$, the model goes over to a trivially gapped theory.

The deformed theory preserves the $U(1)$ symmetry generated by
\begin{align}\label{eqn:U1symmetry}
    V_\alpha: z_1\to e^{-i\alpha/2}z_1, \ \ \ \ z_2\to e^{i\alpha/2}z_2.
\end{align}
While conventional 't Hooft anomalies are not possible with just a $U(1)$ symmetry in 0+1d, an SPT phase is possible. Our claim is that the SPT phase differs between $\gamma\to\pm \infty$. 
Indeed, coupling to a background gauge field $A$ for $U(1)$,
the partition function on $S^1$ with inverse temperature $\beta$ is 
\begin{align}
    Z_\gamma[A]=\int\mathcal{D}z~e^{-S_\gamma[A]}=\sum_{m=-s,-s+1,\dots,s}e^{-\beta\gamma m-im\oint A}~, 
\end{align}
and we find
\begin{align}
    \lim_{\gamma\to\infty}\frac{Z_\gamma[A]}{Z_{-\gamma}[A]}=\exp \left(2is\oint A\right).
\end{align}
Thus, the SPT phase for $U(1)$ has changed, or we can alternatively say that there is an anomaly in the space of coupling constants whenever $s>0$. 
Unlike the 't Hooft anomalies that we discussed in~\eqref{anomaliesspin}, the jump in the SPT phase (or the anomaly the space of coupling constants) occurs for any representation and not just half-integer spin representations.   

The physical upshot is that there must be some coupling $\gamma_\ast$ for which the theory is non-trivial in the IR. Concretely, in quantum mechanics, this means that there must be level crossing for some value of $\gamma_\ast$. Of course, the model is exactly soluble, and so we know that this occurs at $\gamma_\ast=0$ (where the symmetry is enhanced to $SO(3)\times\mathbb{Z}_2^T$). However, this same anomaly will persist when we couple to bulk QFT, and will allow us to reach non-trivial conclusions about defect RG flows.

\subsection{Pinning Field Defects} 
We now briefly review the pinning field conformal defect, which will play an important role in the next subsection. 
We begin with the action of the critical $O(N)$ Wilson-Fisher model $S_{O(N)}$ and deform it by an external magnetic field, which is a vector $\vec h$ in field space: 
\begin{equation}\label{pinning}
S=S_{O(N)}+\int d\tau ~\vec h \cdot \vec\phi(\vec x=0,\tau)~.\end{equation}
Here, $\vec{\phi}=(\phi_1,\dots,\phi_N)$ is an $N$-component vector of real scalar fields. The external localized field $\vec h$ (pinning field) is very small in the ultraviolet, $\vec h_{\mathrm{UV}}\to 0$, but it grows to a fixed point value $\vec h_*$ in the infrared. The defect clearly breaks the $O(N)$ symmetry of the Wilson Fisher model down to $O(N-1)$. 
Since the defect is trivial at short distances, from~\eqref{strictin} it is evident that $s_{\mathrm{IR}}<0$ and hence the infrared limit is a nontrivial conformal defect. 

We will denote the infrared conformal defect obtained in this way by $\mathcal{D}[\hat h]$, which depends only on the direction of the initial magnetic field $\hat h = \vec h/|\vec h|$. A lot is known about $\mathcal{D}[\hat h]$ from analytic results at large $N$, from the $\epsilon$ expansion, and from explicit simulations.  
We will quote various concrete results below. 
One general thing that is useful to keep in mind is that $\mathcal{D}[\hat h]$ admits exactly marginal deformations corresponding to the generators of $O(N)$ which are broken by the defect. When used to deform the action, their effect is to simply rotate the direction of the magnetic field. These exactly marginal operators are called ``tilt'' operators.

\subsection{A Magnetic Defect in the Ising Model: Pinning Fields}\label{subsec:Ising}

Consider the following line defect in the bulk Ising theory,
\begin{equation}\label{pinningiz} S=S_{\rm Ising}+S_{\sps}+h \int d\tau   \phi(\vec x=0,\tau) S_z(\tau)~.\end{equation}
Here $S_{\sps}$ is the first order action~\eqref{bis}, which we analyzed in detail. 

\subsubsection{Symmetries and anomalies}
The coupling~\eqref{pinningiz} preserves the $\mathbb{Z}_2$ symmetry of the bulk Ising model if we combine it with a transformation
 on the defect,
 \begin{align}
     U:\phi \to -\phi, \ \ \ \ z_1\to-iz_2, \ \ \ \ z_2\to -iz_1.
 \end{align}
 Note that, in the quantum mechanics model \eqref{bis} defined without coupling to the bulk Ising theory, the transformation $z_1\to -iz_2$ and $z_2\to-iz_1$ acts via the operator $e^{i\pi S_x}$ on the $2s+1$-dimensional Hilbert space of states, i.e.\ it maps $|m\rangle\to(-1)^s|-m\rangle$.

 In addition, time reversal symmetry is preserved, under which 
 \begin{align}
     \mathcal{T}': \phi(\tau)\to\phi(-\tau), \ \ \ \ z_1(\tau)\to iz_1^\ast(-\tau), \ \ \ \ z_2(\tau)\to -iz_2^\ast(-\tau).
 \end{align}
 This definition of time reversal is related to the one given in Equation \eqref{Taction} by $\mathcal{T}'=\mathcal{T}U$. In particular, $\mathcal{T}'|m\rangle = (-1)^m|m\rangle$ so that $\mathcal{T}^{'2}=+1$ on the states of the model $S_{\sps}$, for both integer and half-integer spin $s$.

Thus, our symmetry class is 
$\mathbb{Z}_2^T\times \mathbb{Z}_2$ and we know that on the endpoints we have  $\mathcal{T}'^2=1$ so there are no Kramers doublets under $\mathcal{T}'$.
However, there is still an anomaly in this symmetry group. While $U^2=1$ on local operators, $U^2=(-1)^{2s}$ on the endpoints.
If we ignore time reversal symmetry this $(-1)^{2s}$ phase can be easily removed (e.g.\ by redefining  $U\to e^{i\pi s} U$); indeed $\mathbb{Z}_2$ does not have nontrivial projective representations because $H^2(\mathbb{Z}_2,U(1))=\mathbb{Z}_1$. 
However, if we try to remove the phase from $U^2$, then we will destroy the relation 
$\mathcal{T}'U = U \mathcal{T}'$. Therefore there is an anomaly in~\eqref{pinningiz}.

Abstractly, the defect anomalies of $\mathbb{Z}_2\times\mathbb{Z}_2^T$ are parametrized by $H^2(\mathbb{Z}_2\times\mathbb{Z}_2^T,U_T(1))\cong \mathbb{Z}_2\times\mathbb{Z}_2$, with explicit representatives given by 
\begin{align}\label{eqn:explicit2cocycles}
    \omega_{A,B}(U^aT^b,U^cT^d)=(-1)^{Aac+Bbd}.
\end{align} By the preceding discussion, the class realized by \eqref{pinningiz} is $\omega_{2s,0}$. 

For half-integer $s$, the anomaly implies a degeneracy in the space of lowest-dimension endpoint operators of the defect. The anomaly also forbids a flow to a trivial defect, consistent with our analysis below.

\subsubsection{Relation with pinning field defects}
The theory~\eqref{pinningiz} does not lead to any couplings between different $S_z$ eigenstates and we can solve separately for the states $S_z=-s,-s+1,\dots, +s$. 
Each one of these (other than $S_z=0$) behaves as a pinning field defect $\mathcal{D}[\pm]$ and hence, using $\mathcal{D}_s$ to label the UV defect in \eqref{pinningiz}, the infrared limit is  
\begin{align}
    \mathcal{D}_s\xrightarrow{\mathrm{IR}} \begin{cases}
        \mathcal{D}[+]^{\oplus (s+\frac12)}\oplus \mathcal{D}[-]^{\oplus(s+\frac12)}, & s\in\mathbb{Z}_{\geq 0}+\frac12 \\
         \mathcal{D}[+]^{\oplus s}\oplus \mathds1\oplus  \mathcal{D}[-]^{\oplus s}, & s\in \mathbb{Z}_{\geq0},
    \end{cases}
\end{align}
where $\mathcal{D}^{\oplus n}$ denotes the direct sum of $\mathcal{D}$ with itself $n$ times.

These are all $\mathbb{Z}_2\times\mathbb{Z}_2^T$-invariant defects, but many of them are fine tuned. First of all, we can add a potential to~\eqref{pinningiz} which is any even function $V(S_z^2)$ of $S_z$, thus preserving the $\mathbb{Z}_2\times\mathbb{Z}_2^T$ symmetry. The $2s+1$ states are still decoupled and $V(S_z^2)$ behaves as the unit operator in each of the $S_z$ eigenstates. Therefore these are relevant deformations of scaling dimension 0. 
For $s\in \mathbb{Z}_{\geq 0}$ we can use such a potential to reduce all the way to the trivial defect. For $s\in \mathbb{Z}_{\geq 0}+\frac12$ we will necessarily be left at the very least with 
\begin{equation}\label{dsum}
    \mathcal{D}[+]\oplus \mathcal{D}[-].
\end{equation}
The anomaly we found above implies that it is not possible to destroy~\eqref{dsum} (in the sense of having it flow to a trivial defect) by any $\mathbb{Z}_2\times\mathbb{Z}_2^T$-symmetric deformation.

\subsubsection{Stable spontaneous symmetry breaking}

By an old argument of Peierls \cite{peierls1936ising}, there is no spontaneous breaking of discrete symmetries in 1d classical systems  due to the proliferation of domain walls. 

Here we consider a quantum problem where a qubit is interacting with a bath of fluctuating bosons and ask if the qubit degeneracy is protected if we impose certain symmetries. See \cite{Cuomo:2023qvp} for a recent paper which investigates similar questions for surface defects.

It is in principle possible for a line defect embedded in a CFT to exhibit stable spontaneous breaking of a discrete symmetry. Here, ``stable'' means that the line defect should not admit any relevant symmetry-preserving operators. (That is, in particular, that domain wall operators cannot proliferate and destroy the semi-simple structure of the defect.) See some clarification of this terminology in Footnote~\ref{SSBline}.

The defect \eqref{dsum} exhibits spontaneous symmetry breaking, but whether or not the symmetry breaking is \emph{stable} depends on what symmetries we impose.
For example, an  interesting deformation that preserves $U$ but breaks $\mathcal{T}'$ is $S_x$. Similarly a deformation that preserves $\mathcal{T}'$ but breaks $U$ is $S_y$. The scaling dimension of the deformations $S_x$ and $S_y$ of the conformal defect~\eqref{dsum} has been deduced recently to be $\Delta<1$ \cite{Zhou_2024,coordinate} and both of these deformations lead to the flow 
\begin{equation}
\label{SXdef}\mathcal{D}[+]\oplus \mathcal{D}[-]\xrightarrow{\mathrm{IR}}1~.\end{equation}
This flow is non-perturbative and is consistent with~\eqref{strictin} since 
\begin{align}
 s(\mathcal{D}[+])=s(\mathcal{D}[-])\approx \log 0.6 \Rightarrow s(\mathcal{D}[+]\oplus \mathcal{D}[-])\approx \log(0.6+0.6)>0=s(1).
\end{align}
Hence, it is in principle possible to have the direct sum of two pinning fields flow to a completely trivial infrared line, as long as we break the $\mathbb{Z}_2\times\mathbb{Z}_2^T$ symmetry.

However, there is no relevant deformation that is $\mathbb{Z}_2^T\times \mathbb{Z}_2$ invariant. Therefore the direct sum~\eqref{dsum}
can be viewed as an example of discrete symmetry breaking on a line defect, i.e.\ a defect with two superselection sectors which is stable under the renormalization group despite the nontrivial  interactions.

Above we have made a specific choice of $\mathcal{T}'$, but there are many other choices  that preserve~\eqref{pinningiz}. This is due to the fact that the system \eqref{pinningiz} admits a continuous symmetry, 
\begin{align}
    V_\alpha: z_1\to e^{i\alpha/2}z_1, \ \ \ \ z_2\to e^{-i\alpha/2}z_2,
\end{align}
 which acts on the defect only and does not touch the bulk fields at all. 
We can use this unitary to define a family of time reversal symmetries 
$\mathcal{T}'_\alpha = \mathcal{T}' V_\alpha$. We will make some comments on the particular choice $\mathcal{T}'_{\pi}$. Then our time reversal matrix acting on~\eqref{dsum} becomes
$\mathcal{T}'_\pi = K$ and the action on the endpoints now satisfies $U^2=-1$ and $\mathcal{T}'_\pi U = - U \mathcal{T}'_\pi$. Both of these minus signs can be removed by redefining $U\to i U$. Therefore the class in $H^2(\mathbb{Z}_2^T\times \mathbb{Z}_2, U_T(1))$ is now trivial. 
And indeed, the deformation of the action by $S_x(\tau)$ is now invariant both under $U$ and also under  $\mathcal{T}'_\pi$. Deforming the system by $S_x(\tau)$ drives the system to a trivial phase as in~\eqref{SXdef}.

The summary is that whether the ``spontaneous symmetry breaking" line defect~\eqref{dsum} is stable depends on the precise symmetries we impose. This depends on the lattice regularization, for instance. If we insist on preserving all the symmetries of~\eqref{pinningiz} then the symmetry breaking line defect is stable.  If our microscopic realization preserves only $\mathcal{T}'_\pi$ and $U$, then the symmetry breaking state is unstable to domain wall proliferation.  

\subsubsection{A lattice model}

Let us conclude our discussion of the 2+1d Ising model by framing the discussion above in the context of a lattice model.
Consider the lattice Hamiltonian
\begin{gather}\label{eqn:defectIsinglattice}
    H_{\mathcal{D}} = - J \sum_{\langle i,j \rangle, i,j \neq 0} Z_i Z_j - h_\perp \sum_{i\neq 0} X_i - J_d\sum_{\langle i,0\rangle} Z_i Z_0,
\end{gather}
with $X_i$ and $Z_i$ the Pauli-X and -Z operators supported at the site $i$. The critical point of this model is at $h_\perp \approx 3.044 J$ \cite{de1998critical,Blote:2002ieo,Hashizume:2018zbg}.
This Hamiltonian has in particular the following $\mathbb{Z}_2\times\mathbb{Z}_2^T$ symmetry,
\begin{gather}\label{eqn:latticeZ2xZ2T}
U = i\prod_i X_i, \quad \mathcal{T}' = K Z_0,
\end{gather}
where $K$ denotes complex conjugation. We observe that
\begin{gather}
U^2 = -1, \quad  \mathcal{T'}^2 = 1, \quad   \mathcal{T}' U  = U \mathcal{T'}, \quad
\end{gather}
which realizes the cohomologically non-trivial 2-cocycle $\omega_{1,0}$ from \eqref{eqn:explicit2cocycles}.
Thus, we have a non-trivial projective representation of $\mathbb{Z}_2 \times \mathbb{Z}_2^T$, and in particular the defect is non-trivial in the IR. In fact, it flows to the direct sum of pinning field defects in Equation \eqref{dsum}. One signature of the anomaly which should be accessible to simulations is an exact ground state degeneracy of \eqref{eqn:defectIsinglattice}, e.g.\ when placed on a square lattice with periodic boundary conditions. We stress that this degeneracy is robust to interactions that respect the symmetry, like $\delta H_{\mathcal{D}}=\lambda\sum_{\langle i\neq j,0\rangle} X_i Y_0 Z_j$.

The symmetries in \eqref{eqn:latticeZ2xZ2T} forbid several defect local operators which would mix the ground states and thus lead to the breaking of the two fold degeneracy, like $X_0$ and $Y_0$. Indeed, one can check that the first operator that can be added to the defect consistent with the symmetries \eqref{eqn:latticeZ2xZ2T} has dimension $\Delta > 1$, and thus the defect is IR stable \cite{coordinate}.
However, there is another choice of $\mathbb{Z}_2\times\mathbb{Z}_2^T$,
 \begin{gather}
     U = \prod_i X_i, \quad \mathcal{T} = K,
 \end{gather}
 which is realized non-anomalously.
  If we enforce just this symmetry, we find there is no obstruction to breaking the two fold degeneracy of $H_{\mathcal{D}}$.

\subsection{A Magnetic Defect in the XY model: The Halon problem}\label{subsec:XY}

We now turn to the $O(2)$ model. Our starting point is similar to~\eqref{pinningiz}, but now with a coupling that preserves $O(2)$,
\begin{equation}\label{pinningi} S=S_{O(2)}+S_{\mathrm{spin}\text{-}  s}+h \int d\tau \left[\Phi(\vec x=0,\tau) S_-(\tau)+ {\rm c.c.}\right], \quad S_\pm =  S_x \pm i S_y.
\end{equation}
Here, we have introduced  a complex scalar $\Phi=\phi_1+i\phi_2$.

\subsubsection{Symmetries}

The $O(2)$ symmetry  in the bulk is generated by rotations of $\Phi$ and complex conjugation. In the presence of the impurity, we must also transform the defect degrees of freedom. In particular, we can take
\begin{align}
\begin{split}
    U_\alpha &: \Phi \to e^{i\alpha}\Phi, \ \ \ \ \ z_1\to e^{-i\alpha/2}z_1, \ \ \ \ \ z_2\to e^{i\alpha/2} z_2, \\
     C&: \Phi \to \Phi^\ast, \ \ \ \ \ \ \ \ z_1 \to -i z_2, \ \ \ \ \ \ \ \ \ \ \ z_2\to -i z_1.
\end{split}
\end{align}
Note that, in the quantum mechanics theory \eqref{bis}, the transformation $z_1\to e^{-i\alpha/2}z_1$ and $z_2\to e^{i\alpha/2}z_2$ acts on the $|m\rangle$ states via the operator $e^{i\alpha S_z}$, and the transformation $z_1\to-iz_2$ and $z_2\to-iz_1$ acts via the operator $e^{i\pi S_x}$.

In addition, the system possesses the following time reversal symmetry, 
\begin{align}
    \mathcal{T}':\Phi\to \Phi^\ast, \ \ \ \ z_1\to iz_2^\ast, \ \ \ \ z_2\to iz_1^\ast
\end{align}
which, on the $|m\rangle$ states, acts as $\mathcal{T}e^{i \pi S_z}$, with $\mathcal{T}$ defined in \eqref{eqn:Tactionmstates}. This is a natural choice since it commutes with the $O(2)$ symmetry, so that in particular our symmetry class is $\mathbb{Z}_2^T\times O(2)$.

\subsubsection{Defect 't Hooft anomalies}

To compute the anomalies, we start by noting that, by an application of the K\"unneth formula \cite[Equation (4.13)]{Inamura:2021wuo}, 
\begin{align}\label{eqn:kunneth}
\begin{split}
    H^2(\mathbb{Z}_2^T\times O(2),U_T(1)) &\cong  \mathbb{Z}_2\times{_2}H^2(O(2),U(1))\times H^1(O(2),U(1))/2H^1(O(2),U(1)) \\
    &\cong \widetilde{\mathbb{Z}}_2^T\times\widetilde{\mathbb{Z}}_2^{O(2)}\times\widetilde{\mathbb{Z}}_2^{T/O(2)},
\end{split}
\end{align}
where ${_2}H^2(G,U(1))$ refers to the 2-torsion subgroup of $H^2(G,U(1))$, i.e.\ the classes $\nu$ for which $\nu^2=1$. 
Therefore it is possible that the defect introduces anomalies (there are of course no bulk anomalies since the model can be gapped symmetrically by a mass term). 

The first  factor  $\widetilde{\mathbb{Z}}_2^T$ encodes the pure time-reversal symmetry anomaly of the system, which can be diagnosed by checking whether $\mathcal{T}^{'2}=\pm 1$. In the present case, a straightforward computation reveals that $\mathcal{T}^{'2}=+1$ so that the contribution from  $\widetilde{\mathbb{Z}}_2^T$  to the anomaly is absent.

The second factor $\widetilde{\mathbb{Z}}_2^{O(2)}$ encodes the pure $O(2)$ anomaly of the system. In a system with a $\mathbb{Z}_2^T\times G$ symmetry, the pure $G$ anomaly of the system cannot be arbitrary, but rather must belong to the 2-torsion subgroup. This can be seen by noting that, for $\omega\in H^2(G\times\mathbb{Z}_2^T,U_T(1))$ and $g,h\in G$,
\begin{align}\label{eqn:Tidentity}
    \frac{\omega(g,h)}{\omega(h^{-1},g^{-1})}=\frac{\phi(g)\phi(h)}{\phi(gh)}, \ \ \ \ \phi(g)=\frac{\omega(T,g^{-1})\omega(Tg^{-1},T)\omega(g,g^{-1})}{\omega(T,T)},
\end{align}
which implies that the 2-cocycle $\omega(g,h)/\omega(h^{-1},g^{-1})$ is trivial in $H^2(G,U(1))$. On the other hand, it is possible to show that $\omega(g,h)/\omega(h^{-1},g^{-1})$ is cohomologous to $\omega(g,h)^2$, so that in particular $\omega$ resides in the 2-torsion subgroup. 

Since $H^2(O(2),U(1))\cong \mathbb{Z}_2$ is equal to its 2-torsion subgroup, the existence of a commuting time reversal imposes no restriction on the pure $O(2)$ anomaly $\widetilde{\mathbb{Z}}_2^{O(2)}$. To determine whether the pure $O(2)$ anomaly is trivial or not, it is sufficient to compute its restriction to the $\mathbb{Z}_2\times\mathbb{Z}_2=\langle U_\pi,C\rangle$ subgroup. In the present situation, when $s$ is half-integer, it is not possible to redefine $U'_\pi = \alpha U_\pi$ and  $C'=\beta C$ by phases so that $U'_\pi$ and $C'$ generate a $\mathbb{Z}_2\times\mathbb{Z}_2$ representation with trivial 2-cocycle, since $U_\pi C = (-1)^{2s} CU_\pi$. On the other hand, when $s$ is an integer, the 2-cocycle is trivial on the nose. Thus, there is a non-trivial anomaly in $\widetilde{\mathbb{Z}}_2^{O(2)}$ if and only if $s$ is a half-integer.

Now, suppose one works with a representative satisfying $\omega(g,h)/\omega(h^{-1},g^{-1})=1$, which is possible since the 2-cocycle is trivial in cohomology. In this situation, Equation \eqref{eqn:Tidentity} says that $\phi$ is a homomorphism $G\to U(1)$; the image of $\phi$ under the natural quotient map $H^1(G,U(1))\to H^1(G,U(1))/2H^1(G,U(1))$ contributes to the third factor on the right-hand side of Equation \eqref{eqn:kunneth}. This can be interpreted as a kind of mixed anomaly between $G$ and $\mathbb{Z}_2^T$. In the case that $G=O(2)$, homomorphisms $G\to U(1)$ are completely determined by whether they assign $+1$ or $-1$ to charge conjugation, and so it is sufficient to compute the mixed anomaly between time reversal symmetry and charge conjugation. An elementary calculation shows that, in the defected $O(2)$ model, $\phi(C)=1$ so that there is no mixed anomaly.

To summarize, the anomaly of $\mathbb{Z}_2^T\times O(2)$ in the model \eqref{pinningi} is described by the class 
\begin{align}\label{eqn:TxO(2)anomaly}
    (0,2s,0)\in\widetilde{\mathbb{Z}}_2^T\times \widetilde{\mathbb{Z}}_2^{O(2)}\times\widetilde{\mathbb{Z}}_2^{T/O(2)}.
\end{align}
That is, there is a pure $O(2)$ anomaly, but no mixed anomaly between time-reversal and $O(2)$, and no pure time reversal anomaly. This answer is consistent with the fact that there are  no anomalies for integer $s$: indeed,  we can add to the defect action $V=S_z^2$ so that the ground state is the $S_z=0$ component, and the rest are integrated out and hence the line becomes trivial. 

For $s=\sfrac12$, it will be convenient for later when we compare to the $\pi$-flux vortex line to note the following explicit matrix representation,
\begin{equation}\label{Halsymm} 
U_\alpha =e^{{i\frac\alpha2}  \sigma_z}~,\quad C=i\sigma_x~,\quad \mathcal{ T}' = \sigma_y   \sigma_z K = i\sigma_x K~. 
\end{equation} 
This is known as $\mathrm{Pin}_-(2)$.

Making the relatively safe assumption that there are no non-trivial topological line operators in the critical $O(2)$ model, this anomaly implies that, for half-integer $s$, the defect~\eqref{pinningi}  is either coupled to the bulk and is a nontrivial conformal defect in the infrared, or it is TQM$_{k\geq 2}$, i.e.\ decoupled from the bulk with a Hilbert space which is at least 2 dimensional and is in a projective representation of $O(2)$. (We will give a more detailed proposal shortly.) However, for integer $s$, we cannot arrive at any conclusions by appealing to standard defect 't Hooft anomalies alone.

\subsubsection{An anomaly in the space of defect couplings}

Equation \eqref{eqn:TxO(2)anomaly} says that the spin impurity in the $O(2)$ model does not have any of the standard defect 't Hooft anomalies (the ones described in Section \ref{subsec:defectanomalies}) for its $\mathbb{Z}_2^T\times O(2)$ symmetry when $s$ is an integer. But the model \emph{does} possess an anomaly in the space of defect couplings (like the ones described in Section \ref{subsec:anomaliesinspaceofdefectcouplings}) for all values of $s$.
To reveal this anomaly, we deform the action in \eqref{pinningi} as 
\begin{align}\label{eqn:Szdeformed}
    S\to S+\gamma \int d\tau S_z(\tau).
\end{align}
This clearly breaks $C$, but it preserves the $U(1)$ symmetry $U_\alpha$. It also preserves $\mathcal{T}'' = \mathcal{T}'C$, which acts by complex conjugation on  the coefficients in the $|m\rangle$ basis. The
symmetry is therefore $SO(2) \rtimes \mathbb{Z}_2^T$.
It is realized with no anomaly on the endpoints, and the infrared limit with $\gamma\neq 0$ is a trivial line.

As in Section \ref{subsubsec:spinsanomalycouplings},
the SPT phase jumps when comparing $\gamma\to\pm\infty$,
\begin{equation}\label{ratio} \lim_{\gamma\to\infty}\frac{Z_{\gamma}[A]}{Z_{-\gamma}[A]} = \exp\left( 2i s \int A \right)~.
\end{equation}
For half-integer $s$, we see that half-integer $U(1)$ charge accumulates on the trivial defect at $\gamma\to\pm\infty$. 
This is striking as no bulk excitations possess fractional $U(1)$ charge. This fractionalization has been seen clearly in numerics~\cite{Chen:2018xqa}. 
One could also think of the parameter space traced out by $\gamma$ as a circle $\mathbb{S}^1=\mathbb{R}\cup \{ \infty\}$, and then the discussion above can be phrased as indicating an anomaly in the space of defect couplings.
We can conclude the following from~\eqref{ratio}: \\

 \noindent\textbf{Claim:} \emph{For any $s$, there is at least one value of the coupling for which the deformed spin impurity of the $O(2)$ model in \eqref{eqn:Szdeformed} defines a non-trivial defect in the IR (though it may be TQM). Similarly, a degenerate ground space must occur somewhere in parameter space when the impurity is placed on a compact manifold, such as $S^2$ or $\mathbb{T}^2$.} \\
 
 \noindent  We  remark that this conclusion is robust to any deformation (such as an arbitrary potential $V(S_z)$) which preserves the $U(1)$ symmetry. For half-integer $s$, the standard defect 't Hooft anomalies in Equation \eqref{eqn:TxO(2)anomaly} guarantee that the symmetric impurity at $\gamma_\ast=0$ is one such diabolical point. For integer $s$, it is less clear that $\gamma=0$ defines a non-trivial defect. We formulate a conjecture about this in the next subsubsection.

 \subsubsection{Examples with small spin $s$ and a conjecture}\label{subsubsec:smallsexamples}

Above we concluded that the $s=\sfrac12$ spin impurity at $\gamma_\ast =0$ must be either TQM$_{k>1}$ or conformal. Which of these occurs can be definitively settled using the $g$-theorem. The ultraviolet ($h=0$) defect entropy is $s_{\rm UV} = \log (2s+1)$. Therefore, if the defect is TQM in the IR, then 
$\log(2s+1)>\log(\dim R_\omega)$ where $\log(\dim R_\omega)>\log 2 $ for half-integer $s$. This leads to a contradiction when $s=\sfrac12$, and we conclude:  \\

\noindent \textbf{Claim:} \emph{Assume the absence of non-trivial topological line operators in the critical $O(2)$ model. Then the line defect in Equation \eqref{pinningi} flows to a nontrivial conformal defect in the infrared for $s=\sfrac12$. It possesses a degenerate space of lowest-dimension endpoint operators. Its defect entropy is bounded as $0<s_{\mathrm{IR}}<\log 2$.}  \\

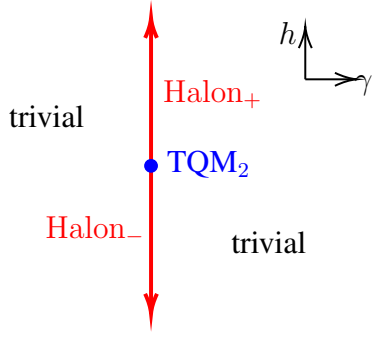
\begin{figure}
    \begin{center}
        \tikzset{every picture/.style={line width=0.75pt}} 

\begin{tikzpicture}[x=0.75pt,y=0.75pt,yscale=-1,xscale=1]

\draw [color=red
,draw opacity=1 ][line width=1.5]    (77,151) -- (77,7.92) ;
\draw [shift={(77,4.92)}, rotate = 90] [color= red
,draw opacity=1 ][line width=1.5]    (9.95,-2.99) .. controls (6.32,-1.27) and (3.01,-0.27) .. (0,0) .. controls (3.01,0.27) and (6.32,1.27) .. (9.95,2.99)   ;
\draw [shift={(77,154)}, rotate = 270] [color=red
,draw opacity=1 ][line width=1.5]    (9.95,-2.99) .. controls (6.32,-1.27) and (3.01,-0.27) .. (0,0) .. controls (3.01,0.27) and (6.32,1.27) .. (9.95,2.99)   ;
\draw    (154.79,35.79) -- (178.67,35.79) ;
\draw [shift={(180.67,35.79)}, rotate = 180] [color=
{rgb, 255:red, 0; green, 0; blue, 0 } 
][line width=0.75]    (10.93,-3.29) .. controls (6.95,-1.4) and (3.31,-0.3) .. (0,0) .. controls (3.31,0.3) and (6.95,1.4) .. (10.93,3.29)   ;
\draw  [draw opacity=0][fill=blue
,fill opacity=1 ] (73.52,79.46) .. controls (73.52,77.54) and (75.08,75.98) .. (77,75.98) .. controls (78.92,75.98) and (80.48,77.54) .. (80.48,79.46) .. controls (80.48,81.38) and (78.92,82.94) .. (77,82.94) .. controls (75.08,82.94) and (73.52,81.38) .. (73.52,79.46) -- cycle ;
\draw    (154.79,35.79) -- (154.79,11.09) ;
\draw [shift={(154.79,9.09)}, rotate = 90] [color= 
{rgb, 255:red, 0; green, 0; blue, 0 } 
][line width=0.75]    (10.93,-3.29) .. controls (6.95,-1.4) and (3.31,-0.3) .. (0,0) .. controls (3.31,0.3) and (6.95,1.4) .. (10.93,3.29)   ;

\draw (178.33,32.07) node [anchor=north west][inner sep=0.75pt]    {$\gamma $};
\draw (140,5.4) node [anchor=north west][inner sep=0.75pt]    {$h$};
\draw (116,111) node [anchor=north west][inner sep=0.75pt]  [color={rgb, 255:red, 0; green, 0; blue, 0 }  ,opacity=1 ] [align=left] {trivial};
\draw (4,47.33) node [anchor=north west][inner sep=0.75pt]  [color={rgb, 255:red, 0; green, 0; blue, 0 }  ,opacity=1 ] [align=left] {trivial};
\draw (83.67,70.33) node [anchor=north west][inner sep=0.75pt]  [color=blue
,opacity=1 ] [align=left] {TQM$_2$};
\draw (81.67,34.4) node [anchor=north west][inner sep=0.75pt]  [
color= red
,opacity=1 ]  {$\mathrm{Halon}_{+}$};
\draw (22.33,103.07) node [anchor=north west][inner sep=0.75pt]  [color = red
,opacity=1 ]  {$\mathrm{Halon}_{-}$};

\end{tikzpicture}
        \caption{The phase diagram for the deformed spin-$\sfrac12$ impurity \eqref{eqn:Szdeformed} in the critical $O(2)$ model. The defect is trivial for all $\gamma\neq 0$. Along the $\gamma=0$ locus, the defect becomes decoupled (TQM$_2$) at $h=0$ and otherwise flows to the conformal halon impurity.}\label{fig:halonphasediagram}
    \end{center}
\end{figure}

\noindent The fact that $s_{\mathrm{IR}}>0$ follows from the fact that the defect can be deformed to a trivial line using \eqref{eqn:Szdeformed}. We call this fixed point the ``Halon'' following \cite{Chen:2018xqa}. There are very strong numerical indications that this fixed point exists~\cite{Chen:2018xqa}, as well as results using the $\epsilon$ expansion \cite{Whitsitt:2017pmf}, in agreement with the theorem above. The phase diagram is depicted in Figure \ref{fig:halonphasediagram}, and the RG flow lines are essentially identical to those in Figure \ref{fig:RGflow}. As one can see, for all $h$, the diabolical value of $\gamma$ guaranteed by the anomaly in the space of defect couplings, Equation \eqref{ratio}, is $\gamma_\ast=0$.

Let us next consider the $s=1$ spin impurity of the critical $O(2)$ model. There is now a further relevant deformation $u$,
\begin{align}\label{eqn:s=1spinimpurity}
\begin{split}
    &S=S_{O(2)}+S_{{\rm spin}-1} \\
    & \hspace{.5in}+\int d\tau\left( h(\Phi(\vec{x}=0,\tau) S_-(\tau)+ \Phi^\ast(\vec{x}=0,\tau) S_+(\tau))+\gamma S_z(\tau)+u S^2_z(\tau) \right).
\end{split}
\end{align}
Let us fix $h>0$ and study the phase diagram of the defect in the $(\gamma,u)$ plane. The perturbation $S_z^2$ is relevant and thus must be fine tuned to some specific point $u_*$ that depends on the regularization scheme. Far away from this point, we can have two phases. 
For $u\gg u_\ast$, the $S_z=0$ state is energetically favorable and the defect is eventually driven to a trivial phase. When $u\ll u_\ast$, the physics depends on the sign of $\gamma$. At $\gamma=0$, the $S_z=\pm 1$ states are equally energetically favorable and we expect a TQM$_2$ phase. There is no anomaly which protects this degeneracy, so when we turn on $\gamma$ we expect to flow to a trivial phase where either $S_z=+1$ or $S_z=-1$ is the ground state, depending on the sign of $\gamma$.

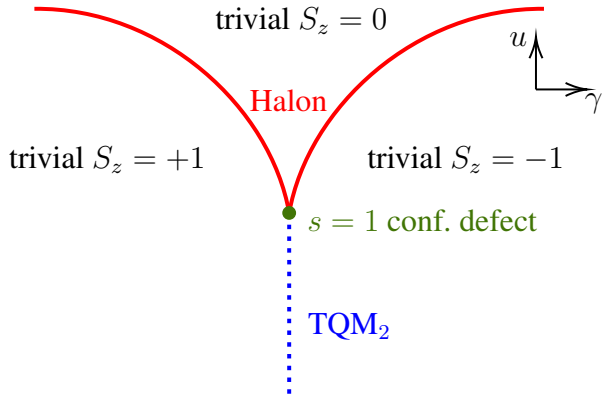
\begin{figure}
    \begin{center}
        \tikzset{every picture/.style={line width=0.75pt}} 

\begin{tikzpicture}[x=0.75pt,y=0.75pt,yscale=-1,xscale=1]

\draw    (262.13,57.46) -- (286,57.46) ;
\draw [shift={(288,57.46)}, rotate = 180] [color={rgb, 255:red, 0; green, 0; blue, 0 }  ][line width=0.75]    (10.93,-3.29) .. controls (6.95,-1.4) and (3.31,-0.3) .. (0,0) .. controls (3.31,0.3) and (6.95,1.4) .. (10.93,3.29)   ;
\draw    (262.13,57.46) -- (262.13,32.76) ;
\draw [shift={(262.13,30.76)}, rotate = 90] [color={rgb, 255:red, 0; green, 0; blue, 0 }  ][line width=0.75]    (10.93,-3.29) .. controls (6.95,-1.4) and (3.31,-0.3) .. (0,0) .. controls (3.31,0.3) and (6.95,1.4) .. (10.93,3.29)   ;
\draw [color=red
,draw opacity=1 ][line width=1.5]    (137.67,119.79) .. controls (147,66.76) and (201.5,16.09) .. (266.17,16.76) ;
\draw [color= blue
,draw opacity=1 ][line width=1.5]  [dash pattern={on 1.69pt off 2.76pt}]  (137.67,119.79) -- (137.67,211.42) ;
\draw [color= red
,draw opacity=1 ][line width=1.5]    (137.67,119.79) .. controls (128.33,66.76) and (73.83,16.09) .. (9.17,16.76) ;
\draw  [draw opacity=0][fill={rgb, 255:red, 65; green, 117; blue, 5 }  ,fill opacity=1 ] (134.19,119.79) .. controls (134.19,117.87) and (135.75,116.32) .. (137.67,116.32) .. controls (139.59,116.32) and (141.15,117.87) .. (141.15,119.79) .. controls (141.15,121.72) and (139.59,123.27) .. (137.67,123.27) .. controls (135.75,123.27) and (134.19,121.72) .. (134.19,119.79) -- cycle ;

\draw (285.33,58.07) node [anchor=north west][inner sep=0.75pt]    {$\gamma $};
\draw (247.33,27.07) node [anchor=north west][inner sep=0.75pt]    {$u$};
\draw (175.67,84.76) node [anchor=north west][inner sep=0.75pt]   [align=left] {trivial $S_z = -1$};
\draw (-5,84.76) node [anchor=north west][inner sep=0.75pt]   [align=left] {trivial $S_z = +1$};
\draw (116.33,55.76) node [anchor=north west][inner sep=0.75pt]  [color=red
,opacity=1 ] [align=left] {Halon};
\draw (145.67,168.42) node [anchor=north west][inner sep=0.75pt]  [color=
blue
,opacity=1 ] [align=left] {TQM$_2$};
\draw (145.33,116.42) node [anchor=north west][inner sep=0.75pt]  [color={rgb, 255:red, 65; green, 117; blue, 5 }  ,opacity=1 ] [align=left] {$s=1$ conf.\ defect};
\draw (99,14.42) node [anchor=north west][inner sep=0.75pt]   [align=left] {trivial $S_z = 0$};

\end{tikzpicture}
        \caption{The phase diagram of the $s=1$ spin impurity, Equation \eqref{eqn:s=1spinimpurity}, at fixed $h>1$. Taken from Figure 1 of \cite{chen2019even}.}\label{fig:s=1O(2)phasediagram}
    \end{center}
\end{figure}

The analysis of the extreme regions of the phase diagram above implies that, for every $\gamma$, we should expect to encounter a phase transition as we tune the parameter $u$. The nature of these transitions was studied in \cite{chen2019even}. There, it was found that at $\gamma=0$, one encounters a fine-tuned (infrared unstable) multicritical conformal line as one tunes $u$. Away from $\gamma=0$, one finds the $s=\sfrac12$ critical Halon impurity interpolating between the trivial phases. This discussion is summarized in Figure \ref{fig:s=1O(2)phasediagram}.

Notice the coarse topological feature that, for each $u$, one  encounters at least one diabolical value $\gamma_\ast$ as one moves from left to right on the phase diagram. This is guaranteed by the anomaly in the space of defect couplings, Equation \eqref{ratio}. As this example illustrates, there can be more than one diabolical point, though they must appear in a manner which is symmetric under $\gamma\to -\gamma$ due to charge conjugation symmetry. The nature of the diabolical point(s) also varies as one tunes $\gamma$: for $u\ll 0$ it is a TQM defect, for $u\sim 0$ it is a multicritical $s=1$ conformal line, and for $u\gg 0$ one encounters a pair of diabolical conformal Halon lines.

As one cranks up the spin $s$ of the impurity, there are more and more relevant operators and the analysis of the phase diagram becomes increasingly complicated. Nevertheless, we expect the following generalization of the results described thus far: \\

\noindent\textbf{Conjecture:} The spin impurity of the critical $O(2)$ model, Equation \eqref{pinningi}, flows to a fine-tuned, infrared unstable conformal line operator for all $s$.\\

\noindent  The defect entropy of the spin-$s$ conformal line is necessarily bounded from above by $\log(2s+1)$ and, if the conjecture above is true, is also bounded from below by $0$. It would be useful to obtain evidence for this proposal, e.g.\ by performing a large $s$ expansion, as was done in \cite{Cuomo:2022xgw} for the $O(3)$ model (and in~\cite{Krishnan:2024wrc} for the original Kondo problem) or by performing simulations.

\subsubsection{Bulk deformations}

In Section \ref{sec:vortexlines}, we will present a conjecture that the $s=\sfrac12$ Halon spin impurity (with $\gamma_\ast=0$) in the critical $O(2)$ model is particle/vortex dual to the $\pi$-flux vortex line in the Abelian-Higgs model. In order to prepare for this discussion, we now analyze the fate of spin impurities in the $O(2)$ model when the bulk is deformed away from criticality. 

Let $m$ be the mass of the bulk complex scalar $\Phi$. When $m^2\gg 0$, the theory in the bulk becomes trivially gapped, and $\Phi=0$ with very small fluctuations. Thus, in the limit of large positive mass, the coupling between the spin degrees of freedom and the bulk scalar effectively vanishes, and the combined system goes over in the IR to a decoupled spin-$s$ quantum mechanical qubit embedded in a trivially gapped bulk, i.e.\ it becomes TQM$_{2s+1}$.

When $m^2\ll 0$, the bulk $U(1)$ symmetry is spontaneously broken, and the bulk theory goes over to a free compact scalar describing the Goldstone mode. In this phase, the fate of the spin impurity is more involved. 

We can parametrize the bulk scalar as $\Phi(\tau,\vec{x})=ve^{-i\theta(\tau,\vec{x})}$, where $v$ sets the vacuum expectation value, and if we plug in to \eqref{eqn:TxO(2)anomaly} we find 
\begin{align}
    S=S_{O(2)}+S_{\sps}+\int d\tau \left(hve^{-i\theta(\tau,\vec{0})}S_-(\tau)+hve^{i\theta(\tau,\vec{0})}S_-(\tau)+\gamma S_z(\tau)\right).
\end{align}
We would like to integrate out the defect degrees of freedom $z(\tau)$ and $\lambda(\tau)$. 

\begin{figure}
    \centering
    \tikzset{every picture/.style={line width=0.75pt}} 

\begin{tikzpicture}[x=0.75pt,y=0.75pt,yscale=-1,xscale=1]

\draw [line width=1.5]    (10.83,54) -- (173.87,54) ;
\draw [shift={(7.83,54)}, rotate = 0] [color={rgb, 255:red, 0; green, 0; blue, 0 }  ][line width=1.5]    (11.37,-3.42) .. controls (7.23,-1.45) and (3.44,-0.31) .. (0,0) .. controls (3.44,0.31) and (7.23,1.45) .. (11.37,3.42)   ;
\draw [line width=1.5]    (173.87,54) -- (337.25,54) ;
\draw [shift={(340.25,54)}, rotate = 180] [color={rgb, 255:red, 0; green, 0; blue, 0 }  ][line width=1.5]    (11.37,-3.42) .. controls (7.23,-1.45) and (3.44,-0.31) .. (0,0) .. controls (3.44,0.31) and (7.23,1.45) .. (11.37,3.42)   ;
\draw    (173.93,21.96) -- (173.93,40.29) ;
\draw [shift={(173.93,42.29)}, rotate = 270] [color={rgb, 255:red, 0; green, 0; blue, 0 }  ][line width=0.75]    (10.93,-3.29) .. controls (6.95,-1.4) and (3.31,-0.3) .. (0,0) .. controls (3.31,0.3) and (6.95,1.4) .. (10.93,3.29)   ;
\draw  [draw opacity=0][fill={rgb, 255:red, 80; green, 227; blue, 194 }  ,fill opacity=1 ] (168.33,54) .. controls (168.33,50.94) and (170.81,48.47) .. (173.87,48.47) .. controls (176.92,48.47) and (179.4,50.94) .. (179.4,54) .. controls (179.4,57.06) and (176.92,59.53) .. (173.87,59.53) .. controls (170.81,59.53) and (168.33,57.06) .. (168.33,54) -- cycle ;

\draw (331,29.4) node [anchor=north west][inner sep=0.75pt]    {$m^{2}$};
\draw (142.17,4) node [anchor=north west][inner sep=0.75pt]   [align=left] {conformal};
\draw (227,32) node [anchor=north west][inner sep=0.75pt]   [align=left] {TQM$_2$};
\draw (28,28) node [anchor=north west][inner sep=0.75pt]    {topological $\mathbb{Z}_{2}^{( 1)}$};

\end{tikzpicture}
     \caption{The phase diagram of the $s=\sfrac12$ spin impurity  in the $O(2)$ model at $\gamma=0$ as a function of the bulk mass $m^2$.}
    \label{fig:O(2)phasediagram}
\end{figure}
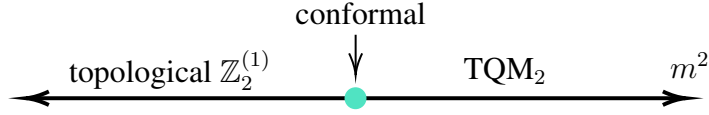

Let us specialize to $s=\sfrac12$ for simplicity. This task essentially amounts to computing the partition function of a qubit with time-dependent Hamiltonian given by 
\begin{align}
    H(\theta)=\left(\begin{array}{cc} \gamma & hv e^{i\theta(\tau)} \\ hve^{-i\theta(\tau)} & -\gamma \end{array}\right), \label{eq:ham_imp_ah_gap}
\end{align}
where we have abbreviated $\theta(\tau) \equiv \theta(\tau,\vec{0})$. 
Since the spectrum is always gapped we can use the adiabatic theorem, 
and computing the Berry and dynamical phases, we find that the time evolution of a state $|\Psi_\pm(t)\rangle$ with initial conditions given by $|\Psi_\pm(-T)\rangle = |\psi_\pm(\theta(-T))\rangle$ is 
\begin{align}
    |\Psi_\pm(t)\rangle = e^{\mp i \sqrt{\gamma^2+(hv)^2}(t+T)}\exp\left(-i \frac{\alpha_\pm}{2\pi} \int_{-T}^t \dot\theta dt   \right)|\psi_\pm(\theta(t))\rangle
\end{align}
where we have defined $\ket{\psi_\pm(\theta)}$ to be instantaneous eigenstates of the Hamiltonian  \eqref{eq:ham_imp_ah_gap} and
\begin{align}
    \alpha_\pm \equiv  \pi\left(1\pm\frac{\gamma}{\sqrt{\gamma^2+(hv)^2}}\right).
\end{align}
If we rotate back to Euclidean signature and compute the trace of the Euclidean time evolution operator to integrate out the defect fields, we find (up to an unimportant overall $\theta$-independent number) that 
\begin{align}
    Z \sim \int \mathcal{D}\theta e^{-S_{\mathrm{eff}}[\theta]}e^{i\frac{\alpha_-}{2\pi} \int \dot\theta d\tau}.
\end{align}
That is, deep in the symmetry breaking phase, the line defect goes over to a topological $U(1)$ one-form symmetry line corresponding to the winding symmetry of the Goldstone boson. 
In particular, for the fully symmetric defect at $\gamma=0$, it is easy to see that $\alpha_-\to \pi$ corresponding to the insertion of the generator of the $\mathbb{Z}_2^{(1)}\subset U(1)^{(1)}$ subgroup.  As one deforms by $\gamma$, the topological line operator one obtains goes from $\alpha_-\to 0$ as $\gamma\to-\infty$ to $\alpha_-\to 2\pi$ as $\gamma\to +\infty$. 
This entire discussion can be summarized by Figure \ref{fig:O(2)phasediagram}.

\subsubsection{A lattice model}\label{subsubsec:xylattice}

We have conjectured above that the $\gamma=0$ spin impurities in the $O(2)$ model are conformal for any spin $s$. To facilitate numerical tests of this proposal, we briefly describe a lattice model for such defects, following
\cite{zhang2013phase,stoudenmire2014corner,whitsitt2017spectrum}:
\begin{align}\label{eqn:O(2)latticemodelspinimpurity}
    H_{\mathcal{D}}=-J\sum_{\langle v,v'\rangle}(S^x_vS^x_{v'}+S^y_vS^y_{v'})-J'\sum_{\langle v,v'\rangle} S^z_v S^z_{v'}+\Delta \sum_v (S^z_v)^2+h(\widetilde{S}^x S_0^x+\widetilde{S}^y S_0^y) + \cdots,
\end{align}
where $\cdots$ stands for further relevant operators that need to be taken into account to tune to the multicritical point. 
Here, we have placed a spin-1 qudit with spin variables $S^a_v$ on each site $v$ of a 2d square lattice, as well as a spin-$s$ spin impurity $\widetilde{S}^a$ at the origin of the lattice, site $v=0$. The sum over $\langle v,v'\rangle$ indicates a sum over nearest neighbors. As one tunes the bulk coupling $\Delta/J$, one encounters a phase transition described by the $O(2)$ universality class for a range of $J'$ around $J'\sim 0$. 

This lattice model manifestly preserves $O(2)$, 
\begin{align}
    U_\alpha = e^{i\alpha \widetilde{S}^z} \prod_v e^{i\alpha S^z_v}, \ \ \ \ C=e^{i\pi \widetilde{S}^x}\prod_v e^{i\pi S^x_v}
\end{align}
and it is easy to check that this symmetry is realized projectively when spin of impurity is a half-integer, in accordance with the anomaly. It would be nice to check our various predictions in this model.

\subsection{A Magnetic Defect in the \texorpdfstring{$O(3)$}{O(3)} model: The Boson-Kondo problem}\label{subsec:O(3)bosonkondoproblem}

Finally, the most symmetric version of the models above is a magnetic defect in the $O(3)$ Wilson-Fisher model. 
The action, analogously to the examples above, is
\begin{equation}\label{pinningiii} S=S_{O(3)}+S_{\sps}+h \int d\tau   \phi^a(\vec x=0,\tau) S_a(\tau)~.\end{equation}
This is the ``Boson-Kondo" model. 
The $SO(3)$ symmetry which in the bulk acts by rotating the $\phi^a$ can be extended to the defect by acting on the $S_a$ in the spin-1 (adjoint) representation, as usual. 
Interestingly, the $O(3)$ symmetry of the bulk is broken to $SO(3)$ by the defect~\eqref{pinningiii}. This is because there is no unitary that takes $S^a\to -S^a$. However, there is a time reversal symmetry $\mathcal{T}$ that is preserved by the defect, 
\begin{align}
    \mathcal{T}: \phi^a \to -\phi^a, \ \ \ \ z_1\to z_2^\ast, \ \ \ \ z_2\to -z_1^\ast,
\end{align}
cf.\ Equation \eqref{Taction}.
The symmetry group is therefore $\mathbb{Z}_2^T\times SO(3)$. For half-integer $s$, the analysis in Section \ref{subsubsec:spinsthooftanomalies} shows that the endpoints are in a projective representation satisfying $\mathcal{T}^2=-1$. 
For half-integer $s$ this precludes a trivial  defect with $s_{\mathrm{IR}} =0$.
We can run through precisely the same argument as for the Halon and conclude the following: \\

\noindent\textbf{Claim:} \emph{The $\mathbb{Z}_2^T\times SO(3)$-symmetric spin impurity in Equation \eqref{pinningiii} with $s=\sfrac12$ is a nontrivial conformal line defect of the $O(3)$-model in the infrared. It has a degenerate space of lowest-dimension endpoint operators. Its defect entropy is bounded as $0<s_{\mathrm{IR}}<\log(2)$.} \\

\noindent This is the $s=\sfrac12$ Boson-Kondo defect. 
If we deform the Boson-Kondo defect in \eqref{pinningiii} using $\gamma \int d\tau S_z(\tau)$, as in \eqref{eqn:Szdeformed}, then a $U(1)$ symmetry is preserved which participates in a mixed anomaly with the defect coupling $\gamma$, analogous to the one in Equation \eqref{ratio}. This implies the following.\\

\noindent\textbf{Claim:} \emph{For any spin $s$, there is at least one value of the coupling $\gamma_\ast$ for which the deformed Boson-Kondo defect  is non-trivial in the IR (though it may be TQM).} \\

\noindent Although for e.g.\ $s=1$, the above theorem guarantees that some value of $\gamma_\ast$ defines a non-trivial defect, it is not known whether the fully symmetric impurity $\gamma_\ast=0$ is such a value, and whether it defines a conformal defect or just a TQM defect. We only know that large enough $s$ leads to a conformal defect at $\gamma_\ast=0$ \cite{Cuomo:2022xgw}, and $s=\sfrac12$ does as well by the above theorem. These circumstances embolden us to conjecture that this holds for all $s$, and that the RG flow diagram is once again captured by Figure \ref{fig:RGflow}. Some numerical evidence in support of this conjecture for $s=1$ appears in \cite{hoglund2007anomalous}.
Interestingly, these defects are all infrared stable.

Not much is known about $O(3)$ invariant defects (as opposed to the $SO(3)$ invariant defects discussed above). One could obtain such a defect as a direct sum of two $s=\sfrac12$ Boson-Kondo defects with the sign of $h$ flipped: 
\begin{equation}\label{O3direct} \mathcal{D}_{\mathrm{BK}}[h>0]\oplus \mathcal{D}_{\mathrm{BK}}[h<0].\end{equation}
To understand whether this defect is stable one would need to consider the scaling dimensions of domain wall operators, as we did for the Ising model in Section \ref{subsec:Ising}.

Before moving on, we briefly describe how one might verify some of these predictions on the lattice. We take as our model the spin-1 antiferromagnetic, anisotropic Heisenberg model on a square lattice,
\begin{align}\label{eqn:SO(3)defectlattice}
    H_{\mathcal{D}}=\sum_{i,j}\left(J\,\mathbf{S}_{i,j}\cdot \mathbf{S}_{i+1,j}+J'\,\mathbf{S}_{i,j}\cdot \mathbf{S}_{i,j+1}    \right) + h\,\mathbf{S}_{0,0}\cdot \widetilde{\mathbf{S}}.
\end{align}
Here, $\mathbf{S}_{i,j}$ is the vector of spin-1 operators at the site $(i,j)$ 
while $\widetilde{\mathbf{S}}$ is the vector of spin operators for a spin-$s$ impurity at the origin of the lattice. As one tunes the ratio $J/J'$ in the bulk one eventually passes through a phase transition which is in the $O(3)$ universality class \cite{matsumoto2001ground,Liu:2021nck}. The model preserves an $SO(3)\times\mathbb{Z}_2^T$ symmetry, with the $SO(3)$ acting in the obvious way as simultaneous rotations of the bulk and impurity spins, while time reversal acts as in Equation \eqref{eqn:Tactionmstates}. It follows from the analysis of Section \ref{subsec:quantuminterlude} that the $SO(3)\times\mathbb{Z}_2^T$ symmetry is realized projectively on the Hilbert space with 2-cocycle correlated with the spin $s$ of the impurity, i.e.\ the defect 't Hooft anomaly corresponds to the class
\begin{align}
    (2s,2s)\in\mathbb{Z}_2\times\mathbb{Z}_2\cong H^2(SO(3)\times\mathbb{Z}_2^T,U_T(1))
\end{align}
just as for the continuum defect in Equation \eqref{pinningiii}. 

It is  natural to expect that \eqref{eqn:SO(3)defectlattice} flows to \eqref{pinningiii} at long distances. Thus, one might attempt to verify by a simulation our theorem that $s=\sfrac12$ is a nontrivial conformal defect and the conjectures  about the $s> \sfrac 12$ Boson-Kondo defects, too. 
One can also test other signatures of the anomaly, e.g.\ the degenerate ground space.\footnote{Another way to understand the appearance of the degenerate spectrum on the lattice is as follows. First, the Hamiltonian is invariant under the $SO(3)$ group, so all energy eigenstates form representations of this group. Second, after introducing the defect spin, the total number of spin-$\frac12$ degrees of freedom becomes odd, which makes it impossible to form a total spin singlet. Consequently, the spectrum must only contain degenerate states.}

We can also attempt to construct $O(3)$ invariant defects on the lattice. In order to manifest the full $O(3)$ symmetry, we replace \eqref{eqn:SO(3)defectlattice} with a bilayer antiferromagnet \cite{wang2006high,Liu:2021nck}, 
\begin{align}\label{bilayer}
    H_{\mathcal{D}}=J\sum_{\langle v,v'\rangle}\left( \mathbf{S}^{(1)}_v\cdot \mathbf{S}^{(1)}_{v'}+\mathbf{S}^{(2)}_v\cdot \mathbf{S}^{(2)}_{v'}\right)+J'\sum_v \mathbf{S}^{(1)}_v\cdot \mathbf{S}^{(2)}_v+h\widetilde{\mathbf{S}}\cdot (\mathbf{S}_0^{(1)}+\mathbf{S}_0^{(2)}).
\end{align}
Here, $\mathbf{S}^{(k)}_v$ is the vector of spin-$\sfrac12$ operators at the site $v$ on the layer $k$ and $\langle v,v'\rangle$ are nearest neighbor sites. Again, this model passes through the $O(3)$ critical theory as one tunes the bulk coupling $J/J'$. In addition to the $SO(3)$ symmetry which simultaneously rotates the bulk and impurity spins, there is a $\mathbb{Z}_2$ layer exchange symmetry which combines with $SO(3)$ to form $O(3)$. The defect in~\eqref{bilayer} is coupled to both layers equally and hence by construction preserves $O(3)$. For a discussion of the low energy physics of this model, see \cite{Liu:2021nck}. 

Note that variations of \eqref{bilayer} were simulated in \cite{hoglund2007anomalous}. Our same arguments apply to these models as well.

\section{Vortex Loops in 2+1d \texorpdfstring{$U(1)$}{U(1)} Gauge Theories: the Abelian-Higgs Model}\label{sec:vortexlines}

In the previous section, we studied examples of impurities obtained by coupling a bulk quantum field theory to  quantum mechanical degrees of freedom living on the worldvolume of a defect. It is also common to study another class of defects known as disorder operators, obtained by imposing singular boundary conditions in the path integral along some locus of spacetime. These two classes are not mutually exclusive: some defects admit dual descriptions using either of the two approaches.

In this section, we will study  disorder operators coming from vortex lines. The main example we treat is the 2+1d Abelian-Higgs model, though many of our observations can be easily extended to Abelian gauge theories with more general matter content. 

Our main motivation for specializing to this example is that it is particle/vortex dual to the XY model considered in Section \ref{subsec:XY}. Indeed, we will present evidence that the Halon impurity of the XY model is dual to a certain vortex line in the Abelian-Higgs model, which we refer to as \textbf{spin-flux duality.} This can be viewed as a non-supersymmetric analog of how mirror symmetry \cite{Intriligator:1996ex} acts on line operators  \cite{Assel:2015oxa,Dimofte:2019zzj}. Indeed, it would be interesting to see if spin-flux duality could be derived by applying supersymmetry breaking deformations to mirror symmetric setups, as was done e.g.\ in \cite{Kachru:2016aon,Kachru:2016rui} in the absence of defects.

\subsection{The Problem}

We consider the Abelian-Higgs model with one complex scalar field of unit charge,
\begin{equation}\label{AHM}
\mathcal{L} = -\frac{1}{4e^2}f_{\mu\nu}f^{\mu\nu}+\left|D_\mu\Phi\right|^2 -V(|\Phi|^2)~, \ \ \ \ \ D_\mu\Phi\equiv(\partial_\mu-ia_\mu)\Phi.
\end{equation}
It is well known that by choosing $V(|\Phi|^2)$
appropriately one can obtain a nontrivial fixed point, e.g.\ $V(|\Phi|^2) = \lambda  |\Phi|^4$ with sufficiently large $\lambda>e^2$. This fixed point is in the $O(2)$ universality class, and in particular is dual to the critical XY model \cite{Peskin:1977kp,Dasgupta:1981zz,Seiberg:2016gmd,Karch:2016sxi}.

We will study the vortex loop defect, which is defined as a codimension-2 defect with gauge holonomy \begin{equation}\label{AH}\lim_{l\to 0}\exp\left(i \oint_{C_l} a\right) =e^{i\alpha}~,\end{equation} 
where $C_l$ is any  loop of length $l$ linking with the defect worldline.
Unlike in supersymmetric theories, we do not expect generic $\alpha$ to define a nontrivial conformal defect. 
For generic $\alpha$ we should therefore interpret~\eqref{AH} with some cutoff $\epsilon$, and a renormalization group in the space of holonomies (which is the space of conjugacy classes) occurs. 
We expect $\alpha=\pi~\mathrm{mod}~2\pi$ to be a fixed point of the renormalization group since by charge conjugation symmetry $\alpha=\pi~\mathrm{mod}~2\pi$ cannot flow. For all other values of $\alpha$ the infrared limit is expected to be trivial and hence we focus on studying the infrared limit of 
\begin{equation}\label{AHpi}\lim_{l\to 0} \exp\left(i \oint_{C_l} a\right) =-1~.\end{equation} 
Our main proposal concerning this defect is the following infrared duality: \\

\noindent\textbf{Conjecture:} \emph{The long distance limit of the $\pi$-flux vortex loop~\eqref{AHpi} of the Abelian-Higgs model and the Halon impurity~\eqref{pinningi} of the XY model are related under particle/vortex duality. The deformation $\gamma \int d\tau S_z(\tau)$ of the Halon in \eqref{eqn:Szdeformed} maps to deformations of the holonomy of the vortex line, \eqref{AH}, thought of as a defect coupling constant.} \\

\noindent This conjecture is highly nontrivial and would need to be verified by a simulation. An immediate corollary, by virtue of our theorem about the Halon in Section \ref{subsubsec:smallsexamples}, is that the $\pi$-flux vortex line in the Abelian-Higgs model defines a non-trivial conformal defect. 

Of course, we can start from~\eqref{pinningi} and straightforwardly map it to the Abelian-Higgs model by replacing $\Phi$ with the monopole operator. But that description does not shed light on the infrared limit, which we claim is the vortex loop. We will be able to prove below in Section \ref{subsec:vortexlattice} that coupling the monopole to spin degrees of freedom does lead to a vortex loop on the lattice, though in a regime different from the continuum limit. 

In the rest of this section, we provide evidence for this proposal by matching symmetries, anomalies, and phases across the duality.

\subsection{Symmetries}
We begin by studying the symmetries.
The bulk theory has discrete time reversal and charge conjugation symmetry acting as
\begin{align}
    \mathcal{T}: a\to-a, \quad  C:a\to -a,
\end{align}
where $a = a_0 dt + a_i dx^i$ is a one-form. In particular, the spatial components $a_i$ are odd under time reversal, while the time component $a_0$ is  even. There is also the $U(1)_m^{(0)}$ magnetic zero-form symmetry, which is most conveniently described via the topological operator 
\begin{align}
\widetilde{U}_\eta[\Sigma^{(2)}] = \exp\left(i\frac{\eta}{2\pi}\int_{\Sigma^{(2)}} f\right).
\end{align}
It is elementary to see that the symmetries $\mathcal{T}$, $C$, and $U_\alpha$ together generate the group $\mathbb{Z}_2^T\times O(2)$.

Now, consider introducing the vortex line into the system. Time reversal symmetry and charge conjugation both act on the holonomy as $\alpha\mapsto -\alpha$. Therefore, for a general holonomy $\alpha$, both $\mathcal{T}, \mathcal{C}$ are broken but the combination $\mathcal{T}C$ is preserved. Remarkably, the boundary condition~\eqref{AHpi} with $\alpha=\pi$ preserves $\mathcal{C}$ and $\mathcal{T}$ separately. 

A slightly more involved argument shows that the introduction of the vortex line into the system does not break the $U(1)^{(0)}_m$ symmetry. To see this, we must specify a junction between the topological surfaces which implement $U(1)^{(0)}_m$ and the vortex line. There are two equivalent prescriptions. In one, we excise a small tubular neighborhood $T^{(2)}_\epsilon$ from spacetime around the support $L^{(1)}$ of the vortex line and only integrate $F$ in the complement of this tube, 
\begin{align}
    U_\eta[\Sigma^{(2)}]=\lim_{\epsilon\to 0}\exp\left(  i\frac{\eta}{2\pi} \int_{\Sigma^{(2)}-\Sigma^{(2)}\cap T^{(2)}_\epsilon}f  \right).
\end{align}
In another prescription, we perform the integral over all of $\Sigma^{(2)}$ but add contributions coming from the vortex, 
\begin{align}
    U_\eta'[\Sigma^{(2)}] = \exp\left(-i\frac{\alpha\eta}{2\pi}+i\frac{\eta}{2\pi}\int_{\Sigma^{(2)}}f\right),
\end{align}
where, for the moment, we have assumed that $\Sigma^{(2)}$ only intersects the worldline $L^{(1)}$ of the vortex at a single point. The two prescriptions are equivalent, but let us use the first for simplicity.

\begin{figure}
    \begin{center}
\input{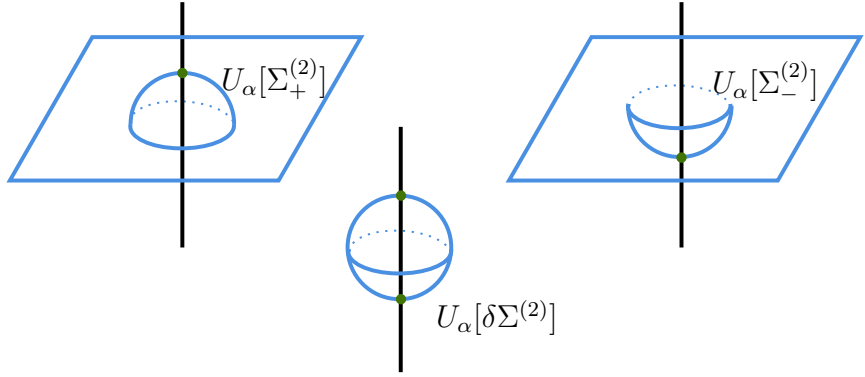}
\caption{The $U(1)^{(0)}_m$ magnetic zero-form symmetry, depicted by the blue surfaces, is preserved by the vortex line, depicted in black.}\label{fig:vortexpreservesurface}
    \end{center}
\end{figure}

To see that the vortex is symmetric, note that the surface operator $U_\alpha[\Sigma^{(2)}]$ is clearly topological with respect to deformations of $\Sigma^{(2)}$ which are supported away from the vortex line. The only thing we must verify is that the same is true of deformations of $\Sigma^{(2)}$ close to the vortex line. 
To this end, consider the quotient $U_\alpha[\Sigma^{(2)}_+]U_\alpha[\Sigma^{(2)}_-]^{-1}$ of two insertions of the magnetic operator on surfaces $\Sigma^{(2)}_+$ and $\Sigma^{(2)}_-$ which are deformed relative to each other in a small region near the vortex. Evaluating this quotient is tantamount to evaluating $U_\alpha[\delta \Sigma^{(2)}]$, where $\delta \Sigma^{(2)}$ is a small 2-sphere  which intersects the vortex line in its north and south poles, and does not enclose any charged operators, as in Figure \ref{fig:vortexpreservesurface}. Invoking Stoke's theorem gives
\begin{align}\label{eqn:northsouth}
    \int_{\delta\Sigma^{(2)}} f = -\oint_N a+\oint_Sa = 0,
\end{align}
where $\oint_N$ and $\oint_S$ are integrals over small circles surrounding the north and south poles of $\delta\Sigma^{(2)}$, respectively; these two contributions  cancel each other since the holonomy of $a$ around the vortex is fixed and there are no enclosed monopoles by assumption. Thus, we find that $U_\alpha[\Sigma^{(2)}_+]=U_\alpha[\Sigma^{(2)}_-]$, i.e.\ that the magnetic symmetry operators can be topologically deformed in the presence of a vortex line. In particular, $U(1)_m^{(0)}$ is preserved.

To summarize, we see that the $\pi$-flux vortex (and more generally the $\alpha$-flux vortex) in the Abelian-Higgs model preserves the same symmetries as the  Halon (and more generally the $\gamma$-deformed Halon) in the $O(2)$ model.

\subsection{Phases}
Before computing the anomalies, it will be helpful to consider phases of the model. Specifically, we study the phase diagram of the $\pi$-flux vortex line as a function of the bulk mass $m^2$. We will see that we reproduce the qualitative features of Figure \ref{fig:O(2)phasediagram}, which depicts the analogous phase diagram in the dual $O(2)$ model.

First, consider tuning to $m^2\gg 0$. In this situation, the bulk scalar $\Phi$ is heavy and can be integrated out, leading to pure Maxwell theory at low energies. Thus, we are lead to ask how vortex lines behave in free Abelian gauge theory. Our claim that the vortex line with flux $\alpha$ can be identified with an emergent $U(1)^{(1)}_e$ electric one-form symmetry line, 
\begin{align}\label{eqn:electric1formline}
    \mathcal{L}^{(e)}_\alpha[L^{(1)}]=\exp\left( i\frac{\alpha}{e^2}\int_{L^{(1)}}\star f\right).
\end{align}

Indeed, this can be seen by starting with an insertion of \eqref{eqn:electric1formline} and performing the path integral over a small tubular neighborhood of $L^{(1)}$ to find that the boundary condition \eqref{eqn:electric1formline} is induced. In particular, the $\pi$-flux vortex line flows to the topological line which generates the $\mathbb{Z}_2^{(1)}\subset U(1)^{(1)}_e$ subgroup.

Free Maxwell theory is dual to the theory of a 2+1d compact boson, with the electric one-form symmetry of the former mapping to the $U(1)^{(1)}$ winding one-form symmetry of the latter. Thus, we can equivalently say that, deep in the $m^2\gg 0$ phase, the Abelian-Higgs model flows to a compact boson,  the $\alpha$-flux vortex line goes over to an emergent $U(1)^{(1)}$ one-form winding symmetry line, and the special case $\alpha=\pi$ flows to the generator of the $\mathbb{Z}_2^{(1)}$ subgroup. This is consistent with what we found in the $m^2\ll 0$ phase of the $O(2)$ model, to which the $m^2\gg 0$ phase of Abelian-Higgs maps under particle/vortex duality. This supports not only that the $\pi$-flux line maps to the Halon, but more generally our claim that the $\mathbb{S}^1$ parameter space traced out by the possible holonomies $\alpha$ should be identified on the $O(2)$ side with the parameter space traced out by the deformation $\gamma$ in Equation \eqref{eqn:Szdeformed}.

Next, let us study the $m^2\ll 0$ phase of the Abelian-Higgs model. In the bulk, the scalar $\Phi$ condenses and, via the Higgs mechanism, we end up with a trivially gapped theory. In a trivially gapped theory, the only possible IR fate of a line defect is that it becomes trivial or TQM, corresponding to scenario \eqref{eqn:trivialdefect} or \eqref{eqn:decoupleddefect}. For example, there are no topological lines for the defect to flow to in a trivially gapped theory, besides a direct sum of some number of copies of the identity line. Our only task is to compute the number of ground states, and confirm that it is $n=2$ to match with the $O(2)$ model in the $m^2\gg 0$ phase (cf.\ Figure \ref{fig:O(2)phasediagram}).

We will argue semiclassically, which is valid deep in the Higgs phase. In particular, we will show that there are two degenerate lowest energy solutions to the classical equations of motion in the presence of the $\pi$-flux vortex line. (This is essentially the problem of a superconductor with a solenoid.) These solutions are  Nielsen-Olesen vortices  \cite{Nielsen:1973cs}  with modified boundary conditions for the gauge field at the core~\eqref{AH}. Since we are interested in a static configuration we just need to minimize energy. If we write the potential as $V(\Phi)=\lambda(|\Phi|^2-v^2)^2$ and assume an ansatz of the form 
\begin{align}
    a_\theta(t,r,\theta) \equiv a(r), \ \ \ \  a_t = a_r = 0, \ \ \ \ \Phi(t,r,\theta) = v(r) e^{i\phi(r,\theta)},
\end{align} then the energy density and the equations of motion are 
\begin{align}
    &\mathcal{E} = \int 2\pi r dr\left[\frac{1}{2e^2}\left(\frac{\partial_r a}{2\pi r}\right)^2 + \left(\partial_r v(r)\right)^2 + v^2(r) \left(\partial_r \phi\right)^2 + \frac{v^2(r)}{r^2} \left(\partial_\theta \phi +  a(r)\right)^2 + V(v(r))  \right] \notag\\
     &v''(r) + \frac{1}{r}v'(r) = 2 v(r) (\partial_r \phi)^2 +\frac{2v(r)}{r^2} (\partial_\theta \phi + a(r))^2 + \lambda v(r) (v^2(r) - v^2) \notag\\
     &\partial_r(r v^2 \partial_r \phi) + \frac{v^2}{r}\partial_\theta^2 \phi = 0, \quad 
    a''(r) - \frac{1}{r}a(r)  = 2\pi e^2 v^2(r)\left(\partial_\theta \phi + a(r)\right) .
\end{align}
We assume that $\lambda\gg e^2$, so that we are in a Type II superconducting phase. Then we can asssume that $v(r) \equiv v$ and the equation for $\phi$ can be solved by taking $\phi(r,\theta) = n \theta$ with $n \in \mathbb{Z}$. Then we have the following equation for $a(r)$,
\begin{align}
    a''(r)-\frac{1}{r}a'(r)=-2e^2v^2(n-a(r)).
\end{align}
Imposing the initial condition $a(0)=\alpha/2\pi$ corresponding to \eqref{AH} and demanding a real solution gives 
\begin{align}
    a_{n,\alpha}(r) = n+\frac{evr}{\sqrt{2}\pi} (\alpha-2\pi n) K_1(\sqrt{2}evr)
\end{align}
where $K_1(x)$ is a Bessel function. By checking the asymptotics, one can verify that this solution interpolates between a holonomy $\alpha$ as $r\to 0$ and a holonomy $2\pi n$ as $r\to\infty$.

If we specialize to $\alpha=\pi$, the solutions corresponding to $n=0,1$ have the lowest energy, and in fact are doubly degenerate, as one can check by observing that they are exchanged by charge conjugation (up to a gauge transformation). Indeed, charge conjugation $C$ acts by flipping the sign of the gauge field, and a gauge transformation using $\lambda=n\theta$ sends $a\to n+a$, so 
\begin{align}
    a_{0,\pi} \xrightarrow{C}-a_{0,\pi} \xrightarrow{\lambda=\theta} 1-a_{0,\pi} = a_{1,\pi}.
\end{align}
Similar comments apply to the scalar $\Phi$.
Therefore, we find semiclassically that there is a doublet of ground states in the defect Hilbert space of the $\pi$-flux vortex deep in the Higgs phase, consistent with the assertion that \eqref{AHpi} flows to a TQM$_2$ line, $\mathds{1}\oplus \mathds{1}$, i.e.\ with two superselection sectors at low energies. (This is a relative of the superconducting qubit.)

\subsection{Anomalies}

Next we need to understand if the defect~\eqref{AHpi} introduces anomalies into the system. Of course, in the absence of the defect, the model~\eqref{AHM} does not have anomalies as it can be trivially gapped by condensing $\Phi$. 

\subsubsection*{Defect 't Hooft anomalies}
There is a shortcut for determining the defect 't Hooft anomalies. One of the useful features of anomalies is that they are robust against continuous deformations; in particular, they can be computed in any phase of the model. Recall that, in the previous subsection, we found that when $m^2\gg 0$, the $\pi$-flux vortex line goes over to the $\mathbb{Z}_2^{(1)}$ electric one-form symmetry line of free Maxwell theory, which in turn is dual to the $\mathbb{Z}_2^{(1)}$ winding one-form symmetry line of the compact boson, which is a deformation of the Halon in the critical $O(2)$ model. Thus, the defect 't Hooft anomalies of the $\pi$-flux vortex agree with those of the Halon, which were computed in Equation \eqref{eqn:TxO(2)anomaly}. 

\begin{figure}
    \centering
    \begin{tikzpicture}[>=latex, line join=round, line cap=round,scale=.7]
  \def\R{0.4}  
  \def\H{6}   
  \def\h{3}    

  \draw[ultra thick] (0,\H) ellipse (\R cm and 0.25cm);

  \draw[ultra thick] (-\R,\H) -- (-\R,0);
  \draw[ultra thick] ( \R,\H) -- ( \R,0);

  \draw[ultra thick, fill=black] 
    ( \R,0) arc (0:-180:\R cm and 0.4cm) -- cycle;

  \draw[ultra thick, dashed] (0,\h) ellipse (\R + 0.5 cm and 0.25cm);

  \node[right] at (\R+0.5,\h)
    {$\displaystyle \exp\left(i \oint a\right) = -1$};

  \draw[->, ultra thick] (-1.5,1) -- (-.5,0);
  \node[left] at (-1.5,1) {Half Monopole};
\end{tikzpicture}
    \caption{The endpoints of the $\pi$ holonomy defect are half-monopoles.}
    \label{fig:halfmono}
\end{figure}
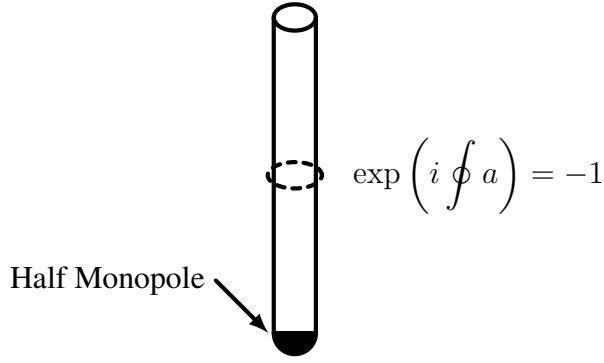

Specifically, there is a $\mathbb{Z}_2$-valued pure $O(2)$ anomaly, but no time reversal anomaly and no mixed anomaly between time reversal and $O(2)$. The pure $O(2)$ anomaly guarantees that the local operators appearing at the endpoint of a $\pi$-flux vortex line transform under a projective representation of $O(2)$, and hence  are degenerate. In particular, we expect a doublet of ``half-monopoles'', i.e.\ monopole operators which have charges $\pm\sfrac12$ under the  $U(1)_m^{(0)}$ symmetry (see Figure \ref{fig:halfmono}).

We can also directly compute the anomalies by understanding how $O(2)\times\mathbb{Z}_2^T$ acts on the endpoints. Thus we represent the endpoints $\mathcal{M}_{n-\frac12}$ of the defect  by Wu-Yang monopoles in the path integral. Then we can immediately see that 
\begin{align}
    U_\eta \cdot \mathcal{M}_{n-\frac12}=e^{i\eta(n-\frac12)}\mathcal{M}_{n-\frac12}
\end{align}
where $U_\eta\cdot \mathcal{M}_{n-\frac12}$ refers to the operation of surrounding the monopole with a spherical surface operator. 
This is just the statement that the endpoint must have half-integer monopole charge.
Furthermore, we see that under charge conjugation and time-reversal symmetry the operators  transform as e.g.\ 
\begin{align}
    C\cdot \mathcal{M}_{\frac12}=\mathcal{M}_{-\frac12}, \ \ \ \ \mathcal{T}\cdot \mathcal{M}_{\frac12} = \mathcal{M}_{-\frac12}.
\end{align}
Thus, on the two fundamental half-monopoles, the symmetries are realized by the matrix representation 
\begin{equation}\label{AHsymm} U_\eta=e^{{i\frac\eta2}  \sigma_z}~,\quad C=\sigma_x~,\quad \mathcal{ T} =\sigma_x K~.  \end{equation} 
Note that this is exactly the matrix representation we found in the Halon, Equation \eqref{Halsymm}, after a redefinition of charge conjugation and time reversal by a factor of $i$, which of course does not change the cocycle in $H^2(\mathbb{Z}_2^T\times O(2),U_T(1))$. Thus, we see by direct computation that the anomalies agree between the Halon and the $\pi$-flux vortex line.

\subsubsection*{Anomalies in the space of defect couplings}
Recall that the Halon has a deformation $\gamma \int d\tau S_z(\tau)$ which drives the defect to a trivial phase with a jump in the $U(1)$ quantum numbers by one unit as we go from $\gamma=-\infty$ to $\gamma=\infty$. 
Similarly, we can deform our vortex loop by the holonomy $\delta \alpha$, and, depending on the sign of the deformation, the flow proceeds to total holonomy $0~\mathrm{mod}~2\pi$. These differ by a properly quantized monopole operator  and hence exhibit the same jump in $U(1)$ quantum numbers. In the language we have been using, both sides possess a mixed anomaly between their bulk $U(1)$ symmetries and their defect coupling constants.

Let us show in more detail that, on the Abelian-Higgs side, there is a mixed anomaly between the $U(1)^{(0)}_m$ magnetic zero-form symmetry and the holonomy $\alpha$, thought of as a defect coupling constant, which mirrors the anomaly \eqref{ratio} of the Halon. This will provide further evidence for the conjecture that the deformation $\gamma$ in the Halon model maps to the holonomy $\alpha$ under particle/vortex duality.

To reveal the anomaly, we must couple to a background gauge field $A$ for the $U(1)_m^{(0)}$ magnetic zero-form symmetry. We claim that the correct coupling in the presence of an $\alpha$-flux vortex line supported on $L^{(1)}$ is
\begin{align}
    S[A]=\int_{M^{(3)}}d^3x~\mathcal{L}+ \frac{i}{2\pi} \int_{M^{(3)}} f\wedge A+\frac{i}{2\pi}\alpha\int_{L^{(1)}}A, \quad \lim_{l\to 0}\exp\left(i\oint_{C_l}a\right)=e^{i\alpha}.
\end{align}
The second term is the standard coupling of the magnetic symmetry.  The third term proportional to $\alpha$ must be added to cancel boundary contributions coming from the vortex line to the variation of the action with respect to gauge transformations of $A$. (See Appendix \ref{subsec:vortexanalog} for a parallel discussion in the case of the compact boson.) 

The addition of this term is directly responsible for the violation of the $2\pi$ periodicity of the defect coupling $\alpha$ in the presence of a non-trivial background gauge field $A$, 
\begin{align}\label{eqn:defectcouplinganomalyvortexline}
    Z_{\alpha+2\pi}[A] = Z_\alpha[A]\exp\left(-i \int_{L^{(1)}}A\right),
\end{align}
which is the hallmark signature of an anomaly in the space of defect couplings.

We can also detect this same anomaly by allowing $\alpha(\tau)$ to have a $\tau$-dependent profile. If we put the theory on $M^{(2)}\times S^1_\tau$, we can imagine wrapping a topological surface along the $M^{(2)}$ factor at some fixed time $\tau$, and then sweeping the operator one full revolution along the thermal circle. The same argument we used to show the invariance of the vortex line (with $\tau$-\emph{independent} holonomy $\alpha$) under the magnetic zero-form symmetry can be adapted to the case of a $\tau$-dependent holonomy. The crucial difference is that now, Equation \eqref{eqn:northsouth} is replaced with
\begin{align}\label{eqn:taudepholonomy}
\begin{split}
    &\int_{\delta \Sigma^{(2)}}f=\oint_Na-\oint_Sa=-\alpha(\tau_N)+\alpha(\tau_S) \\
    & \hspace{.3in}\implies U_\alpha[\Sigma_+^{(2)}]U_\alpha[\Sigma_-^{(2)}]^{-1}=\exp\left(-i\frac{\eta}{2\pi}(\alpha(\tau_N)-\alpha(\tau_S))\right).
\end{split}
\end{align}
In particular, if we integrate Equation \eqref{eqn:taudepholonomy} all the way around the thermal circle, we find that 
\begin{align}
    Z_{\alpha(\tau)}\to \exp\left(-i\frac{\eta}{2\pi}\int d\tau \dot\alpha \right)Z_{\alpha(\tau)}.
\end{align}
In other words, the magnetic zero-form symmetry is violated in the presence of a vortex line with holonomy $\alpha(\tau)$ which winds around the thermal circle, again signaling an anomaly in the space of defect couplings. 

One interesting consequence of this anomaly, which parallels the discussion in Section \ref{subsubsec:physicalconsequences} on boundary-changing local operators in the compact boson theory, is the following. If we give the holonomy a time dependent profile $\alpha_\epsilon(\tau)$ as in \eqref{eqn:spatialprofiletheta}, then we obtain a defect local operator supported on the vortex line with holonomy $\alpha_0$. The anomaly in the space of defect couplings, \eqref{eqn:defectcouplinganomalyvortexline}, ensures that this operator must have charge $k$ under the bulk $U(1)^{(0)}_m$ magnetic zero-form symmetry, and hence must define a non-trivial defect local operator. In particular, it must be a monopole operator.

\subsection{Lattice}\label{subsec:vortexlattice}

Let us conclude by writing down  concrete lattice models which we expect flow to the $\pi$-flux vortex line, and which exhibit some aspects of the discussion above explicitly. In particular, these considerations will lead us to a lattice scale argument in favor of spin-flux duality.

\subsubsection{A model for vortex lines}
We require a lattice model that realizes the $U(1)^{(0)}_m$ magnetic zero-form symmetry. To this end, a Villain formulation of 2+1d free Maxwell theory is useful \cite{Sulejmanpasic:2019ytl,Fazza:2022fss}. In the continuum, the Villain construction corresponds to obtaining Maxwell theory by starting with $\mathbb{R}$ gauge theory and gauging a $\mathbb{Z}^{(1)}$ one-form symmetry. We start by reviewing this model of pure gauge theory, and then explain how to incorporate matter shortly. 

We work on a 2d square lattice endowed with an arbitrary choice of orientation\footnote{For example, one can specify an orientation by placing an arrow on each link which points up on the vertical links and points to the right on all the horizontal links.} and place an
$\mathbb{R}$-valued degree of freedom $a_l$ on each link $l$, calling its conjugate momentum $\pi_l$. Additionally we place a $\mathbb{Z}$-valued degree of freedom $n_p$ on each plaquette $p$ and call its conjugate momentum $\phi_p$. The Hamiltonian is then 
\begin{align}
    H=\sum_l \frac{\pi_l^2}{2\beta a^2} + \sum_p \frac{\beta}{2a^2}[(da)_p+2\pi n_p]^2,\quad \ \ \  (da)_p\equiv \sum_{l\in \partial p}a_l. \label{eqn:plaquetteterm}
\end{align}
 In the expression for $(da)_p$, one should imagine traversing the boundary links of the plaquette $p$ in a counter-clockwise direction and adding $a_l$ with a $+$ or $-$ sign depending on whether one is going in the same or opposite direction as the arrow on $l$ which specifies its orientation. We impose the standard Gauss law constraint
\begin{align}
    \sum_{l\in\mathrm{star}(v)}\pi_l=0,
\end{align}
where, for each lattice site $v$, the sum is over links $l$ which meet $v$. We understand $\pi_l$ as being added with a $+$ sign if the orientation arrow on the link $l$ points towards $v$, and with a $-$ sign if it points away. We also impose a second Gauss law constraint corresponding to the $\mathbb{Z}^{(1)}$ gauging, 
\begin{align}
   e^{2\pi i\left[ \pi_{\langle p,p'\rangle}+ \frac1{2\pi}(\phi_{p'}-\phi_p)\right]}=1,
\end{align}
where $\langle p,p'\rangle$ is the link which is adjacent to two plaquettes $p$ and $p'$ (and we assume $p'$ is to the right of $p$ with respect to the orientation on the link $\langle p,p'\rangle$). This defines our model in the bulk.

There is a charge conjugation symmetry which acts
on operators as 
\begin{align}\label{eqn:latcc}
    Ca_l C^\dagger = -a_l, \ \ C\pi_l C^\dagger =-\pi_l, \ \ Cn_p C^\dagger =-n_p, \ \ C \phi_p C^\dagger =-\phi_p.
\end{align}
Furthermore, as anticipated, this model manifests the $U(1)_m^{(0)}$ magnetic symmetry. Indeed, the generator $Q$ of $U(1)_m^{(0)}$  in this picture is given by 
\begin{align}
    Q=\sum_p n_p, \ \ \ \ \ \ \ \  U_\eta \equiv e^{i\eta Q}.
\end{align}

Now let us add the $\alpha$-flux vortex line at the plaquette $p_0$. This can be implemented by taking the defect Hamiltonian to have modified flux terms at the plaquette $p_0$,
\begin{align}
    H_{\mathcal{D}^\alpha_{p_0}} = \sum_l \frac{\pi_l^2}{2\beta a^2} + \sum_{p} \frac{\beta}{2a^2}[(da)_p+2\pi n_p- \alpha \delta_{p,p_0}]^2. \label{eq:villain-mod1}
\end{align}
The intuition is that this Hamiltonian makes it energetically favorable for gauge field configurations to have flux $e^{i (da)_{p_0}}=e^{i\alpha}$. 

As we stated earlier, in pure Maxwell theory, vortex lines are topological. In this lattice implementation, this corresponds to the fact that there exists a local unitary operator supported near the location of the defect which commutes with all the Gauss law constraints,
\begin{align}
    V=e^{i\alpha\pi_{\langle p_0,p_1\rangle}},
\end{align}
and which has the effect of transporting the vortex line from plaquette $p_0$ to a neighboring plaquette $p_1$. Indeed, one sees that 
\begin{align}
    VH_{\mathcal{D}^\alpha_{p_0}}V^\dagger = H_{\mathcal{D}^\alpha_{p_1}}.
\end{align}

These vortex lines preserve the $U(1)_m^{(0)}$ symmetry without requiring any modification of the operator $U_\eta$ near the location of the defect, 
\begin{align}
    [H_{\mathcal{D}_{p_0}^\alpha},U_\eta]=0.
\end{align} 
On the other hand, charge conjugation flips the flux of the vortex, 
\begin{align}
    CH_{\mathcal{D}_{p_0}^\alpha}C^\dagger = H_{\mathcal{D}_{p_0}^{-\alpha}},
\end{align}
as expected.
If we specialize to $\alpha=\pi$, then we can dress charge conjugation by a shift operator $S_{p_0}=e^{-i\phi_{p_0}}$, which acts as 
\begin{align}
    S_{p_0}n_{p_0} S_{p_0}^\dagger = n_{p_0}-1
\end{align}
to define a notion of charge conjugation 
\begin{align}
    \mathcal{C}=S_{p_0}C
\end{align}
which is preserved in the presence of the $\pi$-flux vortex line, i.e.\ $[\mathcal{C},H_{\mathcal{D}^\pi_{p_0}}]=0$.

It is straightforward to check that the bulk $O(2)$ possesses a defect 't Hooft anomaly in the presence of the $\pi$-flux vortex line. It is simplest to check this on the $\mathbb{Z}_2\times\mathbb{Z}_2$ subgroup generated by $\mathcal{C}$ and $U_\pi$, which satisfy the commutation relations 
\begin{align}\label{eqn:Z2xZ2projrep}
    U_\pi^2=\mathcal{C}^2=1, \quad U_\pi \mathcal{C}=-\mathcal{C}U_\pi.
\end{align}
This realizes a projective representation of $\mathbb{Z}_2\times\mathbb{Z}_2$ corresponding to the non-trivial class in $H^2(\mathbb{Z}_2\times\mathbb{Z}_2,U(1))\cong \mathbb{Z}_2$. Thus, there is a defect 't Hooft anomaly which guarantees that the defect ground space is degenerate. 

To summarize, we have defined $\alpha$-vortex defects on the lattice in pure Maxwell theory. We have shown that they are  topological defects, and that they preserve the $U(1)_m^{(0)}$ magnetic symmetry. Furthermore,  at $\alpha=\pi$, the symmetry  enhances to $O(2)$, which is realized with a defect 't Hooft anomaly. These are exactly the properties predicted from our continuum analysis, and we see that they are manifest in the above lattice formulation.

The incorporation of matter leaves essentially all of the preceding discussion intact. We simply add the following matter term to the Hamiltonian, 
\begin{align}
    \widetilde{H}_{\mathcal{D}_{p_0}^\alpha} = H_{\mathcal{D}_{p_0}^\alpha}+\sum_{\langle v,v'\rangle}\left(\Phi^\ast_{v'}e^{iA_{\langle v,v'\rangle}}\Phi_v+\mathrm{c.c.}\right)+\sum_v\left(K\, \Pi_v^\ast\, \Pi_v+ V(\Phi_v^\ast\, \Phi_v)\right),  
\end{align}
while appropriately modifying one of the Gauss law constraints, 
\begin{align}
    \sum_{l\in\mathrm{star}(v)}\pi_l = i \sum_v \left[\Phi_v\, \Pi^\ast_v  - \Phi^*_v\, \Pi_v \right].
\end{align}
Here, $\Phi_v$ is a complex degree of freedom placed on the site $v$, and $\Pi_v$ is its conjugate momentum, satisfying $\left[\Phi_v, \Pi^\ast_{v'}\right] = i \delta_{v,v'}$. The coupling $K$ is a constant, and $V(\Phi_v^\ast\, \Phi_v)$ is some potential for the matter fields.

We expect that if $\widetilde{H}_{\mathcal{D}^\pi_{p_0}}$ is tuned to its critical point in the bulk, then the lattice model will flow to the conformal $\pi$-flux vortex line of the Abelian-Higgs model. It would be interesting to use this lattice model, perhaps in conjunction with the lattice model \eqref{eqn:O(2)latticemodelspinimpurity} for the Halon, to perform numerical checks of our proposed spin-flux duality.

\subsubsection{Coupling a spin-impurity to monopole operators}
In the continuum, particle-vortex duality between the $O(2)$ model and the Abelian-Higgs model exchanges 
\begin{align}\label{eqn:particlemonopole}
    \Phi(x) \leftrightarrow \mathcal{M}(x).
\end{align}
Here, $\Phi$ is the scalar appearing in the $O(2)$ model, and $\mathcal{M}(x)$ is a monopole operator of Abelian-Higgs. The latter is a disorder operator, defined in the path integral by integrating only over gauge field configurations which support a Dirac monopole at $x$, i.e.\ gauge fields $a$ for which
\begin{align}
    \frac{1}{2\pi}\int_{S^2}f=1
\end{align}
for any two-sphere $S^2$ surrounding the point $x$.

A naive application of Equation \eqref{eqn:particlemonopole}  would thus predict that particle-vortex duality should map the Halon of the $O(2)$ model to a spin impurity which is coupled to the fundamental monopole operator of Abelian-Higgs, i.e.\ 
\begin{align}\label{eqn:spinimpurityAH}
    S=\int d^3x~ \mathcal{L} + h\int d\tau [\mathcal{M}(\vec{x}=0,\tau)S_-(\tau)+\text{c.c.}]+\gamma \int d\tau S_z(\tau),
\end{align}
where $\mathcal{L}$ is the Lagrangian in \eqref{AHM}. Thus, if one could establish that the defect \eqref{eqn:spinimpurityAH} is infrared dual to the $\alpha$ vortex line in \eqref{AH} (with $\alpha=\pi$ corresponding to $\gamma=0$), then spin-flux duality would follow (at least if one assumes particle-vortex duality in the bulk).

Let us try to argue for the equivalence of \eqref{AH} and \eqref{eqn:spinimpurityAH} on the lattice. We go back to free Maxwell for ease of exposition, as the addition of matter does not affect the arguments which follow. We model \eqref{eqn:spinimpurityAH} with the following lattice Hamiltonian,
\begin{gather}
    H'_{\mathcal{D}^\gamma_{p_0}} = \sum_l \frac{\pi_l^2}{2\beta a^2} + \sum_p \frac{\beta}{2a^2}\left[(da)_p + 2\pi n_p\right]^2 + h\left(S_+ e^{- i \phi_{p_0}} + S_- e^{i\phi_{p_0}}\right)+\gamma S_z,
\end{gather}
where here, $S_+,S_-,S_z$ are the spin operators of a spin-$\sfrac12$ impurity coupled to the monopole operators $e^{\pm i\phi_{p_0}}$ at the plaquette $p_0$. (That $e^{\pm i \phi_{p_0}}$ indeed correspond to monopole operators in the continuum follows because they are charged under the operator $U_\eta$ implementing the $U(1)_m^{(0)}$ magnetic symmetry.)

First, let us analyze the symmetries and anomalies, and see that they match those of   vortex lines. The bulk $U(1)_m^{(0)}$ symmetry is preserved if one performs a simultaneous rotation of the internal spin of the impurity, 
\begin{align}
    U_\eta' = \exp \left(i \eta\big(S_z+ \sum_p n_p\big)\right).
\end{align}
Moreover, at $\gamma=0$, the symmetry enhances to $O(2)$ if one defines charge conjugation as 
\begin{align}
    \mathcal{C}'=S_xC
\end{align}
where $C$ was defined in \eqref{eqn:latcc}. We can again examine the $\mathbb{Z}_2\times\mathbb{Z}_2$ subgroup generated by $U'_\pi$ and $\mathcal{C}'$ to see that these symmetries are realized with a defect 't Hooft anomaly, 
\begin{gather}
    (\mathcal{C}')^2=1, \ \ \ \ (U'_\pi)^2=-1, \ \ \ \ \mathcal{C}' U'_\pi = - \mathcal{C}'U'_\pi.
\end{gather}
Indeed,  the symmetries realize exactly the same cocycle in $H^2\left(\mathbb{Z}_2 \times \mathbb{Z}_2, U(1)\right)$ as \eqref{eqn:Z2xZ2projrep} after redefining $U'_\pi \to iU'_\pi$.

Now, our claim is that, in the limit that $h\to \infty$, the Hamiltonian $H'_{\mathcal{D}_{p_0}^{\gamma=0}}$ goes over   to \eqref{eq:villain-mod1} with $\alpha=\pi$. To show this, we write down a low energy effective Hamiltonian in perturbation theory, by
splitting
\begin{align}
    H'_{\mathcal{D}^{\gamma=0}_{p_0}}=H_{\mathrm{bulk}}+H_{\mathrm{imp}}, \ \ \ \ \ \ H_{\mathrm{imp}}=h(S_+e^{-i\phi_{p_0}}+S_-e^{i\phi_{p_0}}) + \gamma S_z,
\end{align}
and treating the bulk Hamiltonian as a perturbation around the impurity Hamiltonian. The eigenvectors/eigenvalues of the latter are of the form
\begin{gather}
    |\Psi\rangle\otimes\ket{n_{p_0},\pm}, \quad E_{n_{p_0},\pm} = \pm \sqrt{h^2 + \gamma^2},
\end{gather}
where $|\Psi\rangle$ is an arbitrary wavefunction for the system on the links and plaquettes other than $p_0$, and we have defined 
\begin{align}
   \ket{n_{p_0},+}&\equiv \cos \frac{\theta}{2} \,\ket{n_{p_0},\uparrow} + \sin  \frac{\theta}{2}  \, \ket{n_{p_0}+1,\downarrow} \\
    \ket{n_{p_0},-}&\equiv -\sin   \frac{\theta}{2}  \,\ket{n_{p_0},\uparrow} + \cos  \frac{\theta}{2}  \, \ket{n_{p_0}+1,\downarrow}, \quad \tan \theta = \frac{h}{\gamma}.
\end{align}
Thus, at low energies, the states $|\Psi\rangle\otimes |n_{p_0},+\rangle$ are gapped out and we obtain a low energy effective Hamiltonian which acts within the subspace $\mathcal{H}_{\mathrm{eff}}$ spanned by the states $|\Psi\rangle \otimes |n_{p_0},-\rangle$. 

Let $P = \ket{n_{p_0}, -} \otimes \bra{n_{p_0,+}}$ be the projector onto this subspace $\mathcal{H}_{\mathrm{eff}}$. Then to first order in perturbation theory, the low energy effective Hamiltonian is given by the expression 
\begin{align}
    H_{\mathrm{eff}} = PH_{\mathrm{bulk}}P + \mathcal{O}\left(h^{-1}\right).
\end{align}
To write down an expression for $H_{\mathrm{eff}}$, we introduce a new operator $\widetilde{n}_{p_0}$ which is defined to act as
\begin{gather}
    \widetilde{n}_{p_0} |\Psi\rangle \otimes \ket{n_{p_0}, \pm} = n_{p_0}|\Psi\rangle \otimes \ket{n_{p_0},\pm}, \quad [\widetilde{n}_{p_0}, \phi_{p_0}] = i.
\end{gather}
In terms of $\widetilde{n}_{p_0}$, we find that
\begin{align}
\begin{split}
    &H_{\mathrm{eff}} = \sum_l \frac{\pi_l^2}{2\beta a^2} + \sum_{p\neq p_0} \frac{\beta}{2a^2}\left[(da)_p + 2\pi n_p \right]^2  \\&\hspace{.75in}
    +\frac{\beta}{2a^2}\left[(da)_{p_0} + 2\pi \tilde{n}_{p_0} +  2 \pi \sin^2 \frac{\theta}{2} \right]^2 + \frac{\beta\pi^2}{2a^2}+\mathcal{O}\left(h^{-1}\right).
\end{split}
\end{align}
Thus, up to an overall zero-point energy, the low energy effective Hamiltonian $H_{\mathrm{eff}}$ is unitarily equivalent to the vortex line $H_{\mathcal{D}_{p_0}}^{\alpha}$ in Equation \eqref{eq:villain-mod1}. For $\gamma = 0$ we get $\theta = \frac{\pi}{2}$ and $\alpha = \pi$. More generally, we find that $\alpha = \pi - \frac{\pi \gamma}{\sqrt{h^2+\gamma^2}}$.

We caution that this does not actually prove the equivalence of \eqref{AH} and \eqref{eqn:spinimpurityAH} in the continuum. Indeed, our derivation required taking $h$ much larger than the rest of the couplings in the Hamiltonian, whereas taking the continuum limit requires working in essentially the opposite regime. Rather, this derivation should be thought of as providing a heuristic lattice scale picture for the equivalence.

\section{Future directions}

We conclude this paper with a number of directions for future research. 

\begin{enumerate}
    \item It is interesting to ask what the spin impurities discussed in Section \ref{sec:magnets} with $s>\sfrac12$ flow to in the critical XY model. We have described various anomalies which furnish some constraints, but they are not powerful enough, outside of $s=\sfrac12$, to uniquely determine the character of the low energy physics. In the $O(3)$ model, it is known from the large $s$ expansion \cite{Cuomo:2022xgw} that spin impurities eventually become conformal for large enough $s$; when taken together with the fact that the $s=\sfrac12$ Boson-Kondo defect is conformal, this gives compelling evidence that spin impurities are conformal for all $s$. Can a similar large $s$ expansion be developed for the $O(2)$ model?
    \item It will surely be fruitful to discover further examples of systems with anomalies in the space of their defect couplings. One  rich well of examples, to which we hope to return in the future, is surfaces in 3+1d non-Abelian gauge theories. 
    \item It would be interesting to develop a general theory of anomalies in the space of defect couplings. For example, all of the models we studied involved symmetries which are non-anomalous in the bulk. Are there any non-trivial interactions between bulk and defect anomalies?
    \item Our various proposals, like spin-flux duality, should be tested numerically. To this end, we hope that the lattice models we have recorded in the main text provide a useful starting point for those who wish to explore these topics further.

\item 
It is interesting to understand the transformation rules of conformal line operators in various 3d dualities.
In the planar limit of Chern-Simons-Matter dualities, the mapping between certain conformal line operators was worked out in~\cite{Gabai:2022vri,Gabai:2022mya,Gabai:2023lax,Ferrando:2025ufj}. 
Inspired by the spin-flux duality, we hope to extend the dictionary of conformal line operators beyond the planar limit.

\end{enumerate}

\section*{Acknowledgements}

We thank Matthew Buican, Yichul Choi, Diego Delmastro, Justin Kulp, Ho Tat Lam, Lorenzo Di Pietro, Nathan Seiberg, Sahand Seifnashri, Adar Sharon, Yifan Wang, and Fei Yan for valuable discussions. We especially thank Max Metlitski for many invaluable discussions and comments on a draft. BR gratefully acknowledges NSF grant PHY-2210533. Z.K. gratefully acknowledges  NSF Award Number 2310283.

\appendix

\section{Further Examples of Anomalies in the Space of Defect Couplings}\label{app:furtherexamples}

\subsection{An Anomaly in the Boundary Theta Angle of 2+1d Abelian Gauge Theories}\label{subsec:2+1dboundaries}

Much of the discussion in Section \ref{subsec:anomaliesinspaceofdefectcouplings} can be extended to a large class of theories in higher dimensions. For concreteness, we telegraphically sketch how the main ideas apply to Abelian gauge theories in 2+1d.

For concreteness, we work with free Maxwell theory, 
\begin{align}
    S[a]=\frac{1}{4e^2}\int d^3x f_{\mu\nu}^2,\quad f_{\mu\nu}=\partial_\mu a_\nu - \partial_\nu a_\mu,
\end{align}
though many features persist if we couple to certain matter content (see below). This theory has a $U(1)_e^{(1)}$ electric one-form symmetry generated by topological line operators of the form
\begin{align}
    U(1)_e^{(1)}: \ \mathcal{L}_\alpha^{(e)}[L^{(1)}]=\exp\left(\frac{i}{e^2}\alpha\int_{L^{(1)}} \star f\right),
\end{align}
a magnetic $U(1)^{(0)}_m$ zero-form symmetry generated by topological surfaces of the form 
\begin{align}
    U(1)^{(0)}_m: \ \mathcal{L}^{(m)}_\eta[\Sigma^{(2)}]=\exp \left(\frac{i}{2\pi}\eta\int_{\Sigma^{(2)}}f\right),
\end{align}
and a charge conjugation symmetry which acts as
\begin{align}
\mathbb{Z}_2^C: \ a\to -a.
\end{align}

Denoting spacetime coordinates with $\tau$, $x$, $y$, we place a Neumann boundary condition at $x=0$, 
\begin{align}
    f_{x\mu}\vert_{x=0} = 0, \ \  \ \ \ \ \mu=\tau,y,
\end{align}
and deform the action by a boundary theta angle, 
\begin{align}
S_\eta[a]=\frac{1}{2e^2}\int f\wedge \star f - \frac{i}{2\pi} \int_{x=0}\eta\, f.
\end{align}
We note that $\eta$ is a $2\pi$-periodic variable.

Neumann boundaries do not respect the $U(1)^{(0)}_m$ magnetic symmetry; indeed one can easily see that fusing the topological operator $\mathcal{L}^{(m)}_{\eta'}$ onto the boundary shifts the boundary theta angle as $\eta\mapsto \eta+\eta'$. However, the $U(1)^{(1)}_e$ electric one-form symmetry is preserved by Neumann boundaries, in the sense that the topological line $\mathcal{L}^{(e)}_\alpha$ is able to terminate topologically on them. Charge conjugation $\mathbb{Z}_2^C$ acts as $\eta\mapsto -\eta$ on the boundary theta angle, and hence only the Neumann boundaries with $\eta=0,\pi$ are $C$-symmetric. See \cite{Choi:2023xjw} for a related discussion.

We claim that there is an anomaly in the space of boundary couplings between the bulk $U(1)^{(1)}_e$ electric one-form symmetry and the $2\pi$-periodicity of the boundary theta angle $\eta$. The easiest way to see this anomaly is to couple to a background two-form gauge field $A$ for $U(1)_e^{(1)}$, 
\begin{align}
    S_\eta[a,A]=\frac1{2e^2}\int_{M^{(3)}}(f-A)\wedge \star (f-A)-\frac{i}{2\pi}\int_{\partial M^{(2)}}\eta (f-A),
\end{align}
which leads to a violation of the $2\pi$-periodicity of $\eta$,
\begin{align}
    Z_{\eta+2\pi}[A]=Z_\eta[A]\exp\left(-i \int_{\partial M^{(2)}}A\right), \ \ \ \ \ \ Z_\eta[A]=\int\mathcal{D}a ~e^{-S_\eta[a,A]}.
\end{align}
One could also allow $\eta$ to vary with the boundary coordinates and check that the $U(1)^{(1)}_e$ electric one-form symmetry is broken, following the discussion in Section 4.1.1 of \cite{Cordova:2019jnf}.

Just as in the free boson, this anomaly is robust to deformations which break the $U(1)^{(1)}_e$ symmetry down to a $\mathbb{Z}_p^{(1)}$ subgroup, like the addition of dynamical charge $p$ matter. Neither the magnetic zero-form symmetry nor charge conjugation are needed to enjoy the physical consequences of the anomaly. 

One implication is that line operators supported on the boundary, which are defined by allowing $\eta$ to smoothly interpolate from $\eta_0$ to $\eta_0-2\pi k$ near a 1-dimensional locus, should be non-trivial because they carry charge under the bulk $U(1)^{(1)}_e$ symmetry (or a $\mathbb{Z}_p^{(1)}$ subgroup if it is partially broken). Indeed, one can easily check that if one uses $\eta_\epsilon(\tau)$ as defined in Equation \eqref{eqn:spatialprofiletheta}, then one obtains 
\begin{align}
    Z_{\eta_\epsilon(\tau)}\xrightarrow{\epsilon\ll 0} \int\mathcal{D}a ~e^{ik\int_{x,\tau=0} dy a_y}~e^{-S_{\eta_0}[a]}.
\end{align}
That is, we obtain a boundary Wilson line whose charge is equal to the number of times that the boundary theta angle winds.

Another implication of the anomaly is that the dimensional reduction on an interval with $\mathrm{N}[0]$ imposed at one end of the interval and $\mathrm{N}[\eta]$ imposed at the other should flow to a non-trivial 2d QFT for some value of $\eta$. In special cases, we can use the charge conjugation symmetry to argue that $\eta=\pi$ is one such special value. Indeed, working in free Maxwell for simplicity, one can check that there is a mixed anomaly between $U(1)^{(1)}_e$ and $C$ on the boundary. It can be diagnosed by coupling to a background gauge field for $U(1)^{(1)}_e$ and performing a charge conjugation transformation, wherein one finds 
\begin{align}
    Z_\pi[A]\xrightarrow{C} Z_\pi[-A]\exp\left(i\int_{\partial M^{(2)}}A\right).
\end{align}
The interval reduction of free Maxwell is 2d Abelian gauge theory with theta angle $\eta$, which is known to spontaneously break charge conjugation symmetry at $\eta=\pi$.

\subsection{An Analog of Vortex Lines in the 1+1d Compact Boson}\label{subsec:vortexanalog}

In this section, we briefly discuss a simple toy model which provides helpful orientation for our treatment of vortex lines in 2+1d Abelian gauge theories in Section \ref{sec:vortexlines}.

Let us again consider the 1+1d compact boson, 
\begin{align}
    S[\phi] = \frac{R^2}{8\pi} \int d\phi \wedge *d\phi 
\end{align}
with period $\phi\cong \phi+2\pi$. We insert a line operator in the theory at $x=0$ which is defined by imposing the following singular boundary conditions,
\begin{align}\label{eqn:vortexlineanalog}
    \lim_{x\to 0+}\left(\phi(\tau,x)-\phi(\tau,-x)\right) = \alpha~\mathrm{mod}~2\pi,
\end{align}
where $\lim_{x\to 0^+}$ denotes the limit as $x$ approaches $0$ from the right. This is analogous to vortex lines, where one imposes a boundary condition on the integral of the gauge field around the line; in the present context, the integral is 0-dimensional and degenerates to the difference between the values of $\phi$ on either side of the line.

Another way to express this line defect, which we will often use as a shorthand, is
\begin{align}
    d\phi = \alpha \delta(x)+\cdots.
\end{align}
This formulation echoes how Gukov-Witten operators are often written, see e.g.\ Equation (2.5) of \cite{Gukov:2006jk}. Of course, in the free boson theory, the line defect \eqref{eqn:vortexlineanalog} is simply a $U(1)^{(0)}_m$ momentum symmetry line, as it implements the shift symmetry of $\phi$, however, this identification does not survive in the presence of interactions. 

Now, we would like to show that there is a mixed anomaly between the bulk $U(1)_w^{(0)}$ winding symmetry and the $2\pi$-periodic defect coupling constant $\alpha$. To reveal this anomaly, we must couple to a background gauge field $A$. In the absence of the line defect, it is well-known (see e.g.\ \cite{Turner:2019wnh} for a pedagogical explanation) that this is achieved by adding the following terms to the action, 
\begin{align}
    S[A] = \frac{R^2}{8\pi} \int d \phi \wedge *d\phi+ \frac{i}{2\pi} \int_{M^{(2)}} d\phi \wedge A.
\end{align}
This is invariant under background gauge transformations because 
\begin{align}
    S[A+d\lambda]-S[A] = \frac{i}{2\pi}\int_{M^{(2)}}d\phi \wedge d\lambda = -\frac{i}{2\pi} \int_{M^{(2)}}\phi \wedge dd\lambda =0.
\end{align}
However, in the presence of \eqref{eqn:vortexlineanalog}, we incur boundary contributions to the variation coming from the line defect when we integrate by parts,  
\begin{align}
    \int_{M^{(2)}}d\phi \wedge d\lambda = \int d\tau \left[\phi(\tau,0^-)-\phi(\tau,0^+)\right]\partial_\tau \lambda = -\alpha\int_{x=0} d\lambda .
\end{align}
Thus, to ensure gauge invariance, we must add a counterterm supported on the line to arrive at
\begin{align}
\begin{split}
   & S[A] = \frac{R^2}{8\pi}\int_{M^{(2)}}d\phi \wedge\star d\phi + \frac{i}{2\pi }\int_{M^{(2)}}d\phi \wedge A + \frac{i}{2\pi}\alpha\int_{x=0}A, \\
   & \hspace{1in} d\phi\vert_{x=0}  = \alpha \delta(x) 
\end{split}
\end{align}

We immediately see that the $2\pi$-periodicity of $\alpha$ is violated in the presence of a background gauge field $A$ for the $U(1)^{(0)}_w$ winding symmetry, 
\begin{align}
    Z_{\alpha+2\pi}[A] = Z_\alpha[A]\exp\left(  -i \int_{x=0}A\right)
\end{align}
signaling an anomaly in the space of defect couplings. The physical consequences of this anomaly are similar to those discussed in Section \ref{subsec:anomaliesinspaceofdefectcouplings}. For example, it shows that, in the presence of deformations which preserve at least a $\mathbb{Z}_n$ subgroup of $U(1)^{(0)}_w$, the line defect \eqref{eqn:vortexlineanalog} must define a non-trivial defect in the IR for at least one value of the coupling $\alpha_\ast$.

\bibliographystyle{JHEP}
\bibliography{main}

\end{document}